\documentclass[a4paper,UKenglish,cleveref]{lipics-v2021}
\usepackage{thm-restate}

\usepackage{xcolor}
\usepackage{xspace}
\usepackage{mathtools}
\crefname{observation}{observation}{observations}
\Crefname{observation}{Observation}{Observations}
\crefname{claim}{claim}{claims}
\Crefname{claim}{Claim}{Claims}
\usepackage{booktabs}
\usepackage{tabularx}
\usepackage{environ}
\usepackage[most]{tcolorbox}
\usepackage[boxed,vlined,linesnumbered,ruled]{algorithm2e}

\usepackage{tikz}
\usetikzlibrary{fit,positioning,arrows.meta}
\SetKwInput{KwInput}{Input}
\SetKwInput{KwOutput}{Output}
\SetKwInput{KwInvariant}{Invariant}
\DontPrintSemicolon
\SetKwFor{ForEach}{for each}{do}{end for}

\definecolor{anti-flashwhite}{rgb}{0.95,0.95,0.96}
\definecolor{ForestGreen}{rgb}{0.1333,0.5451,0.1333}
\definecolor{DarkRed}{rgb}{0.8,0,0}
\definecolor{Red}{rgb}{1,0,0}
\definecolor{BrickRed}{rgb}{0.8,0.25,0.33}
\definecolor{PineGreen}{rgb}{0.0,0.47,0.44}
\definecolor{Brown}{rgb}{0.59,0.29,0.0}
\definecolor{bblue}{rgb}{0.1,0.4,0.6}
\colorlet{mix}{red!50!black}

\newcommand{\cI}{\mathcal{I}}
\newcommand{\cS}{\mathcal{S}}
\newcommand{\cM}{\mathcal{M}}
\newcommand{\D}{\textsf{D}}

\makeatletter
\newcommand{\problemtitle}[1]{\gdef\@problemtitle{#1}}
\newcommand{\probleminput}[1]{\gdef\@probleminput{#1}}
\newcommand{\problemquestion}[1]{\gdef\@problemquestion{#1}}

\NewEnviron{problem}{
  \problemtitle{}\probleminput{}\problemquestion{}
  \BODY
  \par\addvspace{.5\baselineskip}
  \noindent
  \begin{tabularx}{\textwidth}{@{\hspace{\parindent}} l X c}
    \multicolumn{2}{@{\hspace{\parindent}}l}{\@problemtitle} \\
    \textbf{Input:} & \@probleminput \\
    \textbf{Question:} & \@problemquestion
  \end{tabularx}
  \par\addvspace{.5\baselineskip}
}
\makeatother

\newcommand{\flsvs}{\textsc{$\ell$-Shallow Vertex Splitting to Forest}\xspace}
\newcommand{\sflsvs}{\textsc{$\ell$-Shallow Vertex Splitting to Pseudoforest}\xspace}
\newcommand{\clsvs}{\textsc{$\ell$-Shallow Vertex Splitting to Cluster}\xspace}

\newcommand{\lsvf}{\textsc{$\ell$-Shallow Split-Vertex to Forest}\xspace}
\newcommand{\lsvpf}{\textsc{$\ell$-Shallow Split-Vertex to Pseudoforest}\xspace}
\newcommand{\lsvc}{\textsc{$\ell$-Shallow Split-Vertex to Cluster}\xspace}

\title{Matroid Parity Meets Vertex Splitting}
\titlerunning{Matroid Parity Meets Vertex Splitting}

\author{Hitendra Kumar}{Indian Institute of Technology Dharwad, India}{hitendra.kumar1729@gmail.com}{}{}
\author{Pritesh Kumar}{The Institute of Mathematical Sciences, HBNI, Chennai, India}{priteshk@imsc.res.in}{}{}
\author{Madhumita Kundu}{Ben-Gurion University, Be'er Sheva, Israel}{madhumit@post.bgu.ac.il}{}{}
\author{Saket Saurabh}{The Institute of Mathematical Sciences, HBNI, Chennai, India}{saket@imsc.res.in}{}{}
\authorrunning{H. Kumar, P. Kumar, M. Kundu, and S. Saurabh}
\Copyright{Hitendra Kumar, Pritesh Kumar, Madhumita Kundu, and Saket Saurabh}


\category{}
\relatedversion{}

\hideLIPIcs

\newcommand{\RevisionStart}{}
\newcommand{\NewStart}{}
\newcommand{\NewEnd}{\ifhmode\unskip\fi\ignorespaces}
\newcommand{\NewText}[1]{{#1}}
\newcommand{\RevisionEnd}{}
\tikzset{split vertex/.style={circle,draw=black,fill=white,inner sep=1.8pt},
  split edge/.style={draw=black,line width=.6pt},
  split marked/.style={draw=black,dashed,line width=.8pt},
  split arrow/.style={-{Stealth[length=2mm]},draw=black}}

\newcommand{\ManuscriptPath}{}
\IfFileExists{01-introduction.tex}{}{\renewcommand{\ManuscriptPath}{IPCO-full-reorganized-v01/}}

\begin{document}
\nolinenumbers

\maketitle

\begin{abstract}
A vertex split replaces a vertex by two non-adjacent descendants and partitions its incident edges between them. We study exclusive $\ell$-shallow splitting, where each original vertex has at most $\ell+1$ descendants, with the aim of obtaining a forest, a pseudoforest, or a cluster graph. We consider two natural objectives: minimizing the total number of split operations and minimizing the number of original vertices that are split.

Our main conceptual contribution is a connection between vertex splitting and matroid parity. For both forests and pseudoforests, we give a characterization of feasible splitting sequences that leads, via weighted linear matroid parity, to polynomial-time algorithms for minimizing the number of split vertices when $\ell=2$. These algorithms are complemented by NP-hardness for every fixed $\ell\ge3$; for pseudoforests, hardness holds even on graphs of maximum degree four.

The total number of split operations behaves differently: for both sparse target classes, this objective is polynomial-time solvable for every $\ell$, using matroid intersection. For cluster graphs, both objectives admit linear-time algorithms when $\ell=1$ and are NP-hard for every fixed $\ell\ge2$. Together, these results identify how the target class, the local splitting bound, and the choice of objective determine the complexity of vertex splitting.

\end{abstract}

\newpage

\section{Introduction}
\label{sec:intro}

Graph modification problems ask how to transform a given graph into a prescribed graph class using a small number of operations. Common operations include deleting vertices and deleting or adding edges. These problems have a long history, including the work of Lewis and Yannakakis on vertex deletion~\cite{LewisYannakakis1980} and of Natanzon, Shamir, and Sharan on edge modification~\cite{NatanzonShamirSharan2001}. Familiar examples include \textsc{Feedback Vertex Set}, where vertices are deleted to obtain a forest, and \textsc{Cluster Editing}, where edges are added or deleted to obtain a disjoint union of cliques. The latter belongs to the family of cluster graph modification problems studied by Shamir, Sharan, and Tsur~\cite{ShamirSharanTsur2004}. For a broader account of edge modification problems and their parameterized complexity, we refer to the survey of Crespelle, Drange, Fomin, and Golovach~\cite{CrespelleDFG2023}.

Vertex splitting offers another way to modify a graph: it replaces a vertex by two new vertices and distributes its incident edges between them. In the {\em exclusive version} studied here, the two new vertices are non-adjacent, and each edge incident with the original vertex is assigned to exactly one of them. Thus, splitting preserves the original edges while changing how they meet at vertices. We seek a sequence of splits that transforms the input graph into a forest, a pseudoforest, or a cluster graph, subject to a bound on the number of descendants of each original vertex. We consider two objectives: minimizing the total number of split operations and minimizing the number of original vertices that are split at least once.

The bound on descendants makes even feasibility a substantive question. Without this restriction, every graph can be transformed into a disjoint union of edges and isolated vertices by separating all incidences at each vertex. This graph belongs to all three target classes. With a bound on descendants, however, such a transformation may be impossible. We must choose both which vertices to split and how to distribute their incident edges.

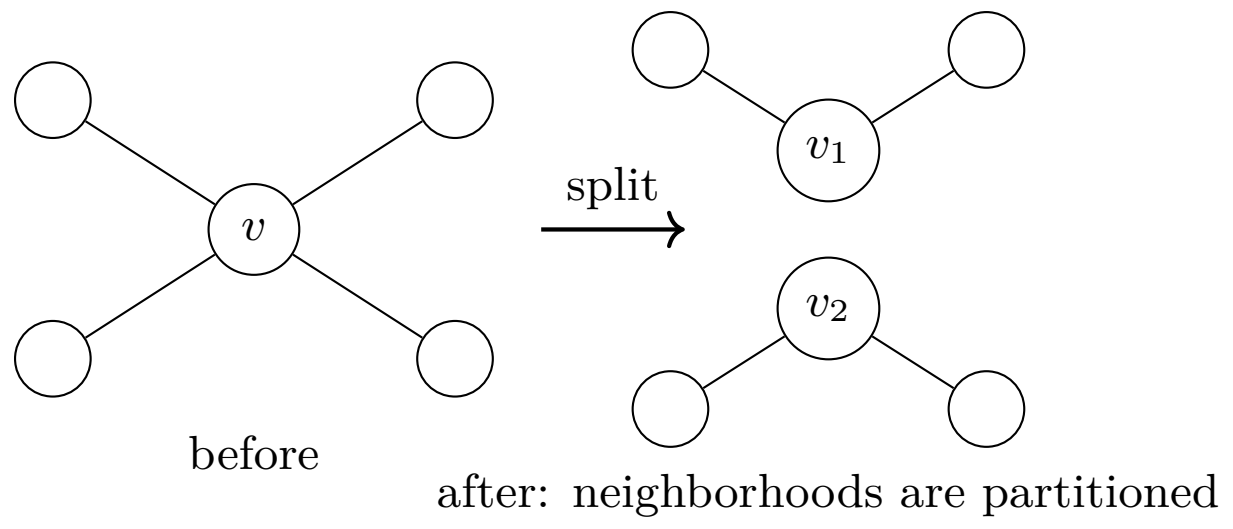
\begin{figure}[htbp]
\centering
\begin{tikzpicture}[scale=0.95, every node/.style={font=\small}]
  \node[circle,draw,minimum size=6mm] (v) at (0,0) {$v$};
  \node[circle,draw,minimum size=5mm] (a) at (-1.4,0.9) {};
  \node[circle,draw,minimum size=5mm] (b) at (-1.4,-0.9) {};
  \node[circle,draw,minimum size=5mm] (c) at (1.4,0.9) {};
  \node[circle,draw,minimum size=5mm] (d) at (1.4,-0.9) {};
  \draw (v)--(a); \draw (v)--(b); \draw (v)--(c); \draw (v)--(d);
  \node at (0,-1.55) {before};
  \draw[->,thick] (2.0,0) -- (3.0,0) node[midway,above] {split};
  \node[circle,draw,minimum size=6mm] (v1) at (4.0,0.55) {$v_1$};
  \node[circle,draw,minimum size=6mm] (v2) at (4.0,-0.55) {$v_2$};
  \node[circle,draw,minimum size=5mm] (aa) at (2.9,1.25) {};
  \node[circle,draw,minimum size=5mm] (bb) at (2.9,-1.25) {};
  \node[circle,draw,minimum size=5mm] (cc) at (5.1,1.25) {};
  \node[circle,draw,minimum size=5mm] (dd) at (5.1,-1.25) {};
  \draw (v1)--(aa); \draw (v1)--(cc);
  \draw (v2)--(bb); \draw (v2)--(dd);
  \node at (4.0,-1.85) {after: neighborhoods are partitioned};
\end{tikzpicture}
\caption{One exclusive split replaces a vertex by two non-adjacent descendants and partitions the incident edges between them.  Repeating this operation gives several descendants of the same original vertex; the $\ell$-shallow restriction allows at most $\ell+1$ such descendants.}
\label{fig:exclusive-split}
\end{figure}

\smallskip 

\noindent
{\bf The model and the two objectives.}
Fix an integer $\ell\geq0$. The vertices obtained from an original vertex through successive splits are its descendants. A splitting sequence is $\ell$-shallow if every original vertex has at most $\ell+1$ descendants at the end of the sequence. Equivalently, at most $\ell$ split operations are performed on an original vertex and its descendants in total. For a target class $\Pi$, the problem \textsc{$\ell$-Shallow Vertex Splitting to $\Pi$} asks whether a graph $G$ can be transformed into a member of $\Pi$ using at most {\em $k$ operations} in an $\ell$-shallow sequence. The problem \textsc{$\ell$-Shallow Split-Vertex to $\Pi$} instead asks whether such a sequence exists in which at most {\em $k$ original vertices} are split.

The two objectives coincide when $\ell=1$. For larger $\ell$, several operations on descendants of the same original vertex contribute several units to the first objective but only one unit to the second. Consider two triangles sharing a vertex $v$. Two operations are necessary and sufficient to obtain a forest. When $\ell=2$, both operations can be performed on $v$ and its descendants, so only one original vertex is split. When $\ell=1$, the two operations must involve different original vertices. Thus, minimizing the number of split vertices requires us to determine how far the necessary operations can be concentrated on a small set of original vertices. We study forests, pseudoforests, and cluster graphs. A pseudoforest has at most one cycle in each connected component, and a cluster graph is a disjoint union of cliques. Forests and pseudoforests allow us to examine how permitting one cycle per component affects splitting. Cluster graphs lead to a different structure, governed by partitions of the edge set into cliques.

Our main {\em conceptual contribution} is a connection between vertex splitting and matroid parity. For forests and pseudoforests, we characterize feasible splitting sequences through assignments of edges to their endpoints. Counting these assignments leads to matroid intersection. Charging each vertex only once, even when it receives two assignments, leads to weighted linear matroid parity. This distinction yields different complexity thresholds for the two objectives.

\subsection{Our results}

For both forests and pseudoforests, minimizing the number of split vertices is polynomial-time solvable when $\ell\leq2$, and NP-hard for every fixed $\ell\geq3$. Our algorithms are deterministic: they either report that no \(\ell\)-shallow splitting sequence reaches the target class or return one minimizing the number of original vertices that are split. For pseudoforests, the hardness holds even on graphs of maximum degree four.

Minimizing the total number of split operations is polynomial-time solvable for every $\ell$ for both sparse target classes. Moreover, whenever a feasible sequence exists, the optimum is determined by elementary graph parameters. Write $n=|V(G)|$, $m=|E(G)|$, and let $c(G)$ denote the number of connected components. For forests, the minimum number of operations is $m-n+c(G)$. For pseudoforests, it is $m-n+t(G)$, where $t(G)$ counts the tree components, including isolated vertices. Thus, for each target, either no \(\ell\)-shallow splitting sequence exists, or the minimum number of operations equals the corresponding value above. 

For cluster graphs, both objectives admit linear-time algorithms when $\ell=1$, and both are NP-hard for every fixed $\ell\geq2$. The results are summarized in \Cref{tab:results}.

\begin{table}[htbp]
\centering
\caption{The complexity of exclusive shallow splitting for the three targets. Hardness statements apply to every fixed value of $\ell$ in the stated range.}
\label{tab:results}
\begin{tabularx}{\textwidth}{@{}lXX@{}}
\toprule
Target & Total number of split operations & Number of split vertices \\
\midrule
Forests & Polynomial time for every $\ell$ & Polynomial time for $\ell\leq2$; NP-hard for $\ell\geq3$ \\
Pseudoforests & Polynomial time for every $\ell$ & Polynomial time for $\ell\leq2$; NP-complete for $\ell\geq3$ \\
Cluster graphs & Linear time for $\ell=1$; NP-hard for $\ell\geq2$ & Linear time for $\ell=1$; NP-hard for $\ell\geq2$ \\
\bottomrule
\end{tabularx}
\end{table}

When $\ell=0$, no split is allowed, and both problems reduce to recognizing the target class.

\subsection{From vertex splitting to matroid optimization}

We now explain how vertex splitting leads to matroid optimization. For forests and pseudoforests, we first describe a simple form of splitting sequence that suffices for both objectives. This description reduces minimizing the number of operations to matroid intersection. We then explain how matroid parity allows us to minimize the number of split vertices when $\ell=2$. Finally, we briefly describe the connection between splitting into cluster graphs and recognizing line graphs.

For the forest and pseudoforest targets, we may treat each connected component independently. We therefore assume throughout their overview that $G$ is connected, and write $n=|V(G)|$ and $m=|E(G)|$. If $G$ is a tree, no splits are needed for either target, so we may further assume that $G$ contains a cycle.

We begin with the forest target. Choose a spanning tree $T$ of $G$. For each edge $e=uv$ outside $T$, choose one endpoint, say $v$, at which to perform a split. Replace $v$ by two vertices: a new vertex $v_e$ incident only with $e$, and a vertex that retains all other edges previously incident with $v$. Thus, $e$ becomes a leaf edge joining $u$ to $v_e$. If several edges are assigned to $v$, detach them one at a time from the vertex retaining the remaining edges.

After processing every edge outside $T$, the resulting graph consists of $T$ with one new leaf attached for each such edge. It is therefore a tree, obtained using exactly $m-n+1$ splits. Each original vertex is split as many times as there are edges assigned to it, so the sequence is $\ell$-shallow precisely when no vertex receives more than $\ell$ assignments.

We prove that every feasible splitting sequence can be replaced by a sequence of this form using no more operations and splitting only original vertices that the given sequence splits. Thus, the replacement increases neither of the two costs. The restriction on which vertices are split is essential for the split-vertex objective; a bound on the number of operations alone would not suffice.

For pseudoforests, we retain a spanning pseudoforest with exactly one cycle in each component and detach every remaining edge as a leaf edge. The retained subgraph has $n$ edges, so this construction uses exactly $m-n$ splits. We prove the same guarantee: every feasible sequence can be replaced by one of this form without increasing either cost.

These characterizations separate the problem into two choices: which edges to detach, and to which endpoints they should be assigned. For the forest target, let $D$ denote the set of edges to detach. An edge set $D$ is independent in the cographic matroid of the connected graph $G$ precisely when deleting $D$ leaves $G$ connected. Its bases are therefore exactly the complements of spanning trees---the sets of edges we detach in our forest construction.

We must also assign each edge of $D$ to one of its endpoints so that no vertex receives more than $\ell$ edges. To see the matroid behind this condition, first consider $\ell=1$. Such an assignment exists precisely when $(V(G),D)$ is a pseudoforest: in a tree component, root the tree and assign each edge to its child; in a unicyclic component, assign the cycle edges cyclically and the remaining edges away from the cycle. Conversely, an assignment with at most one edge per vertex ensures that every component has at most as many edges as vertices. These edge sets are exactly the independent sets of the bicircular matroid.

For general $\ell$, a feasible assignment can be divided into $\ell$ groups, each assigning at most one edge to any vertex. Thus, the assignable edge sets are precisely those that can be partitioned into $\ell$ independent sets of the bicircular matroid---the independent sets of its $\ell$-fold union. We therefore seek a set of $m-n+1$ edges that is independent in both this matroid and the cographic matroid. Matroid intersection either finds such a set or establishes that no feasible splitting sequence exists. From the set found, we recover the endpoint assignments and construct a sequence using exactly $m-n+1$ operations, which is therefore optimal.

For the pseudoforest target, the retained edges form a spanning pseudoforest with exactly one cycle in each component. These edge sets are precisely the bases of the bicircular matroid, and their complements are the bases of its dual. We therefore replace the cographic matroid by the dual bicircular matroid and seek a common independent set of size $m-n$. The endpoint-assignment constraint is unchanged. Matroid intersection again either establishes infeasibility or yields an optimal splitting sequence, now using $m-n$ operations.

We now turn to minimizing the number of split vertices. This objective requires us to distinguish a vertex receiving no assignments from one receiving one or more. When $\ell=2$, we give each vertex two slots, one for each permitted assignment. An assignment selects an edge together with an available slot at one of its endpoints. A vertex remains unsplit precisely when both of its slots remain unused.

This is where matroid parity enters: we encode the edge assignments so that selecting a pair of unused slots rewards leaving the corresponding vertex unsplit. We construct rational representations of polynomial encoding length and apply the deterministic polynomial-time algorithm for weighted linear matroid parity of Iwata and Kobayashi~\cite{IwataKobayashi2022}. The same construction handles both sparse targets by changing the matroid governing the edge choices.

For both sparse targets, minimizing the number of split vertices is NP-hard for every fixed $\ell\ge3$. For forests, we reduce from \textsc{Feedback Vertex Set} on graphs of maximum degree four. For pseudoforests, we reduce from \textsc{Vertex Cover} on graphs of maximum degree three.

For cluster graphs, a feasible splitting sequence corresponds to a partition of the edges of $G$ into cliques, with each original vertex belonging to at most $\ell+1$ cliques. When $\ell=1$, Krausz's characterization states that such a partition exists precisely when $G$ is a line graph. We reconstruct a graph $H$ whose line graph is $G$ and use its structure to construct an optimal splitting sequence in linear time. This minimizes both objectives, since each original vertex can be split at most once. For every fixed $\ell\ge2$, reductions from the Krausz dimension problem show that both objectives are NP-hard.

\subsection{Related work}

Vertex splitting has been studied in graph drawing, clustering, and the algorithmic study of hereditary graph properties. An important distinction is whether the operation is exclusive or inclusive. In inclusive splitting, a neighbor of the split vertex may become adjacent to both new vertices; in exclusive splitting, it must become adjacent to exactly one. Firbas and Sorge study the inclusive operation and develop a framework for proving hardness of establishing hereditary graph properties by splitting~\cite{FirbasSorge2024}.

Firbas's thesis gives polynomial-time algorithms for obtaining forests, split graphs, and threshold graphs by inclusive splitting, and proves NP-completeness for several other target classes, including cluster graphs~\cite{Splitthesis}. Firbas and coauthors also study cluster vertex splitting in connection with overlapping clustering and clique covers~\cite{FirbasDHSSVW25}. Our setting combines exclusive splitting with a bound on the number of descendants and distinguishes the number of operations from the number of original vertices that are split. For forests and pseudoforests, the resulting structural characterizations allow both objectives to be expressed through matroid optimization.

Section~2 introduces the splitting model and the required matroid tools. Sections~3 and~4 develop the characterizations, algorithms, and hardness results for forests and pseudoforests, respectively. Section~5 treats cluster graphs.

\section{Preliminaries}
\label{sec:prelims}
We fix the splitting model and the notation used to count its two costs. We then recall the matroid tools used below: intersection enforces the local assignment bound, while parity will allow us to charge a vertex only once when it receives two assignments.
We work with finite simple undirected graphs.  Unless explicitly stated otherwise, graphs may be disconnected and may contain isolated vertices.  We use standard graph-theoretic notation and refer to Diestel~\cite{Diestel17} for undefined terminology.
For a graph $G$, let $V(G)$ and $E(G)$ denote its vertex and edge sets.  We write $c(G)$ for the number of connected components of $G$.  For $u\in V(G)$, let $N_G(u)$ be its open neighborhood and let $N_G[u]=N_G(u)\cup\{u\}$.  For $S\subseteq V(G)$, let $N_G(S)=\bigcup_{v\in S} N_G(v)$; this set may intersect $S$.  A graph is \emph{acyclic} if it contains no cycle, and a \emph{forest} is an acyclic graph.

A \emph{pseudoforest} is a graph in which every connected component contains at most one cycle.  A connected pseudoforest with exactly one cycle is called a \emph{pseudotree} or a \emph{unicyclic graph}.  The \emph{pseudoarboricity} of a graph $G$, denoted by $p(G)$, is the minimum number of edge-disjoint pseudoforests into which $E(G)$ can be partitioned; in particular, $p(G)=0$ if and only if $E(G)=\emptyset$.  A \emph{cluster graph} is a graph whose connected components are cliques.

A \emph{single exclusive split} of a vertex $v$ replaces $v$ by two non-adjacent vertices $v_1$ and $v_2$ and assigns each edge incident with $v$ to exactly one of $v_1$ and $v_2$.  Thus $N(v_1)\cap N(v_2)=\emptyset$ and $N(v_1)\cup N(v_2)=N(v)$, after identifying the neighbors with the corresponding vertices that were not split in this step.  The vertices $v_1$ and $v_2$ are called the \emph{descendants} created by the split.

A \emph{splitting sequence} of length $k$ is a sequence of graphs $(G_0,G_1,\ldots,G_k)$ such that $G_{i+1}$ is obtained from $G_i$ by one exclusive split.  Descendancy is extended transitively and reflexively through the sequence.  If $v\in V(G_i)$, then $\D_{\mathcal S}(v)$ denotes the set of descendants of $v$ in the final graph $G_k$.  For an original vertex set $X\subseteq V(G_0)$, let $\D_{\mathcal S}(X)=\bigcup_{v\in X}\D_{\mathcal S}(v)$.

For $i\in\{0,\ldots,k\}$, let ${\sf SplitSet}_{\mathcal S}(G_i)$ denote the set of original vertices having at least two descendants in $G_i$.
A splitting sequence $\mathcal S=(G_0,\ldots,G_k)$ is \emph{$\ell$-shallow} if $|\D_{\mathcal S}(v)|\le \ell+1$ for every original vertex $v\in V(G_0)$.
Equivalently, at most $\ell$ operations are performed on an original vertex and its descendants in total. Throughout the paper, operations at $v$ are counted in this sense. The total-split cost is the length of the sequence; the split-vertex cost is $|{\sf SplitSet}_{\mathcal S}(G_k)|$. These costs coincide when $\ell=1$.

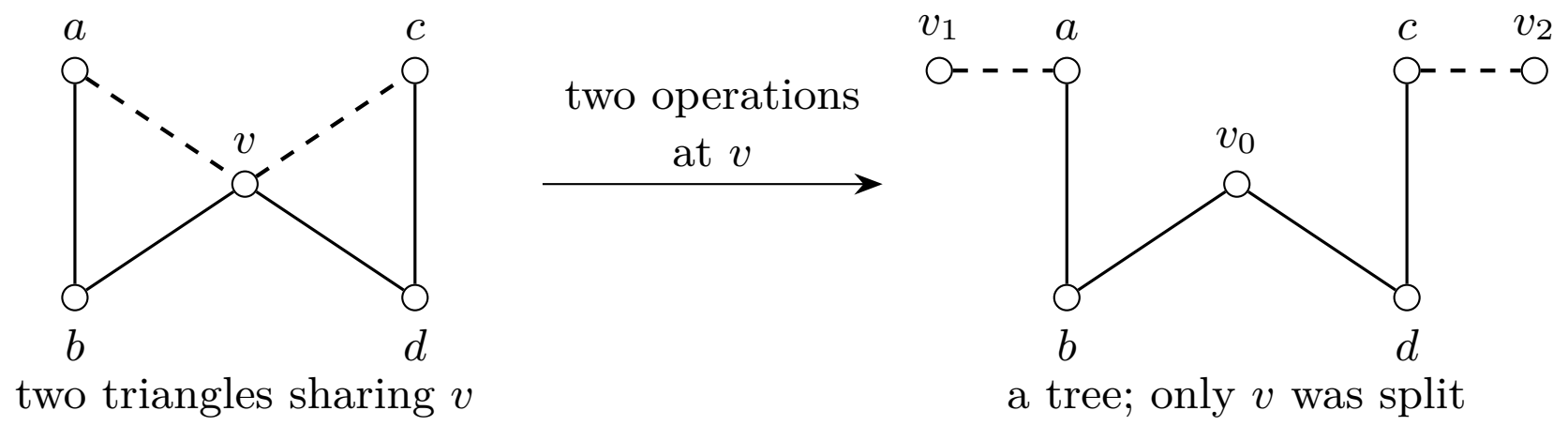
\begin{figure}[htbp]
\centering
\begingroup
\begin{tikzpicture}[x=1cm,y=1cm,font=\small]
\begin{scope}
\node[split vertex,label=above:$v$] (v) at (0,0) {};
\node[split vertex,label=above:$a$] (a) at (-1.2,.8) {};
\node[split vertex,label=below:$b$] (b) at (-1.2,-.8) {};
\node[split vertex,label=above:$c$] (c) at (1.2,.8) {};
\node[split vertex,label=below:$d$] (d) at (1.2,-.8) {};
\draw[split edge] (a)--(b)--(v)--(d)--(c);
\draw[split marked] (a)--(v)--(c);
\node at (0,-1.5) {two triangles sharing $v$};
\end{scope}
\draw[split arrow] (2.1,0)--node[above,align=center]{two operations\\at $v$}(4.5,0);
\begin{scope}[xshift=7cm]
\node[split vertex,label=above:$v_0$] (v) at (0,0) {};
\node[split vertex,label=above:$a$] (a) at (-1.2,.8) {};
\node[split vertex,label=below:$b$] (b) at (-1.2,-.8) {};
\node[split vertex,label=above:$c$] (c) at (1.2,.8) {};
\node[split vertex,label=below:$d$] (d) at (1.2,-.8) {};
\node[split vertex,label=above:$v_1$] (v1) at (-2.1,.8) {};
\node[split vertex,label=above:$v_2$] (v2) at (2.1,.8) {};
\draw[split edge] (a)--(b)--(v)--(d)--(c);
\draw[split marked] (v1)--(a) (c)--(v2);
\node at (0,-1.5) {a tree; only $v$ was split};
\end{scope}
\end{tikzpicture}
\caption{Two operations can cost one split vertex. Dashed edges are retained, with their incidences at $v$ moved to separate leaf descendants. The operation count is two; the split-vertex count is one.}
\label{fig:two-costs}
\endgroup
\end{figure}

A small example separates the two objectives. Let $G$ consist of the triangles $vabv$ and $vcdv$, sharing only $v$ (Figure~\ref{fig:two-costs}). A forest requires at least two operations: it has six edges and, after $s$ splits, $5+s$ vertices, so $6\le 5+s-1$. Detaching $va$ and $vc$ at $v$ attains this bound using one split original vertex when $\ell=2$. Detaching the same edges at $a$ and $c$ uses two split original vertices and is feasible already for $\ell=1$. In contrast, the pseudoforest target needs only one operation: detaching $va$ at $v$ leaves a single cycle.

\RevisionStart
\begin{observation}
\label{obs:merge-descendants}
Let $\mathcal S=(G_0,\ldots,G_k)$ be an $\ell$-shallow
splitting sequence. Distinct descendants of the same original
vertex are nonadjacent and have disjoint neighborhoods.
Moreover, let $x$ and $y$ be two such descendants, and let $H$
be obtained from $G_k$ by replacing $x$ and $y$ with a single
vertex $z$ such that
\[
N_H(z)=N_{G_k}(x)\cup N_{G_k}(y),
\]
with all other vertices and edges unchanged.
Then there exists an $\ell$-shallow splitting sequence of
length $k-1$ from $G_0$ to $H$.
\end{observation}
\RevisionEnd

\begin{proof}
\RevisionStart
Throughout a splitting sequence, each original edge has
exactly one descendant edge. Suppose that $x$ and $y$ are
distinct descendants of an original vertex $v$.
An edge between $x$ and $y$ would correspond to a loop at $v$
in $G_0$. If $x$ and $y$ had a common neighbor with original
vertex $u$, the two incident edges would correspond to
distinct edges of $G_0$ with endpoints $u$ and $v$.
Both possibilities contradict that $G_0$ is a simple graph.
Thus $x$ and $y$ are nonadjacent and have disjoint neighborhoods.
Consequently, replacing them with $z$ preserves the
one-to-one correspondence with the original edges.
\RevisionEnd

\RevisionStart
Give $z$ the common original vertex of $x$ and $y$, and retain
the original vertex of every other vertex of $H$.
For each $v\in V(G_0)$, let
$v_1,\ldots,v_{s_v}$ be the vertices of $H$ whose original
vertex is $v$. For each $i\in[s_v]$, define
\[
E_v^i=
\{e\in E(G_0) :
  \text{the edge corresponding to $e$ in $H$
  is incident with $v_i$}\}.
\]
The sets $E_v^1,\ldots,E_v^{s_v}$ are pairwise disjoint,
and their union is the set of edges incident with $v$ in $G_0$.
Some of these sets may be empty.
\RevisionEnd

\RevisionStart
Label each edge by its corresponding edge of $G_0$.
Starting from $G_0$, process each original vertex $v$ as follows.
If $s_v=1$, leave its single copy unchanged.
Otherwise, for $i=1,\ldots,s_v-1$, split the remaining copy
of $v$ into two vertices: the first receives the incident
edges with labels in $E_v^i$, and the second receives those
with labels in
\[
\bigcup_{j=i+1}^{s_v}E_v^j.
\]
Empty parts are allowed in a split.
This uses exactly $s_v-1$ splits.
After all original vertices have been processed, each edge
has the same endpoints as its corresponding edge in $H$.
Hence the resulting graph is $H$.
\RevisionEnd

\RevisionStart
Since each split increases the number of vertices by one,
the total number of splits is
\[
\sum_{v\in V(G_0)}(s_v-1)
=|V(H)|-|V(G_0)|
=|V(G_k)|-1-|V(G_0)|
=k-1.
\]
At every step, each original vertex $v$ has at most $s_v$
descendants, which is no greater than its number of
descendants in $G_k$. Therefore, the constructed sequence
is $\ell$-shallow.
\RevisionEnd
\end{proof}

We use the following simple componentwise observation repeatedly.  Splitting cannot move an edge from one original connected component to another, and all target classes considered in this paper are closed under disjoint union.  Therefore, for the total-split objective, the optimum value on a disconnected graph is the sum of the optimum values on its connected components.  For the split-vertex objective the same statement holds with the number of split original vertices as the cost.  Isolated vertices contribute zero cost.

\NewText{We write $G_1\cong G_2$ when $G_1$ and $G_2$ are isomorphic.}

\NewText{For a bipartite graph $G=(V_1\cup V_2,E)$, a matching saturates $X\subseteq V(G)$ if every vertex in $X$ is incident with an edge of the matching.}

\begin{theorem}[Hall's theorem~\cite{Hall1935OnRO}]
    Let $G$ be an undirected bipartite graph with bipartition $(V_1, V_2)$.
The graph $G$ has a matching saturating $V_1$ if and only if for all
$X \subseteq V_1$, we have $|N(X)| \ge |X|$.
\end{theorem}

\NewText{A graph is a cluster graph if and only if it contains no induced $P_3$.}

We provide a few definitions from matroid theory; see~\cite{schrijver2003combinatorial} for reference.

\RevisionStart
\begin{definition}[Matroid]
A pair \(\cM  = (U, \mathcal{I})\) is called a \emph{matroid} if \(U\) is a finite set and \(\mathcal{I}\) is a nonempty collection of subsets of \(U\) satisfying:
\begin{enumerate}
    \item If \(I \in \mathcal{I}\) and \(J \subseteq I\), then \(J \in \mathcal{I}\).
    \item If \(I, J \in \mathcal{I}\) and \(|I| < |J|\), then there exists an element \(z \in J \setminus I\) such that \(I \cup \{z\} \in \mathcal{I}\).
\end{enumerate}
\end{definition}
\RevisionEnd

\noindent
\textbf{Independent sets and bases:} Let \(\mathcal{M} = (U, \mathcal{I})\) be a matroid. A subset \(I \subseteq U\) is called \emph{independent} if \(I \in \mathcal{I}\), and \emph{dependent} otherwise. For \(S \subseteq U\), a subset \(B \subseteq S\) is called a \emph{base} of \(S\) if \(B\) is an inclusion-wise maximal independent subset of \(S\). That is, \(B \in \mathcal{I}\) and there is no \(Z \in \mathcal{I}\) such that $B \subsetneq Z \subseteq S.$

\RevisionStart
The common size of the bases of a subset \(S \subseteq U\) is called the \emph{rank} of \(S\), and is denoted by \(r_{\mathcal{M}}(S)\). When the matroid \(\mathcal{M}\) is clear from the context, we simply write \(r(S)\).
\RevisionEnd

A matroid is \emph{linear over a field} $\mathbb F$ if its ground elements index the columns of a matrix over $\mathbb F$ and a set is independent exactly when its columns are linearly independent. Such a matrix is a \emph{linear representation}. Our parity algorithms require an explicit representation, rather than only an independence test; we construct rational matrices of polynomial encoding length.

\begin{definition}[Matroid union]\label{def:matunion}
Let \(\mathcal{M}_1 = (U_1, \mathcal{I}_1), \dots, \mathcal{M}_k = (U_k, \mathcal{I}_k)\) be matroids. The \emph{union} of these matroids is defined as
\[
\mathcal{M}_1 \vee \cdots \vee \mathcal{M}_k = \bigl(U_1 \cup \cdots \cup U_k,\; \mathcal{I}_1 \vee \cdots \vee \mathcal{I}_k\bigr),
\]
where $\mathcal{I}_1 \vee \cdots \vee \mathcal{I}_k
\coloneqq
\left\{ I_1 \cup \cdots \cup I_k \;\middle|\; I_i \in \mathcal{I}_i \text{ for all } i \in [k] \right\}$.

\end{definition}

\NewStart
\begin{proposition}[Matroid union~\cite{schrijver2003combinatorial}]\label{prop:matunionmat}
The union of finitely many matroids is a matroid.
\end{proposition}
\NewEnd

Let $\mathcal{M}_1 = (U, \mathcal{I}_1)$ and $\mathcal{M}_2 = (U, \mathcal{I}_2)$ be two matroids on a common ground set $U$. The \emph{intersection} of $\mathcal{M}_1$ and $\mathcal{M}_2$, denoted by $\mathcal{I}_1 \cap \mathcal{I}_2$ is the set system
\[
\mathcal{I}_1 \cap \mathcal{I}_2 := \{ I \subseteq U \mid I \in \mathcal{I}_1 \text{ and } I \in \mathcal{I}_2 \}.
\]
Note that $\mathcal{I}_1 \cap \mathcal{I}_2$ is not necessarily a matroid.

\NewStart
\begin{theorem}[Matroid intersection~\cite{schrijver2003combinatorial}]
Given two matroids on the same finite ground set, each with a polynomial-time independence test, a maximum-cardinality common independent set can be found in polynomial time.
\end{theorem}
\NewEnd

\RevisionStart
\NewStart
\begin{definition}[Matroid parity]
\label{def:matroid-parity}
An instance of \emph{matroid parity} consists of a matroid $\mathcal M=(U,\mathcal I)$ and a partition $\mathcal P$ of $U$ into two-element sets, called \emph{prescribed pairs}. A collection $\mathcal F\subseteq\mathcal P$ is \emph{feasible} if $\bigcup_{P\in\mathcal F}P\in\mathcal I$. The matroid parity problem asks for a feasible collection containing as many pairs as possible.

In \emph{weighted matroid parity}, each pair $P\in\mathcal P$ has a weight $w(P)$, and the objective is to maximize $\sum_{P\in\mathcal F}w(P)$ over feasible collections $\mathcal F$. In both versions, pairs are selected in full, and their union must be independent in the given matroid.
\end{definition}
\NewEnd

For an explicitly represented rational linear matroid, weighted matroid parity can be solved deterministically in polynomial time~\cite{IwataKobayashi2022}. We use the version that selects an independent union of pairs, not necessarily a basis. The pair weights introduced later encode our unweighted split-vertex objective; the graph input itself has no weights.

\begin{definition}[Cographic Matroid]\label{app:def:cographic}\label{def:cographic}
Let $G$ be a graph. The cographic matroid of $G$, denoted by $\cM_C(G)=(E(G),\cI_C)$, has edge set $E(G)$ as its ground set and
\[
\cI_C=\{I\subseteq E(G): c(G-I)=c(G)\}.
\]
Equivalently, deleting $I$ does not increase the number of connected components; equivalently, $E(G)\setminus I$ contains a spanning forest of $G$.  For connected $G$, this is the usual condition that $(V(G),E(G)\setminus I)$ is connected.
\end{definition}
\RevisionEnd

\NewStart
Writing $n=|V(G)|$ and $m=|E(G)|$, the cographic matroid has rank
\[
r_{\mathcal M_C(G)}(E(G))=m-n+c(G).
\]
Its bases are precisely the complements of spanning forests of $G$~\cite{schrijver2003combinatorial}.

\NewText{We use pseudoforests both as a target class and to describe endpoint assignments in the forest and pseudoforest algorithms.}
\NewEnd

\RevisionStart
\begin{definition}[Bicircular Matroid]\label{app:def:bicircular}\label{def:bicircular}
Let $G$ be a graph. The bicircular matroid of $G$ is the matroid $\mathcal M_B(G)=(E(G),\mathcal I_B)$ in which
\[
\mathcal I_B=\{I\subseteq E(G):(V(G),I)\text{ is a pseudoforest}\}.
\]
\end{definition}
\RevisionEnd

\begin{definition}[Dual Matroid]
Let \( \mathcal{M} = (U, \mathcal{I}) \) be a matroid, and let \( \mathcal{B} \) denote the collection of all bases of \( \mathcal{M}\). The \emph{dual matroid} of \( \mathcal{M}\), denoted by \( \mathcal{M}^* = (U, \mathcal{I}^*) \), is defined on the same ground set \(U\), where a subset \( I \subseteq U \) belongs to \( \mathcal{I}^* \) if and only if there exists a base \( B \in \mathcal{B} \) such that
$I \subseteq U \setminus B.$
Equivalently, the collection of bases of \(\mathcal{M}^*\) is given by
\[
\mathcal{B}^* \;:=\; \{\, U \setminus B \mid B \in \mathcal{B} \,\}.
\]
\end{definition}

\begin{remark}
For any subset \( S \subseteq U \), the rank function \( r_{\mathcal{M}^*} \) of the dual matroid satisfies
\[
r_{\mathcal{M}^*}(S) \;=\; |S| + r_\mathcal{M}(U \setminus S) - r_\mathcal{M}(U).
\]
\end{remark}

We provide a few lemmas from matroid theory; see~\cite{schrijver2003combinatorial} for reference.

\begin{restatable}{lemma}{exactsize}
\label{lem:exactsize}
    Let $\mathcal{M}_1 = (U, \mathcal{I}_1)$ and $\mathcal{M}_2 = (U, \mathcal{I}_2)$ be two matroids on a common ground set $U$. Let
$
r := \max\{\, |I| : I \in \mathcal{I}_1 \cap \mathcal{I}_2 \,\}.
$
Then for every integer $0 \le r' \le r$, there exists a set $I' \in \mathcal{I}_1 \cap \mathcal{I}_2$ such that $|I'| = r'$.
\end{restatable}

\begin{proof}
Let $I \in \mathcal{I}_1 \cap \mathcal{I}_2$ be a common independent set of maximum size $r$. Let $r' \le r$. We show that there exists a set $I' \subseteq U$ such that $|I'| = r'$ and $I' \in \mathcal{I}_1 \cap \mathcal{I}_2$.

Since $I$ is independent in both matroids, and independence in a matroid is hereditary, every subset of $I$ is also independent in both $\mathcal{M}_1$ and $\mathcal{M}_2$. Hence, for any subset $I' \subseteq I$, we have $I' \in \mathcal{I}_1 \cap \mathcal{I}_2$.

Now choose any subset $I' \subseteq I$ with $|I'| = r'$. Such a subset exists since $r' \le r$. Then $I'$ is a common independent set of size $r'$, as required.
\end{proof}

\RevisionStart
\begin{restatable}{lemma}{directsum}
\label{lem:directsum}
Let $G$ be a graph and let $\ell$ be a nonnegative integer. Define $\mathcal M_B^\ell \coloneqq (E(G), \mathcal I_\ell)$, where $\mathcal I_\ell \coloneqq \{F \subseteq E(G) \mid (V(G),F)\text{ has pseudoarboricity at most }\ell\}$. Then $\mathcal M_B^\ell$ is a matroid.
\end{restatable}
\RevisionEnd

\begin{proof}

\RevisionStart
If $\ell=0$, then $\mathcal I_0=\{\emptyset\}$, which is the independent-set family of a matroid. Hence assume that $\ell\ge1$.
\RevisionEnd

Let $\cM_B = (E(G), \cI)$ denote the bicircular matroid of $G$ (see \Cref{def:bicircular}). For a fixed integer $\ell \geq 1$, let $\mathcal{M}_1, \dots, \mathcal{M}_\ell$ be $\ell$ copies of $\cM_B$, where $\mathcal{M}_i = (E(G), \mathcal{I}_i)$ with $\mathcal{I}_i = \cI$ for each $i \in [\ell]$. Let $\mathcal{M}_1 \vee \cdots \vee \mathcal{M}_\ell = (E(G), \widetilde{\mathcal{I}})$ denote the union of the matroids. It follows from \Cref{def:matunion} that the ground set of union remains $E(G)$, and
\[
\widetilde{\mathcal{I}} = \left\{ I_1 \cup \cdots \cup I_\ell \;\middle|\; I_i \in \mathcal{I} \text{ for all } i \in [\ell] \right\}.
\]
By \Cref{prop:matunionmat}, $\mathcal{M}_1 \vee \cdots \vee \mathcal{M}_\ell$ is a matroid. We now establish the equivalence between the matroid union $\mathcal{M}_1 \vee \cdots \vee \mathcal{M}_\ell$ and the matroid $\mathcal{M}_B^\ell$ in the following claim.

\begin{claim}
A set $F \subseteq E(G)$ is independent in $\mathcal{M}_B^\ell$ if and only if $F \subseteq E(G)$ is independent in $\mathcal{M}_1 \vee \cdots \vee \mathcal{M}_\ell$.
\end{claim}

\begin{claimproof}
\RevisionStart
    In the forward direction, let $F \subseteq E(G)$ be independent in $\mathcal{M}_B^\ell$. This implies that the graph \((V(G), F)\) has pseudoarboricity at most $\ell$, that is there exists partition $F_1 \uplus \cdots \uplus F_\ell$ of $F$ such that for each $i \in [\ell]$, $(V(G), F_i)$ is a pseudoforest. For each $i\in [\ell]$, since $(V(G), F_i)$ is a pseudoforest, $F_i$ is independent in $\mathcal{M}_i$. Hence $F$ is an independent set in $\mathcal{M}_1 \vee \cdots \vee \mathcal{M}_\ell$.
\RevisionEnd

\RevisionStart
     Conversely, let $I \subseteq E(G)$ be independent in $\mathcal{M}_1 \vee \cdots \vee \mathcal{M}_\ell$. $I \subseteq E(G)$ is independent in $\mathcal{M}_1 \vee \cdots \vee \mathcal{M}_\ell$ if and only if there exist sets $I_1, \dots, I_\ell$ such that
\[
I = I_1 \cup \cdots \cup I_\ell
\quad \text{and} \quad
I_i \in \mathcal{I}_i \text{ for each } i \in [\ell].
\]
Since each $\mathcal{M}_i$ is the bicircular matroid of $G$, this is equivalent to requiring that $(V(G), I_i)$ is a pseudoforest for each $i \in [\ell]$. Further we construct pairwise disjoint sets $F_1, \dots, F_\ell$ as follows:
\[
F_1 := I_1, \qquad
F_i := I_i \setminus \bigcup_{j < i} I_j \quad \text{for } i = 2, \dots, \ell.
\]
By construction, the sets $F_1, \dots, F_\ell$ are pairwise disjoint and their union is $I$, that is $ \uplus_{i=1}^{\ell} F_i=\cup_{i=1}^{\ell} I_i=I$.
Moreover, for each $i \in [\ell]$, we have $F_i \subseteq I_i$. Since every subgraph of a pseudoforest is again a pseudoforest, it follows that $(V(G),F_i)$ is a pseudoforest for each $i \in [\ell]$. Hence, $(V(G),I)$ has pseudoarboricity at most $\ell$, and therefore $I$ is independent in $\mathcal{M}_B^\ell$.
\RevisionEnd

\end{claimproof}
This completes the proof.
\end{proof}

\section{Shallow Splitting to Forests}
\label{sec:forests}
We study both cost measures for the forest target. We first develop the structural characterization and the total-split algorithm, and then minimize the number of split vertices.

\subsection{Minimizing the total number of splits}
In this subsection, we study the problem of transforming a graph into a forest. We begin by recalling the problem.

\begin{tcolorbox}[colback=anti-flashwhite,colframe=black!35]
\noindent\textcolor{red!50!black}{\flsvs}\\
{\bf Fixed parameter:} an integer $\ell\ge 0$.\\
{\bf Input:} A graph~$G$ and a \NewText{nonnegative} integer $k$.\\
{\bf Question:} Does there exist an $\ell$-shallow splitting sequence $\mathcal{S}\coloneqq (G_0 = G, G_1, \ldots, G_t)$ with $t \le k$, such that $G_t$ is a forest?
\end{tcolorbox}

By the componentwise observation from \Cref{sec:prelims}, it is enough to solve the problem on connected components and add the resulting costs.  Therefore, the structural lemmas below are stated for connected graphs; the theorem for general graphs follows by applying them to every connected component, with isolated vertices contributing zero.

\NewStart
We will keep a spanning forest and assign every remaining edge to an endpoint. Assigning an edge to $v$ will require one split at $v$. We first describe the assignments allowed by the local bound, and then prove that they characterize feasible splitting sequences.
\NewEnd

\RevisionStart
\begin{definition}[$\ell$-incidence bipartite graph]
\label{def:ell-incidence}
Let $G$ be a graph and let $\ell$ be a nonnegative integer, with $[0]=\emptyset$.
The \emph{$\ell$-incidence bipartite graph} of $G$, denoted by $A_{G}^\ell$ is a bipartite graph with bipartition $X \cup Y$, defined as follows.
\begin{itemize}
    \item \textbf{(Edge-part)} $X \coloneqq \{v_{xy} \mid xy \in E(G)\}$.
    \item \textbf{(Vertex-part)} $Y \coloneqq \{v^i \mid v \in V(G),\ i \in [\ell]\}$.
    \item \textbf{(Adjacency)} For every edge $xy \in E(G)$ and every $i \in [\ell]$, the vertex $v_{xy}$ is adjacent to both $x^i$ and $y^i$.
That is, $E(A^{\ell}_G)
=
\{\, v_{xy}x^i,\, v_{xy}y^i \mid xy \in E(G),\ i \in [\ell] \,\}$.
\end{itemize}

\end{definition}
\RevisionEnd

\begin{observation}
\label{obs:edge-vertex-comp}
A graph $G$ is a pseudoforest if and only if for every connected subgraph $C$ of $G$, it holds that $|E(C)| \le |V(C)|$. Moreover, if $G$ is a pseudoforest, then $|E(G)| \le |V(G)|$.
\end{observation}

\begin{proof}
In the forward direction,  let $G$ be a pseudoforest, and let $C$ be any connected subgraph of $G$. Since $C$ is connected, it has a spanning tree $T$. Furthermore, $C$ is also a pseudoforest since $C$ is a subgraph of $G$.
Thus, by definition of pseudoforest, $C$ contains at most one edge that is not in $T$.
Since $|E(T)| = |V(C)| - 1$, it follows that $|E(C)| \le |V(C)|.$

In particular, applying this to each connected component $C_1, \dots, C_r$ of $G$ and summing, we obtain
\[
|E(G)| = \sum_{i=1}^r |E(C_i)| \le \sum_{i=1}^r |V(C_i)| = |V(G)|.
\]

In the reverse direction, suppose that for every connected subgraph $C$ of $G$, we have $|E(C)| \le |V(C)|$.
Let $C$ be any connected component of $G$. We show that $C$ contains at most one cycle. If $C$ is acyclic, then it is a tree and we are done. Otherwise, let $T$ be a spanning tree of $C$. Then $|E(T)| = |V(C)| - 1$. Since $|E(C)| \le |V(C)|$, it follows that $C$ contains at most one edge outside $T$. Hence, $C$ has at most one cycle. Therefore, every connected component of $G$ contains at most one cycle, and thus $G$ is a pseudoforest.
\end{proof}

\begin{lemma}
\label{lem:pseudoforest_matching}
Let $G$ be a graph and $A_G^1= (X \cup Y)$ be the 1-incidence bipartite graph of $G$ where $X$ and $Y$ are the edge-part and vertex-part, respectively. Then $G$ is a pseudoforest if and only if $A_G^1$ admits a matching saturating $X$.
\end{lemma}

\begin{proof}

We first prove the forward direction. Assume that $G$ is a pseudoforest. Let $A_G^1=(X \cup Y)$ be the $1$-incidence bipartite graph of $G$. We show that $A_G^1$ admits a matching saturating $X$ by Hall's theorem.

Let $X' \subseteq X$ be an arbitrary set, and define
$E' \coloneqq \{xy \in E(G) \mid v_{xy} \in X'\}$.
Also let $U \coloneqq \{v \in V(G) \mid \text{$v$ is incident with some edge in } E'\}$.
Then $(U,E')$ is a subgraph of $G$. Since $G$ is a pseudoforest, every subgraph of $G$ is also a pseudoforest, thus in particular $(U,E')$ is also a pseudoforest. By Observation~\ref{obs:edge-vertex-comp}, we have $|E'| \le |U|$. By the definition of $A_G^1$, a vertex $v^1 \in Y$ is adjacent to some vertex of $X'$ if and only if $v$ is incident with some edge in $E'$. Thus
$N_{A_G^1}(X')=\{v^1 \mid v \in U\}$,
and hence
$|N_{A_G^1}(X')|=|U|$. Moreover, by definition,
$|X'|=|E'|$. Therefore,
$|X'|=|E'| \leq |U|=|N_{A_G^1}(X')|$.
Since $X' \subseteq X$ was arbitrary, Hall's condition holds for every subset of $X$. Therefore, by Hall's theorem, $A_G^1$ has a matching saturating $X$.

In the reverse direction, assume that $A^1_G = (X \cup Y)$ has a matching, namely $M$ that saturates $X$, the vertices corresponding to $E(G)$. Orient each edge of $G$ away from the endpoint to whose slot it is matched. For an edge $xy \in E(G)$, if $v_{xy}$ is matched to the vertex corresponding to one of its endpoints, say $x$, then orient $xy$ in $G$ from $x$ to $y$. Since the matching $M$ in $A^1_G$ saturates $X$, every edge of $G$ is oriented exactly once. Each vertex has outdegree at most $1$, because its slot is matched to at most one edge-vertex. Let $C$ be any connected component of $G$ which is oriented. Then $|E(C)|=\sum_{v\in V(C)} d^{+}(v) \le |V(C)|$. Thus it follows from \Cref{obs:edge-vertex-comp} that $G$ is a pseudoforest.

This completes the proof.
\end{proof}

\begin{lemma}
\label{lem:pseudoarboricity_matching}
Let $G$ be a graph and $A_G^{\ell}=(X\cup Y)$ be the $\ell$-incidence bipartite graph of $G$ where $X$ and $Y$ are the edge-part and vertex-part, respectively. Then $G$ has pseudoarboricity at most $\ell$ if and only if $A_G^{\ell}$ admits a matching saturating $X$.
\end{lemma}

\begin{proof}
\RevisionStart
For $\ell=0$, the vertex-part of $A_G^\ell$ is empty, so it has a matching saturating its edge-part exactly when $E(G)=\emptyset$. This is also precisely when $G$ has pseudoarboricity zero. Henceforth assume $\ell\ge1$.
\RevisionEnd

In the forward direction, assume that $G$ has pseudoarboricity at most $\ell$. Then there exists a partition
$E(G)=E_1 \uplus E_2 \uplus \dots \uplus E_\ell$
such that for each $i \in [\ell]$, the graph $G_i \coloneqq (V(G), E_i)$ is a pseudoforest. For each $i \in [\ell]$, by Lemma~\ref{lem:pseudoforest_matching}, the $1$-incidence bipartite graph $A^1_{G_i} = (X_i \cup Y_i)$ admits a matching $M_i$ that saturates $X_i$.

We construct a matching $M$ in $A^\ell_G$ as follows. For each edge $e = xy \in E_i$, if $v_e$ is matched to $x$ in $M_i$, then match $v_e$ to $x^i$ in $A^\ell_G$.
Since the sets $E_1, \dots, E_\ell$ form a partition of $E(G)$, each edge $e \in E(G)$ is matched exactly once in $M$.
Moreover, for each vertex $v \in V(G)$ and each $i \in [\ell]$, the vertex $v^i$ is matched to at most one vertex in $X$, as this follows from the matching $M_i$.
Hence, $M$ is a valid matching in $A^\ell_G$ that saturates $X$.

Conversely, assume that $A_G^{\ell}=(X \cup Y)$ admits a matching $M$ saturating $X$.
We show that $G$ has pseudoarboricity at most $\ell$. For each $i \in [\ell]$, define
$E_i \coloneqq \{\, xy \in E(G) \mid v_{xy} \text{ is matched in } M \text{ to either } x^i \text{ or } y^i \,\}$.

Since $M$ saturates $X$, every vertex $v_{xy} \in X$ is matched exactly once.
Hence every edge of $G$ belongs to exactly one set $E_i$.
Therefore, $E(G)$ is partitioned into $E_1,\dots,E_{\ell}$.

Now, for each $i \in [\ell]$, let $G_i \coloneqq (V(G),E_i)$.
Consider the bipartite subgraph of $A_G^{\ell}$ induced by
$X_i \coloneqq \{\, v_{xy} \mid xy \in E_i \,\}$
and
$Y_i \coloneqq \{\, v^i \mid v \in V(G) \,\}$. By the definition of $A_G^{\ell}$, this induced subgraph is exactly the $1$-incidence bipartite graph $A_{G_i}^1$ of $G_i$.
Moreover, the set
$M_i \coloneqq M \cap E(A_{G_i}^1)$
is a matching saturating $X_i$, because every vertex of $X_i$ is matched by $M$ to a vertex of $Y_i$. Hence, by Lemma \ref{lem:pseudoforest_matching}, the graph $G_i$ is a pseudoforest for every $i \in [\ell]$.

Thus, $E(G)$ is partitioned into $\ell$ edge-disjoint pseudoforests $G_1,\dots,G_{\ell}$.
Therefore, $G$ has pseudoarboricity at most $\ell$.
\end{proof}

The incidence correspondence also gives the independence tests needed by matroid intersection. We record their running time here for both the forest and pseudoforest algorithms.

\RevisionStart
\begin{lemma}
\label{lem:timeformatroid}
Let $G=(V,E)$ be a graph and let $\ell$ be a positive integer. Let $\mathcal{M}_C$ be the cographic matroid of $G$, and let $\mathcal{M}_B^\ast$ be the dual of the bicircular matroid of $G$.
For each matroid $\mathcal{N}\in\{\mathcal{M}_C,\mathcal{M}_B^\ast\}$, a maximum common independent set of $\mathcal{N}$ and $\mathcal{M}_B^\ell$ can be computed in time $O(|V|+\ell\cdot |E|^3)$.
\end{lemma}
\RevisionEnd

\begin{proof}
\RevisionStart
Write $n=|V|$ and $m=|E|$. We first remove isolated vertices in $O(n+m)$ time; this does not change either matroid on the edge set. If $m=0$, the answer is the empty set. Otherwise, the remaining graph has at most $2m$ vertices. We may also replace $\ell$ by $\min\{\ell,m\}$, since every set of at most $m$ edges can be partitioned into at most $m$ pseudoforests.
\RevisionEnd

\RevisionStart
We use the standard matroid-intersection algorithm based on shortest augmenting paths~\cite[Section~41.2]{schrijver2003combinatorial}. Starting with the empty common independent set, this algorithm maintains a common independent set $I$. In each round, it constructs an exchange graph using the tests
\[
I\cup\{f\}\in\mathcal I
\quad\text{and}\quad
(I\setminus\{x\})\cup\{f\}\in\mathcal I
\qquad(f\in E\setminus I,\ x\in I)
\]
for each of the two independence families $\mathcal I$. A shortest augmenting path increases $|I|$ by one; if none exists, $I$ has maximum size. We show that all the tests in one round can be performed in $O(\ell m^2)$ time.
\RevisionEnd

\RevisionStart
For $\mathcal M_B^\ell$, use the incidence graph from \Cref{lem:pseudoarboricity_matching}: its left side consists of the edges of $G$, its right side has $\ell$ copies of each vertex, and each edge is adjacent to the copies of its two endpoints. It has $2\ell m$ incidence edges. Independence is equivalent to the existence of a matching saturating the chosen left-side vertices. A matching saturating $I$ can be computed from scratch using at most $m$ augmenting paths, in $O(\ell m^2)$ time.
\RevisionEnd

\RevisionStart
Fix $f\in E\setminus I$. Starting at $f$, perform an alternating search, using only $I\cup\{f\}$ on the left side: traverse unmatched incidence edges from left to right and matched edges from right to left. If an unmatched right-side vertex is reached, then $I\cup\{f\}$ is independent, and so is every set $(I\setminus\{x\})\cup\{f\}$. Otherwise, an exchange with $x\in I$ is possible exactly when $x$ is reached. Indeed, the path reaching $x$ can be flipped to match $f$ and leave $x$ unmatched. Conversely, the symmetric difference with any matching saturating $(I\setminus\{x\})\cup\{f\}$ contains such a path to $x$, since no path to an unmatched right-side vertex exists. Thus one search answers all tests for $f$ in $O(\ell m)$ time, and all choices of $f$ require $O(\ell m^2)$ time.
\RevisionEnd

\RevisionStart
For the other matroid, write $\mathcal N=\mathcal P^*$, where $\mathcal P$ is the graphic or bicircular matroid of $G$. Since $I$ is independent in $\mathcal N$, the graph $H=(V,E\setminus I)$ has the full rank $r$ of $\mathcal P$. For each $f\in E\setminus I$, compute the connected components of $H-f$, together with their vertex and edge counts, in $O(m)$ time. The graphic rank is the number of vertices minus the number of components. The bicircular rank is the number of vertices minus the number of tree components: a tree component contributes one less than its number of vertices, and a component containing a cycle contributes its number of vertices. These formulas determine whether $H-f$ has rank $r$ or $r-1$.
\RevisionEnd

\RevisionStart
If its rank is $r$, then $I\cup\{f\}$ and all exchanges with $f$ are independent in $\mathcal N$. If its rank is $r-1$, the exchange with $x\in I$ is independent exactly when adding $x$ to $H-f$ restores rank $r$. For the graphic matroid, this happens exactly when the endpoints of $x$ lie in different components. For the bicircular matroid, it happens exactly when at least one endpoint lies in a tree component. Adding an edge within a tree creates one cycle; joining two tree components produces one tree; and joining a tree to a component containing a cycle produces a component containing a cycle. Each operation reduces the number of tree components by one, whereas an edge with both endpoints in components containing cycles does not. Thus each exchange test takes constant time after the component computation, and all tests for $\mathcal N$ require $O(m^2)$ time per round.
\RevisionEnd

\RevisionStart
The exchange graph has $O(m^2)$ arcs, so a breadth-first search finds a shortest augmenting path within the same $O(\ell m^2)$ bound. There are at most $m$ successful rounds and one final unsuccessful round. Recomputing the incidence matching after each augmentation is already included in the bound per round. The total running time, including preprocessing, is therefore $O(n+\ell m^3)$.
\RevisionEnd
\end{proof}

We now turn from assignments to splitting sequences. First, a minimum-length sequence on a connected input can be assumed to end in a tree.

\begin{lemma}\label{lem:foresttotree}
\NewStart
Let $G$ be a connected graph and let $S\subseteq V(G)$. Among the $\ell$-shallow splitting sequences to a forest whose split vertices all belong to $S$, let $\mathcal S=(G_0=G,\ldots,G_k)$ have minimum length. Then $G_k$ is a tree.
\NewEnd
\end{lemma}

\begin{proof}
Suppose, for contradiction, that $G_k$ is not a tree. Since $G_k$ is a forest, it follows that $G_k$ is disconnected.

\RevisionStart
Since $G$ is connected, some original vertex $v$ has descendants in two distinct components of $G_k$. Indeed, otherwise identifying all descendants of each original vertex would leave these components separate, although this identification recovers $G$. Choose two such descendants $v_1$ and $v_2$, and let $H$ be obtained from $G_k$ by identifying them. This joins two trees at one vertex and leaves all other components unchanged. Hence $H$ is a forest.
\RevisionEnd

\RevisionStart
\NewStart
By \Cref{obs:merge-descendants}, $H$ can be obtained by an $\ell$-shallow sequence of length $k-1$. The construction in that observation does not increase the number of descendants of any original vertex. In particular, no vertex outside $S$ is split. This contradicts minimality among the sequences allowed here. Therefore $G_k$ is a tree.
\NewEnd
\RevisionEnd
\end{proof}

\NewStart
For $S\subseteq V(G)$, write
$Y(S)=\{v^i:v\in S,\ i\in[\ell]\}$
for the slots of $S$ in $A_G^\ell$. For $M\subseteq E(G)$, write
$X(M)=\{v_{xy}:xy\in M\}$.
We say that a sequence is \emph{supported in $S$} when every original vertex that is split belongs to $S$; some vertices of $S$ may remain unsplit.

The next lemma is the structural step. Starting from a feasible tree, we merge descendants back and record one detached edge at each merge. Keeping the recorded edges assigned to their original endpoints preserves both the local bound and the set of vertices allowed to split.

\begin{lemma}\label{lem:feasPart}
Let $G$ be connected, let $\ell\ge0$, and let $S\subseteq V(G)$.
There is an $\ell$-shallow splitting sequence to a forest supported in $S$ if and only if $E(G)$ has a partition $T\uplus M$ such that $(V(G),T)$ is a spanning tree and $X(M)$ can be matched into $Y(S)$ in $A_G^\ell$.
Whenever these conditions hold, there is such a sequence of length $|E(G)|-|V(G)|+1$.

In particular, for $S=V(G)$, feasibility is equivalent to a partition into a spanning tree and an edge set of pseudoarboricity at most $\ell$.
\end{lemma}
\NewEnd

\begin{proof}

\NewStart
Suppose first that a feasible sequence supported in $S$ exists. Choose one of minimum length among all such sequences, and denote it by $\mathcal S=(G_0=G,\ldots,G_p)$. By \Cref{lem:foresttotree}, $G_p$ is a tree. We will record each removed edge at an original vertex in $S$.
\NewEnd

\NewStart
We extract the partition by merging vertices with the same original label while recording one removed edge at each merge. The graphs $H_p,H_{p-1},\ldots,H_0$ below are auxiliary labelled graphs used to find $T$ and $M$; they are not intermediate graphs of an exclusive splitting sequence, since this extraction deletes edges. The actual splitting sequence will be constructed in the converse direction. For each $i\in\{0,\ldots,p\}$ and each $v\in V(G)$, let $\mathsf{clone}_i(v)\subseteq V(H_i)$ be the vertices carrying the original label $v$.
\NewEnd

\RevisionStart
Initially, let \(H_p \coloneqq G_p\). For every original vertex \(u\in V(G)\), initialize \(M_u\gets\emptyset\) and \({\sf clone}_p(u)\gets\D_{\mathcal S}(u)\). Each surviving edge retains its original edge label and joins vertices carrying the labels of its original endpoints. In particular, clones of the same original vertex are nonadjacent and have disjoint neighborhoods, since \(G\) is simple and each original edge occurs at most once.
\RevisionEnd

\RevisionStart
We also maintain that \(|{\sf clone}_i(v)|\) equals the number of descendants of \(v\) in \(G_i\). This holds initially. Each reverse step merges two clones of the original vertex split in that step, reducing its clone count by one, and leaves the other counts unchanged. Thus two such clones are available whenever a merge is required. Merging vertices with the same original label preserves the edge-label property.
\RevisionEnd

For each \(i < p\), the graph \(H_i\) is obtained from \(H_{i+1}\) by a \emph{merge operation}. Throughout this process, we maintain the following invariants:
\begin{description}
 \item[(1)] The graph \(H_i\) is a tree.
    \item[(2)] For every \(u, v \in {\sf SplitSet}_{\mathcal{S}}(G_p)\) with \(u \neq v\), it holds that \(M_u \cap M_v = \emptyset\).

\end{description}
It is immediate that these invariants hold for the initial graph \(H_p\) in the sequence we construct.

\RevisionStart
Assume the invariants hold for $H_{i+1}$. We delete one edge and merge two clones to obtain $H_i$, while preserving the tree invariant. Recall the sequence $\cS = (G_0 = G, G_1, \dots, G_p)$. Suppose $G_{i+1}$ is obtained from $G_i$ by splitting vertex $v'_i$. Let $v \in V(G_0)$ be the original vertex corresponding to $v'_i$. Clearly, for $v$, we have $|{\sf clone}_{i+1}(v)|\geq 2$.
Choose any two elements in ${\sf clone}_{i+1}(v)$, say $v_1$ and $v_2$. Since $H_{i+1}$ is a tree, there is a unique path between $v_1$ and $v_2$ in $H_{i+1}$. Let $x$ and $y$ be the neighbors of $v_1$ and $v_2$, respectively, on this path. Delete one of the two edges $v_1x$ and $v_2y$, say without loss of generality $v_1x$, and add the corresponding original edge $vx' \in E(G)$ to $M_v$. Now identify $v_1$ and $v_2$ as a single vertex, denoted by $v_{12}$, and define $H_i$ as follows.
$V(H_i) = \bigl(V(H_{i+1})\setminus\{v_1,v_2\}\bigr)\cup\{v_{12}\},$
and the vertex $v_{12}$ is adjacent in $H_i$ to precisely those vertices that were adjacent in $H_{i+1}$ to $v_1$ or $v_2$, except that the deleted edge $v_1x$ is not retained. Formally,
\RevisionEnd

\[
\begin{aligned}
E(H_i) ={}& \Bigl(E(H_{i+1}) \setminus
\{wu \in E(H_{i+1}) \mid u \in \{v_1,v_2\}\}\Bigr) \\
& \cup \{wv_{12} \mid
w \in N_{H_{i+1}}(v_1)\cup N_{H_{i+1}}(v_2),\; w \neq x\}.
\end{aligned}
\]

\RevisionStart
We also update the clone sets by setting
${\sf clone}_i(v)=\bigl({\sf clone}_{i+1}(v)\setminus\{v_1,v_2\}\bigr)\cup\{v_{12}\},$
and ${\sf clone}_i(w)={\sf clone}_{i+1}(w)$ when $w\neq v$.
\RevisionEnd

\RevisionStart
Note that \(v_1\) and \(v_2\) are not adjacent, as they are descendants of the same vertex. Since they are connected by a path in \(H_{i+1}\), it follows that \(v_{12}\) is adjacent to at least one vertex. We next verify that this operation preserves the invariants.
\RevisionEnd

\begin{claim}\label{clm:Histree}
$H_i$ is a tree.
\end{claim}

\begin{claimproof}
\NewStart
Deleting the path edge $v_1x$ from the tree $H_{i+1}$ gives two trees, one containing $v_1$ and the other containing $v_2$. Identifying $v_1$ and $v_2$ joins these two trees at one vertex. The resulting graph is connected, including when one of the two trees consists of a single vertex.
\NewEnd

\RevisionStart
Moreover, in this operation, exactly one edge is deleted, namely $v_1x$, and exactly one vertex is lost, since the two vertices $v_1$ and $v_2$ are replaced by the single vertex $v_{12}$. Thus,
$|E(H_i)| = |E(H_{i+1})|-1$
and $|V(H_i)| = |V(H_{i+1})|-1$. Since $H_{i+1}$ is a tree, we have
$|E(H_{i+1})| = |V(H_{i+1})|-1$. Therefore, $|E(H_i)| = |V(H_i)|-1$. Thus the fact that $H_i$ is connected and has exactly one fewer edge than vertices, implies that $H_i$ is a tree.
\RevisionEnd
\end{claimproof}

\begin{claim}\label{clm:pairdisj}
For every \(u, v \in {\sf SplitSet}_{\mathcal{S}}(G_p)\) with \(u \neq v\), it holds that \(M_u \cap M_v = \emptyset\).
\end{claim}

\begin{claimproof}
\RevisionStart
Initially, \(M_u=\emptyset\) for every \(u \in {\sf SplitSet}_{\mathcal{S}}(G_p)\), so the claim holds. Now consider any merge operation for a vertex \(v \in {\sf SplitSet}_{\mathcal S}(G_p)\). In this step, exactly one edge is deleted from \(H_{i+1}\), and the corresponding edge of \(G\) is added only to \(M_v\). No other set \(M_u\), where \(u \neq v\), is changed. Moreover, an edge deleted in this step cannot already belong to some other set \(M_u\), since every edge is deleted from \(H_{i+1}\) at most once. Therefore, after this step, the sets remain pairwise disjoint. Hence, throughout the construction, for every \(u,v \in {\sf SplitSet}_{\mathcal{S}}(G_p)\) with \(u \neq v\), we have \(M_u \cap M_v=\emptyset\).
\RevisionEnd
\end{claimproof}

\RevisionStart
We define
$M \coloneqq \bigcup_{{v \in \sf SplitSet}_{\mathcal{S}}(G_p)}M_v$. Clearly, \(M\) is the disjoint union of the sets \(M_v\) for \(v \in {\sf SplitSet}_{\mathcal{S}}(G_p)\). Let \(T\) be the set of corresponding original edges for the edge set of the final tree \(H_0\). By construction, every edge of \(G\) is either deleted at some stage and added to \(M\), or remains in the final tree \(H_0\). Hence \(E(G)=T \uplus M\).
\RevisionEnd

\begin{claim}\label{clm:Spanningtreetrue}
$(V(G),T)$ is a spanning tree of $G$.
\end{claim}

\begin{claimproof}
For each vertex \(v \in V(G)\), recall that \({\sf clone}_i(v)\) denotes the set of vertices in \(H_i\) corresponding to \(v\). Initially, \(|{\sf clone}_p(v)| \geq 1\), and each merge operation is applied only when \(|{\sf clone}_{i+1}(v)| \geq 2\), reducing the size of the clone set by one. Hence, for every \(i\), we have \(|{\sf clone}_i(v)| \geq 1\), and in particular \(|{\sf clone}_0(v)| \geq 1\).

\NewText{The auxiliary process performs one merge for each of the $p$ split operations. Since $H_p=G_p$ has $|V(G)|+p$ vertices, $H_0$ has exactly $|V(G)|$ vertices.} The sets \(\{{\sf clone}_0(v) \mid v \in V(G)\}\) form a partition of \(V(H_0)\). Since each of these sets is nonempty and their union has size \(|V(G)|\), it follows that \(|{\sf clone}_0(v)|=1\) for every \(v \in V(G)\).

Thus, \(V(H_0)\) is in bijection with \(V(G)\). Since \(H_0\) is a tree, the corresponding edge set \(T\) forms a spanning tree of \(G\).
\end{claimproof}

It remains to show that $(V(G),M)$ has pseudoarboricity at most $\ell$.

\begin{claim}\label{clm:pesudoatmostell}
The graph $(V(G),M)$ has pseudoarboricity at most $\ell$.
\end{claim}

\begin{claimproof}
Fix a vertex $v \in {\sf SplitSet}_\mathcal{S}(G_p)$. Since $\mathcal{S}$ is an $\ell$-shallow splitting sequence, we have $|\D(v)| \le \ell+1$. At the beginning of the reverse process, ${\sf clone}_p(v)=\D(v)$, while at the end ${\sf clone}_0(v)$ consists of a single vertex. Moreover, whenever the reverse step from $H_{i+1}$ to $H_i$ involves two vertices of ${\sf clone}_{i+1}(v)$, the size of the clone set of $v$ decreases by exactly one, that is, $|{\sf clone}_i(v)|=|{\sf clone}_{i+1}(v)|-1$. In every other reverse step, the clone set of $v$ remains unchanged. Hence the number of reverse steps involving clones of $v$ is exactly $|{\sf clone}_p(v)|-|{\sf clone}_0(v)|=|\D(v)|-1$. By construction, each such reverse step contributes exactly one edge to $M_v$. Therefore, $|M_v|=|\D(v)|-1 \le \ell$.

\RevisionStart
For each vertex $v \in V(G)\setminus {\sf SplitSet}_\mathcal{S}(G_p)$, the set $M_v$ remains empty. Thus, for every vertex $v \in V(G)$, the set $M_v$ is defined and satisfies $|M_v|\le \ell$. Further, by \Cref{clm:pairdisj}, the family $\{M_v \mid v \in V(G)\}$ is pairwise disjoint. Since every edge added to $M$ during the reverse process is added to exactly one set $M_v$, and $M$ is precisely the set of all such edges, it follows that $M=\biguplus_{v\in V(G)} M_v$.
\RevisionEnd

\RevisionStart
Now, for each vertex $v \in V(G)$, fix an arbitrary ordering $M_v=\{e_v^1,\dots,e_v^{t_v}\}$, where $t_v=|M_v|\le \ell$. Consider the $\ell$-incidence bipartite graph $A_{(V(G),M)}^\ell$. For each vertex $v \in V(G)$ and each index $j \in [t_v]$, if $e_v^j=xy$, then match the vertex $v_{xy}$ in the edge-part to the vertex $v^j$ in the vertex-part. The chosen pair is an edge of the incidence graph because every edge in $M_v$ was deleted at a clone of $v$ and is therefore incident with $v$ in $G$. This defines a matching saturating the edge-part. Indeed, since the sets $M_v$ form a partition of $M$, every edge of $M$ appears exactly once among the sets $M_v$, and hence every vertex in the edge-part is matched exactly once. Also, for a fixed vertex $v$, the vertices $v^1,\dots,v^\ell$ are pairwise distinct, and since $t_v\le \ell$, no two edges in $M_v$ are matched to the same vertex in the vertex-part. Therefore, no two vertices in the edge-part are matched to the same vertex in the vertex-part.
\RevisionEnd

Hence $A_{(V(G),M)}^\ell$ admits a matching saturating its edge-part. By \Cref{lem:pseudoarboricity_matching}, it follows that $(V(G),M)$ has pseudoarboricity at most $\ell$.

\NewStart
Since the chosen sequence is supported in $S$, the sets $M_v$ are empty for $v\notin S$. Thus the matching just constructed uses only $Y(S)$. This proves the required matching condition as well.
\NewEnd
\end{claimproof}

\NewStart
Conversely, suppose $E(G)=T\uplus M$, where $T$ is a spanning tree and a matching sends $X(M)$ into $Y(S)$. Fix such a matching.
\NewEnd

For each edge $xy \in M$, let $v_{xy}$ be the corresponding vertex in the edge-part. Since the matching saturates the edge-part, the vertex $v_{xy}$ is matched to exactly one of
\[
x^1,\dots,x^\ell,y^1,\dots,y^\ell.
\]

We construct a graph $H$ obtained from $G$ by splitting vertices as follows. For each vertex $x \in V(G)$, create one main copy $x_0$, and for each $i \in [\ell]$ that is used by the matching at $x$, create one additional copy $x_i$.

For every edge $xy \in T$, add the edge $x_0y_0$. For every edge $xy \in M$, if $v_{xy}$ is matched to $x^i$, then add the edge $x_i y_0$, and if it is matched to $y^i$, then add the edge $x_0 y_i$.

We claim that $H$ is a tree and that each vertex of $G$ is split at most $\ell$ times. The subgraph induced by the vertices $\{x_0 \mid x \in V(G)\}$ is exactly $(V(G),T)$, and hence it is a tree. Moreover, each vertex $x_i$ has degree at most one, since the matching uses each vertex $x^i$ at most once. Therefore $H$ is obtained from a tree by attaching leaves, and hence it is a tree.
Finally, for each vertex $x$, at most one edge is matched to each $x^i$, and $i \in [\ell]$, so at most $\ell$ additional copies of $x$ are created. Hence each vertex is split at most $\ell$ times.
\NewStart
Only vertices in $S$ receive additional copies. To realize the specified outcome, detach the assigned edges one at a time: an edge matched to $x^i$ is separated onto a new leaf descendant of $x$, while the current main descendant keeps all other incidences. Each edge is detached at exactly one endpoint. This gives an exclusive $\ell$-shallow sequence supported in $S$, with one operation for each edge of $M$.
\NewEnd

This completes the proof of the “if and only if” part of the lemma. We now proceed to prove the moreover statement.

\RevisionStart
\NewStart
The construction uses exactly $|M|=|E(G)|-|V(G)|+1$ operations. Taking $S=V(G)$ and applying \Cref{lem:pseudoarboricity_matching} gives the final assertion of the lemma.
\NewEnd
\RevisionEnd
\end{proof}

\NewStart
\begin{corollary}\label{lem:forest-prescribed-support}
Let $G$ be a graph, let $\ell\ge0$, and let $S\subseteq V(G)$.
An $\ell$-shallow splitting sequence to a forest supported in $S$ exists if and only if some basis $M$ of $\mathcal M_C(G)$ can be assigned to incident vertices in $S$, with at most $\ell$ edges assigned to each vertex.
Given such an assignment, one can construct a sequence using exactly $|M|=|E(G)|-|V(G)|+c(G)$ operations and splitting exactly the vertices that receive assigned edges.
\end{corollary}
\begin{proof}
Apply \Cref{lem:feasPart} to each original connected component with its allowed vertices in $S$. The retained trees form a spanning forest, so their complement is a cographic basis. A matching into $Y(S)$ is precisely an assignment to distinct endpoint slots. Conversely, the leaf construction in that lemma uses one operation for each assigned edge, and no operation at any other vertex. Isolated vertices require no operation.
\end{proof}

For the total-split objective, we now take $S=V(G)$. The stronger statement above will be used again when we minimize the number of split vertices.
\NewEnd

We reduce \flsvs{} to a {\em matroid intersection problem}. For a given instance $(G,k, \ell)$ of \flsvs, we construct two matroids where the first matroid $\cM_C$ is a cographic matroid (see \Cref{def:cographic}) and the second matroid is $\cM_{B}^\ell$ (see \Cref{lem:directsum}).

\begin{lemma}\label{lem:feaMat}
     Let $G$ be connected. Then $G$ admits a feasible $\ell$-shallow splitting to a forest if and only if $\mathcal{M}_C$ and $\mathcal{M}_{B}^\ell$ have a common independent set of size $|E(G)|-|V(G)|+1.$
\end{lemma}

\begin{proof}
\RevisionStart
    In the forward direction, assume that $G$ admits a feasible $\ell$-shallow splitting to a forest. By Lemma \ref{lem:feasPart}, the edge set $E(G)$ can be partitioned into two sets $T$ and $M$ such that
\medskip
\begin{enumerate}
\setlength{\itemsep}{0.1em}
    \item[(1)] $(V(G),T)$ is a spanning tree of $G$, and
    \item[(2)] $(V(G),M)$ has pseudoarboricity at most $\ell$.
\end{enumerate}
\RevisionEnd

Since $(V(G),T)$ is a spanning tree of $G$, this implies that $(V(G),T)= (V(G), E(G)\setminus M)$ is connected. Hence, $M$ is independent in the cographic matroid $\mathcal{M}_C$. Moreover, as $(V(G), M)$ has pseudoarboricity at most $\ell$, it follows that $M$ is independent in $\mathcal{M}_B^\ell$. Finally, since $E(G) = T \uplus M$ and $|T| = |V(G)| - 1$, we obtain $|M| = |E(G)| - |V(G)| + 1$. Thus, $M$ is independent in both $\mathcal{M}_C$ and $\mathcal{M}_B^\ell$, and has size $|E(G)| - |V(G)| + 1$.

Conversely, suppose there exists a set $M \subseteq E(G)$ that is independent in both $\mathcal{M}_C$ and $\mathcal{M}_B^\ell$, and satisfies $|M| = |E(G)| - |V(G)| + 1$. By the definition of these matroids, it follows that $(V(G), M)$ has pseudoarboricity at most $\ell$ and $(V(G), E(G) \setminus M)$ is connected.

Let $T := E(G) \setminus M$. Then $(V(G), T)$ is connected. Moreover, since $E(G) = T \uplus M$ and $|M| = |E(G)| - |V(G)| + 1$, we have $|T| = |E(G)| - |M| = |V(G)| - 1$. Since $(V(G), T)$ is connected and has exactly $|V(G)| - 1$ edges, it follows that $(V(G), T)$ is a spanning tree of $G$.

\RevisionStart
Thus, $E(G)$ admits a partition into $T$ and $M$ such that $(V(G), T)$ is a spanning tree and $(V(G), M)$ has pseudoarboricity at most $\ell$. By Lemma~\ref{lem:feasPart}, $G$ admits a feasible $\ell$-shallow splitting to a forest.
\RevisionEnd
\end{proof}

\RevisionStart
\begin{lemma}
\label{lem:equivalency}
        For connected $G$, $(G,k,\ell)$ is a \textsc{Yes}-instance of \flsvs{} if and only if there exists a set $I \in \mathcal{I}_C \cap \mathcal{I}_\ell$ of size $|E(G)|-|V(G)|+1$ and $|E(G)|-|V(G)|+1 \leq k$.
\end{lemma}
\RevisionEnd

\begin{proof}
\RevisionStart
In the forward direction, let $(G,k,\ell)$ be a \textsc{Yes}-instance of \flsvs. Thus $(G,k,\ell)$ has a feasible solution. Let $\mathcal{S} = (G_0 = G, G_1, \dots, G_q)$ be an arbitrary $\ell$-shallow splitting sequence such that $G_q$ is a forest and $q \le k$. Since each split operation increases the number of vertices by exactly $1$ and does not change the number of edges, it follows that $|E(G_q)| = |E(G)|$ and $|V(G_q)| = |V(G)| + q$. Hence,
$q = |V(G_q)| - |V(G)|$. As $G_q$ is a forest, we have $|E(G_q)| \leq |V(G_q)| - 1$, which implies
$|V(G_q)| \geq |E(G_q)| + 1 = |E(G)| + 1$.
Therefore,
$q = |V(G_q)| - |V(G)| \geq |E(G)| + 1 - |V(G)|$. This implies $|E(G)| + 1 - |V(G)| \le k$. Since $(G,k,\ell)$ has a feasible solution, by Lemma~\ref{lem:feaMat} there exists a set $I \in \mathcal{I}_C \cap \mathcal{I}_\ell$ of size $|E(G)|-|V(G)|+1$.
\RevisionEnd

\RevisionStart
In the reverse direction, assume there exists a set $M \in \mathcal{I}_C \cap \mathcal{I}_\ell$ of size $|E(G)|-|V(G)|+1$ and $|E(G)|-|V(G)|+1 \leq k$. By Lemma \ref{lem:feaMat}, $(G,k,\ell)$ of \flsvs \ has a feasible solution. By Lemma \ref{lem:feasPart}, there exists an $\ell$-shallow splitting sequence $(G_0=G, G_1, \dots, G_p)$ such that $G_p$ is a forest and $p \leq |E(G)| - |V(G)| + 1$. By assumption $|E(G)| - |V(G)| + 1 \leq k$, we have $p \leq k$. Hence $(G,k,\ell)$ is a yes-instance.
\RevisionEnd
\end{proof}

\RevisionStart
The resulting decision procedure first handles $\ell=0$ by testing whether $G$ is already a forest. For $\ell\ge1$ and a connected input, let
\(q=|E(G)|-|V(G)|+1\).  If \(k<q\), then the instance is a \textsc{No}-instance by
\Cref{lem:equivalency}.  Otherwise compute a maximum-size common independent
set of \(\mathcal M_C\) and \(\mathcal M_B^\ell\), and accept exactly when its
size is at least \(q\).
\RevisionEnd

\begin{theorem}
\label{thm:intro-flsvs}
\flsvs{} is polynomial-time solvable.
\end{theorem}

\begin{proof}

\RevisionStart
If $\ell=0$, no split is permitted, so we test whether $G$ is a forest and return the answer. Henceforth assume $\ell\ge1$. For a disconnected input, we solve the problem independently on each connected component and sum the required numbers of splits; this is valid by the componentwise observation in \Cref{sec:prelims}.  Thus it remains to justify the connected case.
\RevisionEnd

Let $q \coloneqq |E(G)| - |V(G)| + 1$. First observe that if $k < q$, then by \Cref{lem:equivalency}, $(G,k,\ell)$ is a \textsc{No} instance.

\RevisionStart
Now suppose that $k \ge q$. By \Cref{lem:equivalency}, it suffices to test whether there exists a set $I \in \mathcal{I}_C \cap \mathcal{I}_\ell$ of size $q$. We construct the matroids $\mathcal{M}_C$ and $\mathcal{M}_B^\ell$, and compute
\[
r \coloneqq \max\{\, |I| : I \in \mathcal{I}_C \cap \mathcal{I}_\ell \,\}
\]
by matroid intersection.
\RevisionEnd

If $r \ge q$, then by \Cref{lem:exactsize}, there exists a common independent set in $\mathcal{M}_C$ and $\mathcal{M}_B^\ell$ of size exactly $q$, and hence \Cref{lem:equivalency} gives a \textsc{Yes}-instance. If $r < q$, then no common independent set of size $q$ exists; by \Cref{lem:feaMat}, $G$ does not admit an $\ell$-shallow splitting sequence to a forest, and the instance is a \textsc{No}-instance.

\RevisionStart
\medskip
\noindent
\textbf{\underline{Running Time:}} Write $n=|V(G)|$ and $m=|E(G)|$. Finding the connected components and testing membership in the target class takes $O(n+m)$ time. For $\ell\ge1$, \Cref{lem:timeformatroid} computes the required common independent set on a component with $n_i$ vertices and $m_i$ edges in $O(n_i+\ell m_i^3)$ time, using its graph representation. Since $\sum_i n_i=n$ and $\sum_i m_i^3\le m^3$, the total running time is $O(n+\ell m^3)$. Including the $\ell=0$ case, the bound is $O(n+m+\ell m^3)$.
This yields a polynomial-time algorithm for \flsvs.
\RevisionEnd
\end{proof}

\subsection{Minimizing the number of split vertices}
We now use the characterization with a prescribed set of allowed split vertices to optimize the second objective.

\begin{tcolorbox}[colback=anti-flashwhite,colframe=black!35]
\noindent\textcolor{red!50!black}{\lsvf}\\
{\bf Fixed parameter:} an integer $\ell\ge 0$.\\
{\bf Input:} A graph~$G$ and a \NewText{nonnegative} integer~$k$.\\
{\bf Question:} Does there exist an $\ell$-shallow splitting sequence \(\mathcal{S}\coloneqq (G_0 = G, G_1, \ldots, G_t)\) with \( |\{v \in V(G) : |\D_\mathcal{S}(v)| \geq 2\}| \leq k\), such that \(G_t\) is a forest?
\end{tcolorbox}

\NewStart
For $\ell=0$, no split is allowed, so we test whether $G$ is a forest. For $\ell=1$, every split original vertex is split exactly once. The two objectives therefore coincide, and \Cref{thm:intro-flsvs} also solves the split-vertex problem.
\NewEnd

\NewStart
\subsubsection{The case \texorpdfstring{$\ell=2$}{ell=2}}
\label{sec:forest-splitvertex-two}

We now minimize the number of original vertices that are split. Each original vertex may undergo at most two operations, but it is counted only once. We first describe the problem by edge assignments. We then define a matroid from the graph and prove that weighted matroid parity solves the assignment problem. Finally, we give a linear representation of this matroid so that we can apply a polynomial-time algorithm.

\paragraph*{From splitting to edge assignments.}
Write $n=|V(G)|$, $m=|E(G)|$, and $q=m-n+c(G)$. Recall the partition $E(G)=T\uplus M$: we retain the spanning forest $T$ and assign every edge of $M$ to one of its endpoints. An endpoint assignment is a map $\phi:M\to V(G)$ satisfying $\phi(uv)\in\{u,v\}$ for each $uv\in M$.

Assigning an edge to $v$ means detaching that edge onto a separate leaf descendant of $v$. Thus our choices must satisfy three requirements:
\begin{enumerate}
    \item $T=E(G)\setminus M$ is a spanning forest, so $|M|=q$;
    \item every edge of $M$ is assigned to one of its endpoints;
    \item at most two edges are assigned to each original vertex.
\end{enumerate}
Among such choices, we want to minimize $|\phi(M)|$, the number of vertices receiving at least one assigned edge. The following lemma explains why this is the correct objective.

\begin{lemma}
\label{lem:forest-two-assignment-objective}
Let $k\ge0$. The graph $G$ has a $2$-shallow splitting sequence to a forest that splits at most $k$ original vertices if and only if there are a partition $E(G)=T\uplus M$ and an endpoint assignment $\phi:M\to V(G)$ such that $T$ is a spanning forest,
$|\phi^{-1}(v)|\le2$ for every $v\in V(G)$, and $|\phi(M)|\le k$.
\end{lemma}
\begin{proof}
Suppose a feasible sequence splits a set $S$ of at most $k$ original vertices. Corollary~\ref{lem:forest-prescribed-support}, applied with $\ell=2$ and this set $S$, gives the required partition and an assignment using only vertices of $S$. Therefore $|\phi(M)|\le |S|\le k$.

Conversely, keep $T$ between one main copy of each original vertex. For every edge $uv\in M$ assigned to $u$, detach its incidence at $u$ onto a new leaf copy of $u$; its other endpoint remains at the main copy of $v$. Different assigned edges receive different leaf copies. This is the construction from Corollary~\ref{lem:forest-prescribed-support}. The result is a spanning forest with leaves attached, and hence a forest. It uses exactly $|\phi^{-1}(v)|\le2$ operations at $v$ and splits exactly the vertices in $\phi(M)$.
\end{proof}

Using the cographic matroid $\mathcal M_C(G)$ from Definition~\ref{def:cographic}, the first requirement is equivalent to choosing an independent set $M$ of size $q$.

We still have to choose an endpoint for each selected edge and enforce the limit of two assignments per vertex. Give each original vertex $v$ two \emph{slots} $v^1,v^2$. A slot at $v$ is a formal position for assigning at most one incident edge to $v$. Let $Y=\{v^i:v\in V(G),\,i\in[2]\}$ be the set of all slots.
These slots form the vertex-part of the $2$-incidence bipartite graph $A_G^2$ from Definition~\ref{def:ell-incidence}. Assigning edges to distinct incident slots enforces the local bound. Each assignment to a slot represents one split operation. In the resulting graph, an original vertex has one main copy and one additional leaf copy for each assigned edge.

We now define a matroid $\mathcal N$ whose elements record edge choices and slot choices. After proving that $\mathcal N$ is a matroid, we specify the pairs used in the matroid parity instance.

Besides assigning edges, we also want to record vertices that receive no assignment. For this purpose, we give each vertex an unused choice that conflicts with an assignment to either of its slots.

The construction therefore has three tasks: choose the $q$ edges to detach, assign at most two to each vertex, and reward vertices receiving none. The last task is what distinguishes the split-vertex objective from the total number of operations.

Introduce the following distinct ground elements:
\begin{itemize}
    \item For each edge $e$, each endpoint $v$ of $e$, and each $i\in[2]$, introduce $x_{e,v,i}$ and $y_{e,v,i}$. The element $x_{e,v,i}$ records the edge $e$, and $y_{e,v,i}$ records the slot $v^i$. We will later require these two elements to be selected together.
    \item For each $v\in V(G)$ and $i\in[2]$, introduce $u_{v,i}$. Selecting $u_{v,1}$ and $u_{v,2}$ together will record the unused choice at $v$.
\end{itemize}
Let $X$ be the set of all $x$-elements. For each slot $v^i$, put
\[
 W_{v,i}=\{u_{v,i}\}\cup\{y_{e,v,i}:e\text{ is incident with }v\}.
\]
This block contains all elements competing for the same slot: the assignment choices and the unused choice. Let $Z$ be the disjoint union of $X$ and all the blocks $W_{v,i}$. For $I\subseteq Z$, let $E_I$ be the set of original edges $e$ for which $I$ contains an element $x_{e,v,i}$ for some $v,i$.

\begin{definition}
\label{def:forest-two-N}
Let $\mathcal I_N$ consist of the sets $I\subseteq Z$ satisfying the following conditions:
\begin{enumerate}
    \item for each edge $e$, the set $I$ contains at most one element of the form $x_{e,v,i}$;
    \item $c(G-E_I)=c(G)$, or equivalently $E_I\in\mathcal I_C$;
    \item $|I\cap W_{v,i}|\le1$ for every $v\in V(G)$ and $i\in[2]$.
\end{enumerate}
Define $\mathcal N=(Z,\mathcal I_N)$.
\end{definition}

\begin{lemma}
\label{lem:forest-two-N-matroid}
The pair $\mathcal N=(Z,\mathcal I_N)$ is a matroid.
\end{lemma}
\begin{proof}
The empty set satisfies all three conditions. Every subset of a set in $\mathcal I_N$ satisfies the first and third conditions. It also satisfies the second, since a subset of a cographic-independent set is cographic-independent. Thus $\mathcal I_N$ is hereditary.

For exchange, let $I,J\in\mathcal I_N$ with $|I|<|J|$. Suppose first that $|I\cap X|<|J\cap X|$. The first condition gives $|E_I|=|I\cap X|$ and $|E_J|=|J\cap X|$. By exchange in $\mathcal M_C(G)$, some $e\in E_J\setminus E_I$ satisfies $E_I\cup\{e\}\in\mathcal I_C$. Add to $I$ the element $x_{e,v,i}$ that belongs to $J$. The first two conditions are preserved, and no slot element has been added, so the third is preserved as well.

Otherwise, $|I\cap X|\ge|J\cap X|$, and hence $|I\setminus X|<|J\setminus X|$. By the third condition, each of these slot sets occupies exactly as many blocks $W_{v,i}$ as it has elements. Therefore some block contains an element of $J$ and no element of $I$. Adding that element to $I$ preserves the third condition and does not change its edge elements. In either case, an element of $J\setminus I$ can be added while remaining in $\mathcal I_N$.
\end{proof}

We now construct a matroid parity instance as in Definition~\ref{def:matroid-parity}, using the matroid $\mathcal N$ and the following prescribed pairs.

\begin{definition}
\label{def:forest-two-parity}
For each edge $e$, each endpoint $v$ of $e$, and each $i\in[2]$, define the \emph{assignment pair} $P_{e,v,i}=\{x_{e,v,i},y_{e,v,i}\}$. For each $v\in V(G)$, define the \emph{unused-vertex pair} $U_v=\{u_{v,1},u_{v,2}\}$.

Let $\mathcal P$ be the family consisting of the assignment pairs $P_{e,v,i}$ and the unused-vertex pairs $U_v$. These pairs partition $Z$, so $(\mathcal N,\mathcal P)$ is a matroid parity instance. A collection $\mathcal F\subseteq\mathcal P$ is feasible precisely when $\bigcup_{P\in\mathcal F}P\in\mathcal I_N$.
\end{definition}

Selecting $P_{e,v,i}$ means assigning $e$ to $v^i$: the edge element and its endpoint-slot element are selected together. This pair represents one split operation. Selecting $U_v$ takes one element from each slot block of $v$ and will record that no edge is assigned to $v$; it represents no split operation.

The matroid $\mathcal N$ controls independence of individual elements, while parity requires whole pairs. A feasible collection need not yet contain the $q$ assigned edges needed for a splitting sequence to a forest. We first prove what its pairs represent, and then choose weights to obtain an optimal splitting sequence.

\begin{lemma}
\label{lem:forest-two-pair-independence}
Let $\mathcal F$ be a collection of assignment pairs, and let $M$ be the set of edges appearing in these pairs. Then $\mathcal F$ is feasible if and only if $M$ is independent in the cographic matroid and $\mathcal F$ assigns each edge of $M$ exactly once to distinct incident slots.
\end{lemma}
\begin{proof}
Let $I=\bigcup_{P\in\mathcal F}P$.

Suppose first that $\mathcal F$ is feasible. Then $I\in\mathcal I_N$. By the first condition in Definition~\ref{def:forest-two-N}, $I$ contains at most one element $x_{e,v,i}$ for each edge $e$. Since every edge of $M$ appears in a pair of $\mathcal F$, it appears in exactly one such pair. Thus $\mathcal F$ assigns each edge of $M$ exactly once. Moreover, $E_I=M$, so the second condition gives $M\in\mathcal I_C$.

If two distinct pairs used the same slot $v^i$, their distinct $y$-elements would both belong to $I\cap W_{v,i}$, contradicting the third condition. Hence the assigned slots are distinct. Each assigned slot is incident with its edge because $P_{e,v,i}$ is defined only when $v$ is an endpoint of $e$.

Conversely, suppose that $M\in\mathcal I_C$ and that $\mathcal F$ assigns each edge of $M$ exactly once to distinct incident slots. Since each edge appears in exactly one pair, $I$ contains at most one element $x_{e,v,i}$ for each edge $e$. Also, $E_I=M\in\mathcal I_C$. Finally, distinct assigned slots imply that each block $W_{v,i}$ contains at most one selected $y$-element. As $\mathcal F$ contains only assignment pairs, $I$ contains no $u$-elements. Therefore $|I\cap W_{v,i}|\le1$ for every $v$ and $i$.

Thus $I$ satisfies all three conditions in Definition~\ref{def:forest-two-N}, so $I\in\mathcal I_N$. Hence $\mathcal F$ is feasible.
\end{proof}

Each selected pair assigns its edge to an incident slot. Since each vertex has two slots, a feasible collection assigns at most two edges to any vertex.

\begin{lemma}
\label{lem:forest-two-unused}
Let $\mathcal F$ be a feasible collection of assignment pairs, and let $v\in V(G)$. Then $\mathcal F\cup\{U_v\}$ is feasible if and only if no edge is assigned to $v$ by $\mathcal F$. Moreover, adding $U_v$ for every vertex $v$ receiving no assigned edge preserves feasibility.
\end{lemma}
\begin{proof}
Let $I=\bigcup_{P\in\mathcal F}P$. Since $\mathcal F$ is feasible, $I\in\mathcal I_N$.

Suppose first that $\mathcal F\cup\{U_v\}$ is feasible. Then $I\cup U_v\in\mathcal I_N$. If an edge $e$ were assigned to $v$ by $\mathcal F$, then $P_{e,v,i}\in\mathcal F$ for some $i\in[2]$. Consequently, $I\cup U_v$ would contain both $y_{e,v,i}$ and $u_{v,i}$, which are distinct elements of $W_{v,i}$. This contradicts the third condition in Definition~\ref{def:forest-two-N}. Hence no edge is assigned to $v$.

Conversely, suppose that no edge is assigned to $v$ by $\mathcal F$. Since $\mathcal F$ contains only assignment pairs, $I\cap W_{v,1}=I\cap W_{v,2}=\emptyset$. Adding $U_v=\{u_{v,1},u_{v,2}\}$ therefore puts exactly one element into each of these two blocks and does not change any other slot block. Thus the third condition in Definition~\ref{def:forest-two-N} remains satisfied. No edge element is added, so the first two conditions also remain satisfied. Therefore $I\cup U_v\in\mathcal I_N$, proving that $\mathcal F\cup\{U_v\}$ is feasible.

Finally, the slot blocks belonging to distinct vertices are disjoint. Adding $U_v$ for every vertex receiving no assigned edge therefore puts one element into each of their previously empty slot blocks. All other slot blocks and all edge elements remain unchanged. The three conditions defining $\mathcal I_N$ are consequently preserved, proving the final assertion.
\end{proof}

A feasible collection therefore records a cographic-independent partial edge assignment, together with unused-vertex choices only at vertices receiving no assigned edge.

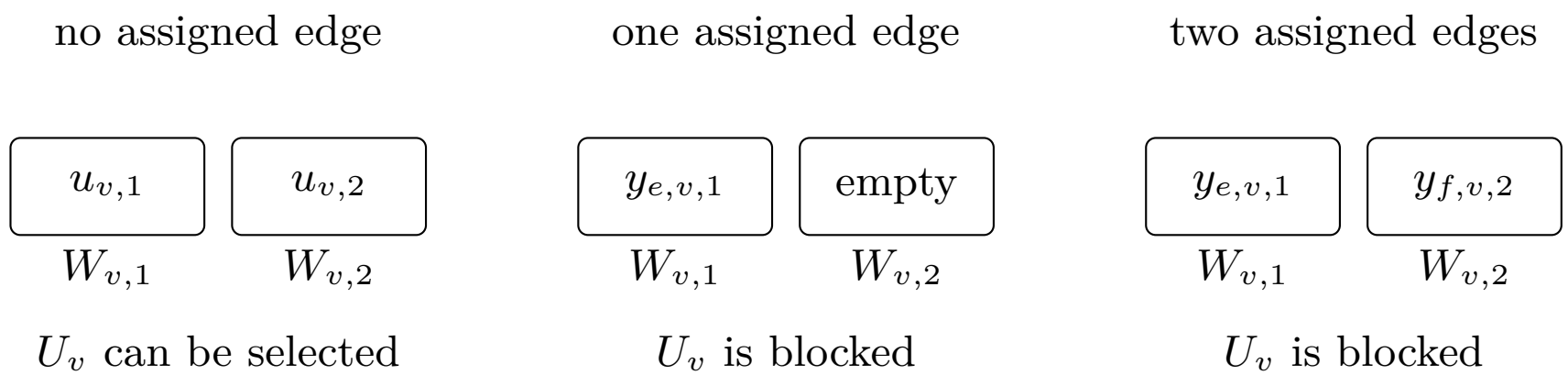
\begin{figure}[htbp]
\centering
\begingroup
\begin{tikzpicture}[font=\small,slot/.style={draw=black,rounded corners=2pt,minimum width=1.4cm,minimum height=.7cm}]
\foreach \xx/\head/\left/\right/\foot in {
0/{no assigned edge}/{$u_{v,1}$}/{$u_{v,2}$}/{$U_v$ can be selected},
4.1/{one assigned edge}/{$y_{e,v,1}$}/{empty}/{$U_v$ is blocked},
8.2/{two assigned edges}/{$y_{e,v,1}$}/{$y_{f,v,2}$}/{$U_v$ is blocked}}
{
 \begin{scope}[xshift=\xx cm]
 \node at (.8,1.1) {\head};
 \node[slot] at (0,0) {\left};
 \node[slot] at (1.6,0) {\right};
 \node at (0,-.6) {$W_{v,1}$};
 \node at (1.6,-.6) {$W_{v,2}$};
 \node at (.8,-1.2) {\foot};
 \end{scope}
}
\end{tikzpicture}
\caption{Each slot block accepts at most one element. An assignment pair selects a slot element together with its edge element (not shown). The pair $U_v$ occupies both slots and therefore records a vertex with no assignment.}
\label{fig:parity-slots}
\endgroup
\end{figure}

\paragraph*{Choosing the weights.}
We must first obtain $q$ assignments, since fewer assignments do not provide the required complement of a spanning forest. Among solutions with $q$ assignments, we want as many vertices as possible to remain unsplit. Set $w(P_{e,v,i})=n+1$ and $w(U_v)=1$. This gives the weighted matroid parity instance $(\mathcal N,\mathcal P,w)$, whose objective is to find a feasible collection of maximum total pair weight.

For a feasible collection $\mathcal F$, let $p(\mathcal F)$ be its number of assignment pairs and $h(\mathcal F)$ its number of unused-vertex pairs. Its weight is
\[
 w(\mathcal F)=(n+1)p(\mathcal F)+h(\mathcal F).
\]

Figure~\ref{fig:parity-slots} explains the unused-vertex choice. With two slots, reserving all slots at an unused vertex is itself a pair. This is the reason ordinary matroid parity fits $\ell=2$; the hardness proof for larger bounds is separate. In the example of Figure~\ref{fig:two-costs}, $n=5$ and $q=2$. Assigning both detached edges to $v$ permits four unused-vertex pairs and has weight $6\cdot2+4=16$. Assigning them to $a$ and $c$ permits three and has weight $15$. Thus the weights distinguish the two split-vertex costs even though the operation counts agree.

\begin{lemma}
\label{lem:forest-two-weights}
Let $\mathcal F^\star$ be a maximum-weight feasible collection. Then $p(\mathcal F^\star)\ge p(\mathcal F)$ for every feasible collection $\mathcal F$. Moreover, if $p(\mathcal F)=p(\mathcal F^\star)$, then $h(\mathcal F^\star)\ge h(\mathcal F)$.
\end{lemma}
\begin{proof}
Let $\mathcal F$ be any feasible collection. Since there are $n$ unused-vertex pairs, $0\le h(\mathcal F)\le n$, and the same bounds hold for $h(\mathcal F^\star)$.

Suppose that $p(\mathcal F)>p(\mathcal F^\star)$. These counts are integers, so $p(\mathcal F)-p(\mathcal F^\star)\ge1$. Therefore,
\[
\begin{aligned}
w(\mathcal F)-w(\mathcal F^\star)
&=(n+1)\bigl(p(\mathcal F)-p(\mathcal F^\star)\bigr)
  +h(\mathcal F)-h(\mathcal F^\star)\\
&\ge(n+1)-n=1.
\end{aligned}
\]
This contradicts the maximality of $w(\mathcal F^\star)$. Hence $p(\mathcal F^\star)\ge p(\mathcal F)$.

Now suppose that $p(\mathcal F)=p(\mathcal F^\star)$. Since $w(\mathcal F^\star)\ge w(\mathcal F)$, cancelling the equal assignment-weight terms gives $h(\mathcal F^\star)\ge h(\mathcal F)$.
\end{proof}

\begin{lemma}
\label{lem:forest-two-feasibility}
Let $\mathcal F^\star$ be a maximum-weight feasible collection. Then $p(\mathcal F^\star)\le q$. Moreover, $G$ admits a $2$-shallow splitting sequence to a forest if and only if $p(\mathcal F^\star)=q$.
\end{lemma}
\begin{proof}
For any feasible collection $\mathcal F$, let $\mathcal A\subseteq\mathcal F$ consist of its assignment pairs. Since $\bigcup_{P\in\mathcal A}P\subseteq\bigcup_{P\in\mathcal F}P$, the collection $\mathcal A$ is feasible by the hereditary property of $\mathcal N$. Lemma~\ref{lem:forest-two-pair-independence} therefore implies that $\mathcal A$ selects $|\mathcal A|$ distinct edges forming an independent set in the cographic matroid. This matroid has rank $q$, so $p(\mathcal F)=|\mathcal A|\le q$. In particular, $p(\mathcal F^\star)\le q$.

\emph{Forward direction.} Suppose that $G$ admits a $2$-shallow splitting sequence to a forest. By Lemma~\ref{lem:forest-two-assignment-objective}, there exist a partition $E(G)=T\uplus M$ and an endpoint assignment $\phi:M\to V(G)$ such that $T$ is a spanning forest and $|\phi^{-1}(v)|\le2$ for every $v\in V(G)$. Thus $M$ is a cographic basis and $|M|=q$.

For each vertex $v$, assign the edges of $\phi^{-1}(v)$ to distinct slots among $v^1,v^2$. Select the pair $P_{e,v,i}$ whenever edge $e$ is assigned to slot $v^i$. By Lemma~\ref{lem:forest-two-pair-independence}, this gives a feasible collection with $q$ assignment pairs. Lemma~\ref{lem:forest-two-weights} implies that $\mathcal F^\star$ has the maximum possible number of assignment pairs. Hence $p(\mathcal F^\star)\ge q$. Together with $p(\mathcal F^\star)\le q$, this yields $p(\mathcal F^\star)=q$.

\emph{Backward direction.} Suppose that $p(\mathcal F^\star)=q$. By Lemma~\ref{lem:forest-two-pair-independence}, the assignment pairs of $\mathcal F^\star$ select a cographic-independent set $M$ of $q$ edges and assign each edge exactly once to distinct incident slots. Since the cographic matroid has rank $q$, $M$ is a basis. Consequently, $T=E(G)\setminus M$ is a spanning forest.

Define $\phi(e)=v$ whenever $P_{e,v,i}\in\mathcal F^\star$. The uniqueness of the selected pair for each edge makes $\phi$ well-defined, and $v$ is an endpoint of $e$. Since the assigned slots are distinct and each vertex has two slots, $|\phi^{-1}(v)|\le2$ for every vertex $v$. Lemma~\ref{lem:forest-two-assignment-objective} therefore gives a $2$-shallow splitting sequence to a forest.
\end{proof}

The preceding lemmas explain the assignments and their weights. We now show how a maximum-weight parity solution determines an optimal splitting sequence. This argument uses only the independent sets of $\mathcal N$.

\begin{lemma}
\label{lem:forest-two-objective}
Let $\mathcal F^\star$ be a maximum-weight feasible collection with $p(\mathcal F^\star)=q$. Then the minimum number of original vertices that must be split in a $2$-shallow splitting sequence to a forest is $n-h(\mathcal F^\star)$. A sequence attaining this minimum can be constructed from $\mathcal F^\star$ in polynomial time.
\end{lemma}
\begin{proof}
Let $M$ be the set of edges selected by the assignment pairs of $\mathcal F^\star$, and define $\phi(e)=v$ whenever $P_{e,v,i}\in\mathcal F^\star$. By Lemma~\ref{lem:forest-two-pair-independence}, $\phi$ is a well-defined endpoint assignment, $M$ is cographic-independent, and $|\phi^{-1}(v)|\le2$ for every vertex $v$. Since $|M|=p(\mathcal F^\star)=q$, the set $M$ is a cographic basis. Hence $T=E(G)\setminus M$ is a spanning forest. Put $S=\phi(M)$.

We first show that $h(\mathcal F^\star)=n-|S|$. By Lemma~\ref{lem:forest-two-unused}, no unused-vertex pair $U_v$ with $v\in S$ can belong to $\mathcal F^\star$. The same lemma shows that the assignment pairs can be combined with all pairs $U_v$ for $v\notin S$. If any such pair were missing from $\mathcal F^\star$, adding it would preserve feasibility and increase the weight by $1$, contradicting maximality. Thus the unused-vertex pairs in $\mathcal F^\star$ are exactly those indexed by $V(G)\setminus S$, proving the equality.

Apply the construction in Lemma~\ref{lem:forest-two-assignment-objective} to $T$, $M$, and $\phi$. It produces a $2$-shallow splitting sequence to a forest that splits exactly the vertices of $S$. Its cost is therefore $|S|=n-h(\mathcal F^\star)$. The sets and assignment can be read from the selected pairs, and the construction uses $q\le m$ operations, so it takes polynomial time.

To prove optimality, consider any $2$-shallow splitting sequence to a forest that splits $r$ original vertices. By Lemma~\ref{lem:forest-two-assignment-objective}, there is an endpoint assignment of $q$ complementary edges using at most $r$ vertices. Assign these edges to distinct incident slots and let $\mathcal F'$ consist of the corresponding assignment pairs together with the unused-vertex pairs at every vertex receiving no assignment. Lemmas~\ref{lem:forest-two-pair-independence} and~\ref{lem:forest-two-unused} imply that $\mathcal F'$ is feasible. Moreover, $p(\mathcal F')=q$ and $h(\mathcal F')\ge n-r$.

Since $p(\mathcal F')=p(\mathcal F^\star)$, Lemma~\ref{lem:forest-two-weights} gives $h(\mathcal F^\star)\ge h(\mathcal F')\ge n-r$. Therefore $n-h(\mathcal F^\star)\le r$. The constructed sequence attains this bound and is consequently optimal.
\end{proof}

\paragraph*{A linear representation.}
The reduction above is stated entirely in terms of the matroid $\mathcal N$. To compute its maximum-weight parity solution, we next show that $\mathcal N$ is linear and construct a matrix representing it.

\begin{lemma}
\label{lem:forest-two-representation}
A rational matrix $B$ representing $\mathcal N$ can be constructed in polynomial time. It has $q+2n$ rows, $8m+2n$ columns, and polynomial binary encoding length.
\end{lemma}
\begin{proof}
Use the standard polynomial-time construction of a rational representation of the cographic matroid~\cite{schrijver2003combinatorial}. Let $A$ be the resulting $q\times m$ matrix, with column $a_e$ for $e\in E(G)$. Thus $M\in\mathcal I_C$ exactly when the columns $(a_e)_{e\in M}$ are linearly independent. The matrix $A$ has polynomial binary encoding length.

For each slot $v^i$, let $b_{v,i}\in\mathbb Q^{2n}$ be the unit vector for that slot. Define $B$ with columns indexed by $Z$, using the following vectors in $\mathbb Q^q\oplus\mathbb Q^{2n}$:
\[
 c_{x_{e,v,i}}=(a_e,0),\qquad
 c_{y_{e,v,i}}=(0,b_{v,i}),\qquad
 c_{u_{v,i}}=(0,b_{v,i}).
\]
The first coordinate block records the edges; the second records the slots. Different ground elements remain different columns even when their vectors are equal.

Let $I\subseteq Z$. The $x$-columns and the slot columns occupy separate coordinate blocks, so their union is independent exactly when each of these two families is independent.

The $x$-columns are independent exactly when no original edge is repeated and $E_I\in\mathcal I_C$. Indeed, two copies of an edge give equal columns and hence dependence. When no edge is repeated, their independence is precisely the condition that $A$ represents $\mathcal M_C(G)$. These are the first two conditions in Definition~\ref{def:forest-two-N}.

All elements of a slot block $W_{v,i}$ have the same unit vector, and different slot blocks have different unit vectors. Thus the slot columns are independent exactly when $|I\cap W_{v,i}|\le1$ for every slot. This is the third condition in the definition. Consequently, $B$ represents $\mathcal N$.

There are four endpoint-slot choices per edge. Each creates two columns, and the unused choices create two further columns per vertex. Hence $B$ has $8m+2n$ columns and $q+2n$ rows. Apart from copies of columns of $A$, its entries are zero or one. The construction and the resulting binary encoding length are therefore polynomial in the size of $G$.
\end{proof}

\paragraph*{The algorithm.}
We use the standard algorithm for weighted linear matroid parity as a black box: given a rational matrix, a partition of its columns into prescribed pairs, and integer pair weights, a maximum-weight collection whose union is linearly independent can be found in time polynomial in the input encoding length~\cite{IwataKobayashi2022}. This version permits any independent union of pairs; it does not require that the selected columns form a basis.

\begin{theorem}
\label{thm:intro-splitvertex-forest-two}
For $\ell=2$, there is a polynomial-time algorithm that decides whether $G$ admits an $\ell$-shallow splitting sequence to a forest and, if so, returns one minimizing the number of original vertices that are split.
\end{theorem}
\begin{proof}
The algorithm has the following steps.
\begin{enumerate}
    \item Compute $q=m-n+c(G)$. If $q=0$, return the empty sequence, since $G$ is already a forest.
    \item Construct $B$ by Lemma~\ref{lem:forest-two-representation}, together with the pair partition $\mathcal P$ from Definition~\ref{def:forest-two-parity} and the weights $w$ above. Apply weighted linear matroid parity to obtain a maximum-weight feasible collection $\mathcal F$.
    \item Read its pair counts $p(\mathcal F)$ and $h(\mathcal F)$. If $p(\mathcal F)<q$, report infeasibility. If $p(\mathcal F)=q$, return the sequence constructed in Lemma~\ref{lem:forest-two-objective}.
\end{enumerate}

For $q=0$, the empty sequence is optimal and meets every nonnegative budget $k$. For $q>0$, Lemma~\ref{lem:forest-two-feasibility} justifies the infeasibility test, and Lemma~\ref{lem:forest-two-objective} proves optimality when $p(\mathcal F)=q$. In this case, a decision instance is accepted exactly when $p(\mathcal F)=q$ and $n-h(\mathcal F)\le k$.

Lemma~\ref{lem:forest-two-representation} constructs $B$ in polynomial time and bounds its encoding length. There are $4m+n$ prescribed pairs, with weights $1$ and $n+1$. Thus the weighted linear matroid parity algorithm runs in polynomial time. Reconstruction is polynomial by Lemma~\ref{lem:forest-two-objective}.

The construction applies to disconnected graphs as stated. Isolated vertices receive no assigned edge, and a bridge cannot belong to a cographic-independent edge set. The case $q=0$, including the empty graph, is handled in the first step.
\end{proof}
\NewEnd

\NewStart
\subsubsection{Hardness for \texorpdfstring{$\ell\ge3$}{ell at least three}}
\NewEnd

\begin{lemma}
\label{lem:fvs_splitvertex_forest}
Let $G$ be a graph of maximum degree at most $4$, $k$ be a nonnegative integer, and let $\ell \ge 3$. Then $G$ has a feedback vertex set of size at most $k$ if and only if $G$ can be transformed into a forest by splitting at most $k$ vertices, where each vertex is split at most $\ell$ times.
\end{lemma}

\begin{proof}
\RevisionStart
For the forward direction, suppose that $G$ has a feedback vertex set $S$ with $|S| \le k$. Then $G-S$ is a forest. Now split each vertex $v \in S$ until every incident edge of $v$ is assigned to a distinct descendant of $v$. Since $\Delta(G)\le 4$, each vertex $v$ has degree at most $4$, and hence this requires at most $\max\{0,\deg_G(v)-1\} \le 3$ splits. As $\ell \ge 3$, this is allowed. After performing this operation for every vertex in $S$, each descendant of a vertex of $S$ has degree at most $1$. Therefore, all cycles are destroyed, and the resulting graph is a forest. Moreover, only the vertices of $S$ are split, so the number of split vertices is at most $k$.
\RevisionEnd

For the backward direction, suppose that $G$ can be transformed into a forest by splitting at most $k$ vertices, where each vertex is split at most $\ell$ times. Let $X$ be the set of vertices of $G$ that are split. Then $|X| \le k$. We claim that $X$ is a feedback vertex set of $G$. Suppose not. Then $G-X$ contains a cycle $C$. Since no vertex of $C$ is split, all vertices and edges of $C$ remain unchanged during the splitting process. Hence the same cycle is present in the final graph, contradicting the fact that the final graph is a forest. Therefore, $G-X$ is acyclic, and so $X$ is a feedback vertex set of size at most $k$.
\end{proof}

\begin{theorem}
\label{thm:intro-splitvertex-forest}
For every fixed integer $\ell \ge 3$, \lsvf{} is NP-hard.
\end{theorem}

\begin{proof}
\RevisionStart
The theorem follows directly from \Cref{lem:fvs_splitvertex_forest}, since \textsc{Feedback Vertex Set} is NP-hard even on graphs of maximum degree at most $4$~\cite[Theorem~24]{Munaro2017}.
\RevisionEnd
\end{proof}

We next turn to pseudoforests, where the dual bicircular matroid replaces the cographic matroid in the two algorithms.

\NewStart
\section{Shallow Splitting to Pseudoforests}
\label{sec:pseudoforest}
We study both cost measures for the pseudoforest target. We first give the total-split algorithm and then use its structural argument to minimize the number of split vertices when $\ell=2$. The final subsection proves hardness for the split-vertex objective when $\ell\ge3$.
\NewEnd

\subsection{Minimizing the total number of splits}
\label{sec:pseudo-total}

In this \NewText{subsection}, we study the problem of transforming a graph into \NewText{a} pseudoforest.

\begin{tcolorbox}[colback=anti-flashwhite,colframe=black!35]
\noindent\textcolor{red!50!black}{\sflsvs}\\
{\bf Fixed parameter:} an integer $\ell\ge 0$.\\
{\bf Input:} A graph~$G$ and a \NewText{nonnegative} integer $k$.\\
{\bf Question:} Does there exist an $\ell$-shallow splitting sequence $\mathcal{S}\coloneqq (G_0 = G, G_1, \ldots, G_t)$ with $t \le k$, such that $G_t$ is a pseudoforest?
\end{tcolorbox}

By the componentwise observation from \Cref{sec:prelims}, disconnected inputs are handled by solving the problem independently on each connected component and summing the costs.  A connected component that is already a pseudoforest requires no split.  Thus the structural part of this section focuses on connected components that are not pseudoforests.

We now introduce the definitions and structural lemmas that underlie the algorithm. These results will be used throughout the section to derive the polynomial-time procedure for \sflsvs.

\begin{definition}[Maximal pseudoforest]
\label{def:pseudo_maximal-pseudoforest}
A pseudoforest $H$ is called a \emph{maximal pseudoforest} if no edge can be added between two nonadjacent vertices of $H$ such that the resulting graph is again a pseudoforest.
\end{definition}

Here maximality is with respect to all missing edges on $V(H)$, not just edges available in an ambient input graph. Thus ``a maximal pseudoforest of $G$ below means a spanning subgraph of $G$ that is maximal in this sense. For a connected graph $G$ containing a cycle, these spanning subgraphs are exactly its bicircular bases: each component is unicyclic.

\RevisionStart
\begin{observation}
\label{obs:pseudo_1maximal-pseudoforest}
A pseudoforest with at least three vertices is a maximal pseudoforest if and only if each of its connected components is a pseudotree.
\end{observation}
\RevisionEnd

\begin{observation}\label{obs:pseudo_2maximal-pseudoforest}
Let $H$ be a pseudoforest with at least three vertices. Then $H$ is a maximal pseudoforest if and only if $|E(H)|=|V(H)|$.
\end{observation}

\begin{lemma}\label{lem:pseudo_minimum-pseudoforest-maximal}
Let $G$ be \NewText{connected and not a pseudoforest}, and let $\mathcal{S}=(G_0,G_1,\dots,G_k)$ be an $\ell$-shallow splitting sequence of minimum length such that $G_0=G$ and $G_k$ is a pseudoforest. Then $G_k$ is a maximal pseudoforest.
\end{lemma}

\begin{proof}
\RevisionStart
Suppose, for contradiction, that $G_k$ is not a maximal pseudoforest. Since the simple graph $G$ is not a pseudoforest, $|V(G)|\ge 4$, and hence $|V(G_k)|\ge 4$. By \Cref{obs:pseudo_1maximal-pseudoforest}, some connected component of $G_k$ is a tree. Let $T$ be such a component.
\RevisionEnd

\RevisionStart
First suppose that $G_k$ is disconnected. There is an original vertex $v$ with a descendant $v_1$ in $T$ and a descendant $v_2$ outside $T$. Otherwise identifying all descendants of each original vertex would leave $T$ separate from the rest, contradicting the connectivity of $G$. Let $H$ be obtained from $G_k$ by identifying $v_1$ and $v_2$. This joins the tree $T$ to either a tree or a pseudotree at one vertex, and therefore $H$ is a pseudoforest. By \Cref{obs:merge-descendants}, $H$ admits an $\ell$-shallow splitting sequence of length $k-1$ from $G$, contradicting the minimality of $\mathcal{S}$.
\RevisionEnd

\RevisionStart
It remains to consider the case that $G_k$ is connected. Then $G_k=T$ is a tree. Since $G$ is not a pseudoforest, $k\ge 1$, so some original vertex has two distinct descendants $v_1,v_2$ in $G_k$. By \Cref{obs:merge-descendants}, these descendants are nonadjacent and have disjoint neighborhoods. Identifying them therefore preserves every edge and decreases the number of vertices by one. The resulting graph $H$ is connected and satisfies $|E(H)|=|V(H)|$, so it is a pseudotree. Again, \Cref{obs:merge-descendants} gives an $\ell$-shallow splitting sequence of length $k-1$ from $G$ to $H$, a contradiction.
\RevisionEnd

\RevisionStart
Therefore $G_k$ is a maximal pseudoforest.
\RevisionEnd
\end{proof}

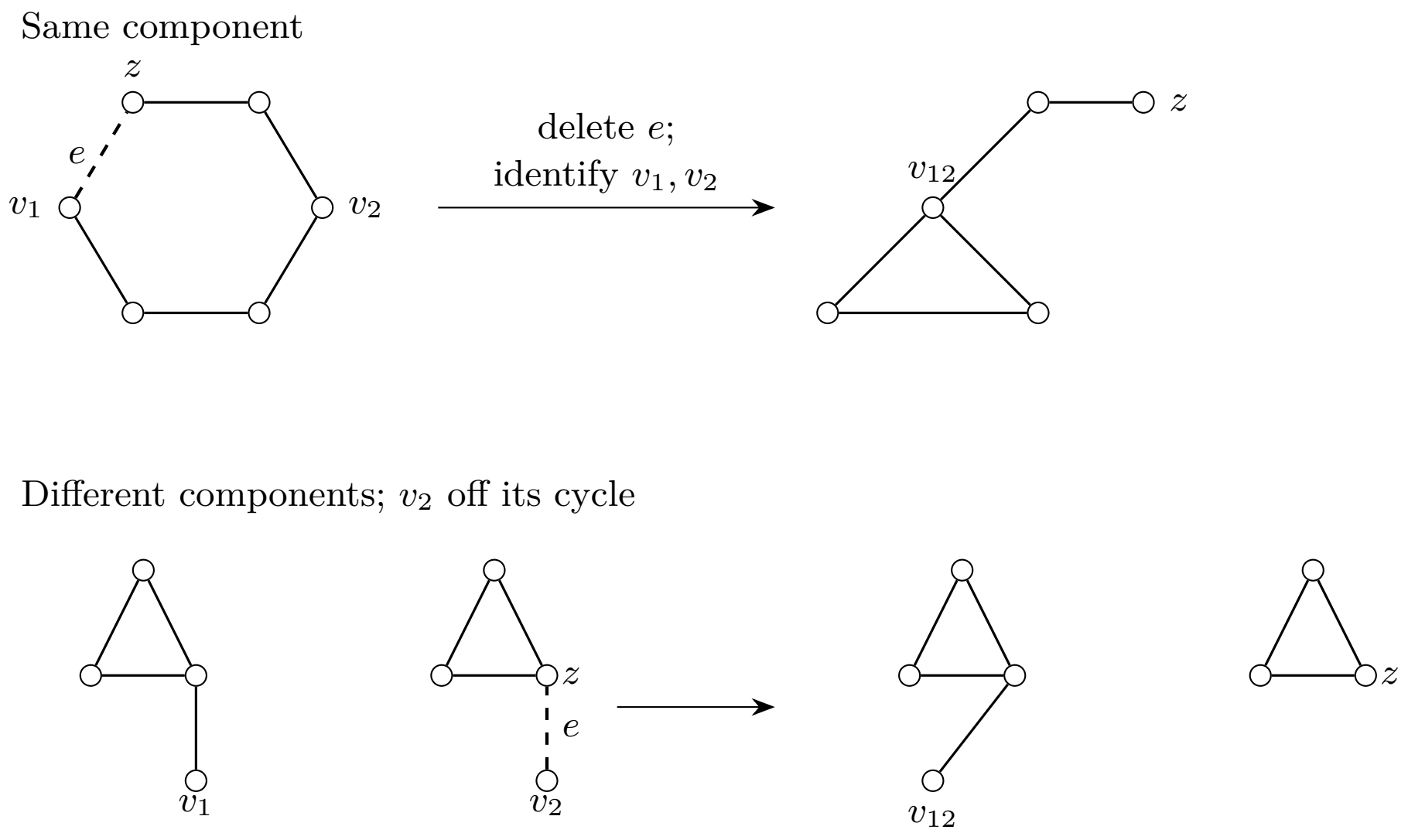
\begin{figure}[htbp]
\centering
\begingroup
\begin{tikzpicture}[x=.9cm,y=.9cm,font=\small]
\node[anchor=west] at (-1.8,1.7) {Same component};
\node[split vertex,label=left:$v_1$] (v1) at (-1.2,0) {};
\node[split vertex,label=above:$z$] (z) at (-.6,1) {};
\node[split vertex] (a) at (.6,1) {};
\node[split vertex,label=right:$v_2$] (v2) at (1.2,0) {};
\node[split vertex] (b) at (.6,-1) {};
\node[split vertex] (c) at (-.6,-1) {};
\draw[split edge] (z)--(a)--(v2)--(b)--(c)--(v1);
\draw[split marked] (v1)--node[left]{$e$}(z);
\draw[split arrow] (2.3,0)--node[above,align=center]{delete $e$;\\identify $v_1,v_2$}(5.5,0);
\node[split vertex,label=above:$v_{12}$] (w) at (7,0) {};
\node[split vertex] (bb) at (8,-1) {};
\node[split vertex] (cc) at (6,-1) {};
\node[split vertex] (aa) at (8,1) {};
\node[split vertex,label=right:$z$] (zz) at (9,1) {};
\draw[split edge] (w)--(bb)--(cc)--(w)--(aa)--(zz);
\begin{scope}[yshift=-4cm]
\node[anchor=west] at (-1.8,1.7) {Different components; $v_2$ off its cycle};
\foreach \xx/\vv in {0/v1,3/v2}{
 \begin{scope}[xshift=\xx cm]
 \node[split vertex] (\vv-a) at (-1,0) {};
 \node[split vertex] (\vv-b) at (-.5,1) {};
 \node[split vertex] (\vv-c) at (0,0) {};
 \node[split vertex] (\vv) at (0,-1) {};
 \draw[split edge] (\vv-a)--(\vv-b)--(\vv-c)--(\vv-a);
 \end{scope}}
\node[below] at (v1) {$v_1$};\node[below] at (v2) {$v_2$};
\node[right] at (v2-c) {$z$};
\draw[split edge] (v1)--(v1-c);
\draw[split marked] (v2)--node[right]{$e$}(v2-c);
\draw[split arrow] (4,-.3)--(5.5,-.3);
\foreach \xx/\vv in {7/a,10/b}{
 \begin{scope}[xshift=\xx cm]
 \node[split vertex] (\vv-a) at (-1,0) {};
 \node[split vertex] (\vv-b) at (-.5,1) {};
 \node[split vertex] (\vv-c) at (0,0) {};
 \draw[split edge] (\vv-a)--(\vv-b)--(\vv-c)--(\vv-a);
 \end{scope}}
\node[split vertex,label=below:$v_{12}$] (w2) at (7,-1) {};
\draw[split edge] (w2)--(a-c);
\node[right] at (b-c) {$z$};
\end{scope}
\end{tikzpicture}
\caption{Representative merge steps in the pseudoforest normalization. The dashed edge is deleted and recorded at the common original vertex. In the lower example, deleting the edge towards the cycle separates a tree containing $v_2$; merging that tree into the component of $v_1$ leaves both components unicyclic.}
\label{fig:pseudo-merge}
\endgroup
\end{figure}

For the pseudoforest target, a \emph{feasible splitting} is an $\ell$-shallow splitting sequence \((G_0 = G, G_1, \dots, G_s)\) such that \(G_s\) is a pseudoforest.

As for forests, we merge descendants while recording one deleted edge at each step. The invariant now requires every component to remain unicyclic. If the two clones lie in different components, the deleted edge must be chosen towards the cycle; if they lie in one component, we delete an edge on a path between them. Figure~\ref{fig:pseudo-merge} illustrates these two cases.

\begin{lemma}
\label{lem:pseudo_feasPart}
Let $G$ be connected and not already a pseudoforest. Then $G$ admits a feasible $\ell$-shallow splitting to a pseudoforest if and only if the edge set $E(G)$ can be partitioned into two sets $Q$ and $M$ such that
\medskip
\begin{enumerate}
\setlength{\itemsep}{0.1em}
    \item[(1)] $(V(G),Q)$ is a maximal pseudoforest of $G$ (and thus, $|Q|=|V(G)|$), and
    \item[(2)] $(V(G),M)$ has pseudoarboricity at most $\ell$.
\end{enumerate}
\medskip
Moreover, whenever $G$ admits such a feasible splitting, there exists an $\ell$-shallow splitting sequence $(G_0=G, G_1, \dots, G_s)$ such that $G_s$ is a pseudoforest and $s \leq |E(G)| - |V(G)|$.
\end{lemma}

\begin{proof}
    In the forward direction, suppose there exists an \(\ell\)-shallow splitting sequence \(\mathcal{S} = (G_0, G_1, \dots, G_k)\) of length \(k\) such that \(G_0 = G\) and \(G_k\) is a  pseudoforest.
    Among all such sequences, consider one of minimum length, renamed as \(\mathcal{S} = (G_0, G_1, \dots, G_p)\), where \(p \le k\).  By \Cref{lem:pseudo_minimum-pseudoforest-maximal}, we may assume that \(G_p\) is, in fact, a maximal pseudoforest.

\RevisionStart
We construct a partition \(E(G) = Q \uplus M\) by gradually reversing the split operations, which we refer to as \emph{merge operations}. Equivalently, we construct a sequence of graphs \(H_p, H_{p-1}, \dots, H_0\) iteratively in this order. For each \(i \in \{0,\ldots,p\}\) and each vertex \(v \in V(G)\), let \({\sf clone}_i(v) \subseteq V(H_i)\) denote the vertices carrying the original label \(v\).
\RevisionEnd

\RevisionStart
Initially, let \(H_p \coloneqq G_p\). For every original vertex \(u\in V(G)\), initialize \(M_u\gets\emptyset\) and \({\sf clone}_p(u)\gets\D_{\mathcal S}(u)\). Each surviving edge retains its original edge label and joins vertices carrying the labels of its original endpoints. In particular, clones of the same original vertex are nonadjacent and have disjoint neighborhoods, since \(G\) is simple and each original edge occurs at most once.
\RevisionEnd

\RevisionStart
We also maintain that \(|{\sf clone}_i(v)|\) equals the number of descendants of \(v\) in \(G_i\). This holds initially. Each reverse step merges two clones of the original vertex split in that step, reducing its clone count by one, and leaves the other counts unchanged. Thus two such clones are available whenever a merge is required. Merging vertices with the same original label preserves the edge-label property.
\NewText{The graphs $H_i$ are auxiliary graphs used to extract the partition. Their merge operations delete edges, whereas the splitting sequence constructed in the backward direction preserves every original edge.}
\RevisionEnd

For each \(i < p\), the graph \(H_i\) is obtained from \(H_{i+1}\) by a \emph{merge operation}. Throughout this process, we maintain the following invariants:
\begin{description}
\item[(1)] The graph \(H_i\) is a maximal pseudoforest.
    \item[(2)] For every \(u, v \in {\sf SplitSet}_{\mathcal{S}}(G_p)\) with \(u \neq v\), it holds that \(M_u \cap M_v = \emptyset\).

\end{description}
It is immediate that these invariants hold for the initial graph \(H_p\) in the sequence we construct.

Assume the invariants hold for $H_{i+1}$. We describe the merge producing $H_i$ and then verify that it preserves them. Suppose that $G_{i+1}$ is obtained from $G_i$ by splitting a vertex $v_i'$,
and let $v\in V(G_0)$ be the original vertex corresponding to $v_i'$. Then
$|{\sf clone}_{i+1}(v)|\geq 2$. Choose any two vertices
$v_1,v_2\in {\sf clone}_{i+1}(v)$.

\textbf{Case 1.} If $v_1$ and $v_2$ belong to different components of $H_{i+1}$, let $C_1$ and $C_2$ be the connected components of $H_{i+1}$ containing $v_1$ and $v_2$, respectively. Since $H_{i+1}$ is a maximal pseudoforest, both $C_1$ and $C_2$ are pseudotrees. We work in the component $C_2$.

If $v_2$ does not lie on the unique cycle of $C_2$, then there is a unique path from $v_2$ to that cycle. Let $v_2z$ be the unique edge incident with $v_2$ that lies on this path. If $v_2$ lies on the unique cycle of $C_2$, then let $v_2z$ be one of the two cycle edges incident with $v_2$. In either case, we delete the edge $v_2z$, and add the corresponding original edge $vz'\in E(G)$ to $M_v$.

\textbf{Case 2.} If $v_1$ and $v_2$ belong to the same component of $H_{i+1}$, choose any simple path $P$ between
$v_1$ and $v_2$ in $H_{i+1}$. Let $z$ and $y$ be the neighbors of $v_1$ and
$v_2$, respectively, on $P$.

We first delete one of the two edges $v_1z$ and $v_2y$, say without loss of
generality $v_1z$, and add the corresponding original edge $vz'\in E(G)$ to $M_v$.

After that, in both cases, we identify $v_1$ and $v_2$ into a single vertex,
denoted by $v_{12}$. Thus $H_i$ is obtained from $H_{i+1}$ by replacing the
two vertices $v_1$ and $v_2$ with the single vertex $v_{12}$.

In both cases,
$V(H_i)=\bigl(V(H_{i+1})\setminus\{v_1,v_2\}\bigr)\cup\{v_{12}\}.$

\RevisionStart
In both cases, the vertex $v_{12}$ is adjacent to precisely those vertices that were adjacent in
$H_{i+1}$ to $v_1$ or $v_2$, except that the deleted edge ($v_2z$ in Case 1 and $v_1z$ in Case 2) is not
retained. Formally,
\[
\begin{aligned}
E(H_i)={}&\bigl(E(H_{i+1})\setminus\{wu\in E(H_{i+1})\mid u\in\{v_1,v_2\}\}\bigr)\\
&\cup\{wv_{12}\mid w\in N_{H_{i+1}}(v_1)\cup N_{H_{i+1}}(v_2),\ w\neq z\}.
\end{aligned}
\]
\RevisionEnd

We also update the clone sets by setting
${\sf clone}_i(v)=\bigl({\sf clone}_{i+1}(v)\setminus\{v_1,v_2\}\bigr)\cup\{v_{12}\},$
and
${\sf clone}_i(w)={\sf clone}_{i+1}(w)\text{ for every }w\neq v.$

\RevisionStart
We next verify that this operation preserves the invariants.
\RevisionEnd

\begin{claim}
$H_i$ is a maximal pseudoforest.
\end{claim}

\begin{claimproof}
We distinguish the two cases from the definition of the merge operation.

\medskip
\noindent\textbf{Case 1.} Suppose that $v_1$ and $v_2$ belong to different connected components of $H_{i+1}$. Let $C_1$ and $C_2$ be the connected components of $H_{i+1}$ containing $v_1$ and $v_2$, respectively. Since $H_{i+1}$ is a maximal pseudoforest, both $C_1$ and $C_2$ are pseudotrees.

\smallskip
\noindent\textit{Subcase 1.1.} Suppose that $v_2$ does not lie on the unique cycle of $C_2$. Then, by the definition of the merge operation, the deleted edge $e$ is the unique edge incident with $v_2$ that lies on the path from $v_2$ to the cycle of $C_2$. Hence $C_2-e$ has exactly two connected components: a tree $T_2$ containing $v_2$, and a pseudotree $C_2'$ containing the unique cycle of $C_2$. After identifying $v_1$ and $v_2$ into $v_{12}$, the component $C_1$ and the tree $T_2$ merge into a connected component $C$. Since $|E(C_1)|=|V(C_1)| \quad\text{and}\quad |E(T_2)|=|V(T_2)|-1$, and the identification decreases the number of vertices by one, we get
$|E(C)|=|V(C)|$.
Thus $C$ is a pseudotree. The component $C_2'$ remains a pseudotree, and all other components are unchanged. Hence every connected component of $H_i$ is a pseudotree.

\smallskip
\noindent\textit{Subcase 1.2.} Suppose that $v_2$ lies on the unique cycle of $C_2$. Then, by the definition of the merge operation, the deleted edge $e$ is a cycle edge incident with $v_2$. Hence $C_2-e$ is a tree. After identifying $v_1$ and $v_2$ into $v_{12}$, the component $C_1$ and the tree $C_2-e$ merge into a connected component $C$. Since
$|E(C_1)|=|V(C_1)| \quad\text{and}\quad |E(C_2-e)|=|V(C_2-e)|-1,$
and the identification decreases the number of vertices by one, we again obtain
$|E(C)|=|V(C)|$.
Thus $C$ is a pseudotree. All other components are unchanged, and hence every connected component of $H_i$ is a pseudotree.

\medskip
\noindent\textbf{Case 2.} Suppose that $v_1$ and $v_2$ belong to the same connected component of $H_{i+1}$. Let $C$ be that component. Since $H_{i+1}$ is a maximal pseudoforest, the component $C$ is a pseudotree.

In this case, the deleted edge $e$ lies on a simple path between $v_1$ and $v_2$. Consider the graph $C-e$.

If $C-e$ is connected, then identifying $v_1$ and $v_2$ into the single vertex $v_{12}$ preserves connectivity. Hence the resulting component, say $C'$, is connected. Since exactly one edge is deleted and exactly one vertex is lost in the identification, we have
$|E(C')|=|E(C)|-1 \quad\text{and}\quad |V(C')|=|V(C)|-1.$
As $C$ is a pseudotree, $|E(C)|=|V(C)|$, and therefore $|E(C')|=|V(C')|$. Hence $C'$ is a pseudotree.

Now assume that $C-e$ is disconnected. Since the edge $e$ lies on the chosen path between $v_1$ and $v_2$, the vertices $v_1$ and $v_2$ belong to different connected components of $C-e$. After identifying $v_1$ and $v_2$ into the single vertex $v_{12}$, these components become joined, and thus the resulting component $C'$ is connected. Again, exactly one edge is deleted and exactly one vertex is lost, so
$|E(C')|=|E(C)|-1 \quad\text{and}\quad |V(C')|=|V(C)|-1.$
Since $|E(C)|=|V(C)|$, it follows that $|E(C')|=|V(C')|$. Therefore $C'$ is a pseudotree.

All other connected components of $H_{i+1}$ remain unchanged, and hence are pseudotrees. Therefore every connected component of $H_i$ is a pseudotree.

In both cases, every connected component of $H_i$ is a pseudotree. Hence $H_i$ is a maximal pseudoforest.
\end{claimproof}

\begin{claim}\label{clm:pseudopairdisj}
For every \(u, v \in {\sf SplitSet}_{\mathcal{S}}(G_p)\) with \(u \neq v\), it holds that \(M_u \cap M_v = \emptyset\).
\end{claim}

\begin{claimproof}
\RevisionStart
Initially, \(M_u=\emptyset\) for every \(u \in {\sf SplitSet}_{\mathcal{S}}(G_p)\), so the claim holds. Now consider any merge operation for a vertex \(v \in {\sf SplitSet}_{\mathcal S}(G_p)\). In this step, exactly one edge is deleted from \(H_{i+1}\), and the corresponding edge of \(G\) is added only to \(M_v\). No other set \(M_u\), where \(u \neq v\), is changed. Moreover, an edge deleted in this step cannot already belong to some other set \(M_u\), since every edge is deleted from \(H_{i+1}\) at most once. Therefore, after this step, the sets remain pairwise disjoint. Hence, throughout the construction, for every \(u,v \in {\sf SplitSet}_{\mathcal{S}}(G_p)\) with \(u \neq v\), we have \(M_u \cap M_v=\emptyset\).
\RevisionEnd
\end{claimproof}

We define
$M \coloneqq \bigcup_{{v \in \sf SplitSet}_{\mathcal{S}}(G_p)}M_v.$

\RevisionStart
Clearly, \(M\) is the disjoint union of the sets \(M_v\) for \(v \in {\sf SplitSet}_{\mathcal{S}}(G_p)\). Let $Q$ be the set of corresponding original edges for the edge set of the final maximal pseudoforest $H_0$. By construction, every edge of $G$ is either deleted at some stage and added to $M$, or remains in the final maximal pseudoforest $H_0$.  Hence $E(G)$ is partitioned into two sets, that is $E(G)=Q \uplus M$. Since the final graph is a maximal pseudoforest, it follows that $(V(G),Q)$ is a maximal pseudoforest of $G$.
\RevisionEnd

For each vertex \(v \in V(G)\), recall that \({\sf clone}_i(v)\) denotes the set of vertices in \(H_i\) corresponding to \(v\). Initially, \(|{\sf clone}_p(v)| \geq 1\), and each merge operation is applied only when \(|{\sf clone}_{i+1}(v)| \geq 2\), reducing the size of the clone set by one. Hence, for every \(i\), we have \(|{\sf clone}_i(v)| \geq 1\), and in particular \(|{\sf clone}_0(v)| \geq 1\).

Moreover, since the sequence \(H_p,H_{p-1},\dots,H_0\) is obtained by reversing the splitting sequence \(\mathcal{S}\), the graph \(H_0\) has exactly \(|V(G)|\) vertices. The sets \(\{{\sf clone}_0(v) \mid v \in V(G)\}\) form a partition of \(V(H_0)\). Since each of these sets is nonempty and their union has size \(|V(G)|\), it follows that \(|{\sf clone}_0(v)|=1\) for every \(v \in V(G)\).

\RevisionStart
Thus, \(V(H_0)\) is in bijection with \(V(G)\). Since \(H_0\) is a maximal pseudoforest, the corresponding edge set \(Q\) forms a maximal pseudoforest of \(G\).
\RevisionEnd

It remains to show that $(V(G),M)$ has pseudoarboricity at most $\ell$.

\begin{claim}
The graph $(V(G),M)$ has pseudoarboricity at most $\ell$.
\end{claim}

\begin{claimproof}
Fix a vertex $v \in {\sf SplitSet}_\mathcal{S}(G_p)$. Since $\mathcal{S}$ is an $\ell$-shallow splitting sequence, we have $|\NewText{\D_{\mathcal S}(v)}| \le \ell+1$. At the beginning of the reverse process, ${\sf clone}_p(v)=\NewText{\D_{\mathcal S}(v)}$, while at the end ${\sf clone}_0(v)$ consists of a single vertex. Moreover, whenever the reverse step from $H_{i+1}$ to $H_i$ involves two vertices of ${\sf clone}_{i+1}(v)$, the size of the clone set of $v$ decreases by exactly one, that is, $|{\sf clone}_i(v)|=|{\sf clone}_{i+1}(v)|-1$. In every other reverse step, the clone set of $v$ remains unchanged. Hence the number of reverse steps involving clones of $v$ is exactly $|{\sf clone}_p(v)|-|{\sf clone}_0(v)|=|\NewText{\D_{\mathcal S}(v)}|-1$. By construction, each such reverse step contributes exactly one edge to $M_v$. Therefore, $|M_v|=|\NewText{\D_{\mathcal S}(v)}|-1 \le \ell$.

\RevisionStart
For each vertex $v \in V(G)\setminus {\sf SplitSet}_\mathcal{S}(G_p)$, the set $M_v$ remains empty. Thus, for every vertex $v \in V(G)$, the set $M_v$ is defined and satisfies $|M_v|\le \ell$. Further, by \Cref{clm:pseudopairdisj}, the family $\{M_v \mid v \in V(G)\}$ is pairwise disjoint. Since every edge added to $M$ during the reverse process is added to exactly one set $M_v$, and $M$ is precisely the set of all such edges, it follows that $M=\biguplus_{v\in V(G)} M_v$.
\RevisionEnd

\RevisionStart
Now, for each vertex $v \in V(G)$, fix an arbitrary ordering $M_v=\{e_v^1,\dots,e_v^{t_v}\}$, where $t_v=|M_v|\le \ell$. Consider the $\ell$-incidence bipartite graph $A_{(V(G),M)}^\ell$. For each vertex $v \in V(G)$ and each index $j \in [t_v]$, if $e_v^j=xy$, then match the vertex $v_{xy}$ in the edge-part to the vertex $v^j$ in the vertex-part. The chosen pair is an edge of the incidence graph because every edge in $M_v$ was deleted at a clone of $v$ and is therefore incident with $v$ in $G$. This defines a matching saturating the edge-part. Indeed, since the sets $M_v$ form a partition of $M$, every edge of $M$ appears exactly once among the sets $M_v$, and hence every vertex in the edge-part is matched exactly once. Also, for a fixed vertex $v$, the vertices $v^1,\dots,v^\ell$ are pairwise distinct, and since $t_v\le \ell$, no two edges in $M_v$ are matched to the same vertex in the vertex-part. Therefore, no two vertices in the edge-part are matched to the same vertex in the vertex-part.
\RevisionEnd

Hence $A_{(V(G),M)}^\ell$ admits a matching saturating its edge-part. By \Cref{lem:pseudoarboricity_matching}, it follows that $(V(G),M)$ has pseudoarboricity at most $\ell$.
\end{claimproof}

Conversely, assume that the edge set $E(G)$ is partitioned into two sets $Q$ and $M$ such that $(V(G),Q)$ is a maximal pseudoforest of $G$ and $(V(G),M)$ has pseudoarboricity at most $\ell$. By \Cref{lem:pseudoarboricity_matching}, the $\ell$-incidence bipartite graph of $(V(G),M)$ admits a matching saturating its edge-part. Fix such a matching.

For each edge $xy \in M$, let $v_{xy}$ be the corresponding vertex in the edge-part. Since the matching saturates the edge-part, the vertex $v_{xy}$ is matched to exactly one of
\[
x^1,\dots,x^\ell,y^1,\dots,y^\ell.
\]

We construct a graph $H$ obtained from $G$ by splitting vertices as follows. For each vertex $x \in V(G)$, create one main copy $x_0$, and for each $i \in [\ell]$ that is used by the matching at $x$, create one additional copy $x_i$.

For every edge $xy \in Q$, add the edge $x_0y_0$. For every edge $xy \in M$, if $v_{xy}$ is matched to $x^i$, then add the edge $x_i y_0$, and if it is matched to $y^i$, then add the edge $x_0 y_i$.

\RevisionStart
We claim that $H$ is a pseudoforest and that each vertex of $G$ is split at most $\ell$ times. The subgraph induced by the vertices $\{x_0 \mid x \in V(G)\}$ is exactly $(V(G),Q)$, and hence it is a maximal pseudoforest. Moreover, each vertex $x_i$ has degree at most one, since the matching uses each vertex $x^i$ at most once. Therefore $H$ is obtained from a maximal pseudoforest by attaching leaves, and hence it is a maximal pseudoforest.
Finally, for each vertex $x$, at most one edge is matched to each $x^i$, and $i \in [\ell]$, so at most $\ell$ additional copies of $x$ are created. Hence each vertex is split at most $\ell$ times.
Therefore, $G$ can be transformed into a pseudoforest by splitting each vertex at most $\ell$ times.
\RevisionEnd

This completes the proof of the “if and only if” part of the lemma. We now proceed to prove the moreover statement.

\RevisionStart
Assume that $G$ admits a feasible $\ell$-shallow splitting to a pseudoforest.
Then the edge set \(E(G)\) is partitioned into two sets \(Q\) and \(M\) such that \((V(G),Q)\) is a maximal pseudoforest of \(G\) and \((V(G),M)\) has pseudoarboricity at most \(\ell\). We use the vertex splits done in the backward direction of the proof. The final graph $H$ is obtained from $G$ by splitting vertices as follows.
For each vertex $x \in V(G)$, we created one main copy $x_0$, and for each $i \in [\ell]$ that is used by the matching at $x$, we created one additional copy $x_i$. The number of splits done is exactly $|M|$. Since $(V(G),Q)$ is a maximal pseudoforest and $E(G)=Q \uplus M$, we have $|M|=|E(G)|-|V(G)|$. Hence, there exists an $\ell$-shallow splitting sequence $(G_0=G, G_1, \dots, G_s)$ such that $G_s$ is a pseudoforest and $s=|E(G)| - |V(G)|$.
\RevisionEnd
\end{proof}

\RevisionStart
Let $\mathcal{M}_B^*$ denote the dual matroid of the bicircular matroid $\mathcal{M}_B$.
\RevisionEnd

\begin{lemma}
\label{lem:pseudo_feaMat}
     Let $G$ be connected and not already a pseudoforest. Then $G$ admits a feasible $\ell$-shallow splitting to a pseudoforest if and only if $\mathcal{M}_B^*$ and $\mathcal{M}_{B}^\ell$ have a common independent set of size $|E(G)|-|V(G)|$.
\end{lemma}

\begin{proof}
\RevisionStart
    In the forward direction, assume that $G$ admits a feasible $\ell$-shallow splitting to a pseudoforest. By Lemma \ref{lem:pseudo_feasPart}, the edge set $E(G)$ can be partitioned into two sets $Q$ and $M$ such that
\medskip
\begin{enumerate}
\setlength{\itemsep}{0.1em}
    \item[(1)] $(V(G),Q)$ is a maximal pseudoforest of $G$, and
    \item[(2)] $(V(G),M)$ has pseudoarboricity at most $\ell$.
\end{enumerate}
\RevisionEnd

\NewText{Since $(V(G),Q)$ is a maximal pseudoforest, $Q$ is independent in $\mathcal M_B$ and has size $|V(G)|$. No pseudoforest on $V(G)$ has more edges, so $Q$ is a bicircular basis. Its complement $E(G)\setminus Q=M$ is therefore a basis of $\mathcal M_B^*$, and in particular is independent in this matroid.} Moreover, as $(V(G), M)$ has pseudoarboricity at most $\ell$, it follows that $M$ is independent in $\mathcal{M}_B^\ell$. Thus, $M$ is independent in both $\mathcal{M}_B^*$ and $\mathcal{M}_B^\ell$.

\RevisionStart
Finally, since $E(G) = Q \uplus M$ and $|Q| = |V(G)|$, we obtain $|M| = |E(G)| - |V(G)|$.  Hence, $M$ is our desired independent set of size $|E(G)| - |V(G)|$.
\RevisionEnd

\RevisionStart
Conversely, suppose there exists a set $M \subseteq E(G)$ that is independent in both $\mathcal{M}_B^*$ and $\mathcal{M}_B^\ell$, and satisfies $|M| = |E(G)| - |V(G)|$. Define $Q=E(G)\setminus M$, so $|Q|=|V(G)|$. Since $G$ is connected and contains a cycle, its bicircular matroid has rank $|V(G)|$: a spanning tree together with one additional edge is a pseudoforest of this size, and no pseudoforest has more edges than vertices. Since $M$ is independent in the dual matroid $\mathcal M_B^*$, its complement $Q$ contains a basis of $\mathcal M_B$. As $|Q|=|V(G)|$, $Q$ itself is such a basis. Thus $(V(G),Q)$ is a pseudoforest, and \Cref{obs:pseudo_2maximal-pseudoforest} implies that it is maximal. Also, since $M$ is independent in $\mathcal{M}_B^\ell$, $(V(G),M)$ has pseudoarboricity at most $\ell$.
Thus, $E(G)$ admits a partition into $Q$ and $M$ such that $(V(G), Q)$ is a maximal pseudoforest and $(V(G), M)$ has pseudoarboricity at most $\ell$. By Lemma~\ref{lem:pseudo_feasPart}, $G$ admits a feasible $\ell$-shallow splitting to a pseudoforest.
\RevisionEnd
\end{proof}

\RevisionStart
\begin{lemma}
\label{lem:pseudo_equivalency}
        For connected $G$ that is not already a pseudoforest, $(G,k,\ell)$ is a \textsc{Yes}-instance of \sflsvs{} if and only if there exists a set $I \in \mathcal{I}_B^* \cap \mathcal{I}_\ell$ of size $|E(G)|-|V(G)|$ and $|E(G)|-|V(G)| \leq k$.
\end{lemma}
\RevisionEnd

\begin{proof}
\RevisionStart
In the forward direction, let $(G,k,\ell)$ be a \textsc{Yes}-instance of \sflsvs. Thus $(G,k,\ell)$ has a feasible solution. Let $\mathcal{S} = (G_0 = G, G_1, \dots, G_q)$ be an arbitrary $\ell$-shallow splitting sequence such that $G_q$ is a pseudoforest and $q \le k$.
Since each split operation increases the number of vertices by exactly $1$ and does not change the number of edges, it follows that $|E(G_q)| = |E(G)|$ and $|V(G_q)| = |V(G)| + q$.
Hence,
$q = |V(G_q)| - |V(G)|$. As $G_q$ is a pseudoforest, we have $|E(G_q)| \leq |V(G_q)|$, which implies
$|V(G_q)| \geq |E(G_q)| = |E(G)|$.
\RevisionEnd

\RevisionStart
Therefore,
$q = |V(G_q)| - |V(G)| \geq |E(G)|- |V(G)|$. This implies $|E(G)| - |V(G)| \leq q \le k$.
Since $(G,k,\ell)$ has a feasible solution, by Lemma~\ref{lem:pseudo_feaMat} there exists a set $I \in \mathcal{I}_B^* \cap \mathcal{I}_\ell$ of size $|E(G)|-|V(G)|$.
\RevisionEnd

\RevisionStart
In the reverse direction, assume there exists a set $M \in \mathcal{I}_B^* \cap \mathcal{I}_\ell$ of size $|E(G)|-|V(G)|$ and $|E(G)|-|V(G)|\leq k$.
By Lemma \ref{lem:pseudo_feaMat}, $(G,k,\ell)$ of \sflsvs{}  has a feasible solution. By Lemma \ref{lem:pseudo_feasPart}, there exists an $\ell$-shallow splitting sequence $(G_0=G, G_1, \dots, G_s)$ such that $G_s$ is a pseudoforest and $s \leq |E(G)| - |V(G)|$. By assumption $|E(G)| - |V(G)| \leq k$, hence we have $s\leq k$.
Therefore, $(G,k,\ell)$ is a yes-instance of \sflsvs.
\RevisionEnd
\end{proof}

\RevisionStart
The resulting decision procedure first handles $\ell=0$ by testing whether $G$ is already a pseudoforest. For $\ell\ge1$ and a connected input, it is again a matroid
intersection test. If \(G\) is already a pseudoforest, the answer is
\textsc{Yes}.  Otherwise let \(q=|E(G)|-|V(G)|\).  If \(k<q\), the answer is
\textsc{No}; if \(k\ge q\), compute a maximum-size common independent set of
\(\mathcal M_B^*\) and \(\mathcal M_B^\ell\), and accept exactly when its size
is at least \(q\).  We omit pseudocode and give the correctness and running-time
argument below.
\RevisionEnd

\begin{theorem}
\label{thm:pseudo-total}
\sflsvs{} is polynomial-time solvable.
\end{theorem}

\begin{proof}
\RevisionStart
If $\ell=0$, no split is permitted, so we test whether $G$ is a pseudoforest and return the answer. Henceforth assume $\ell\ge1$. For a disconnected input, we solve the problem independently on each connected component and sum the required numbers of splits.  Components that are already pseudoforests contribute zero.  Thus it remains to justify the connected case.
\RevisionEnd

If $G$ is already a pseudoforest, then no vertex split is required. Hence, $(G,k,\ell)$ is a \textsc{Yes}-instance for every value of $k$. Therefore, in what follows, we assume that $G$ is not already a pseudoforest.

Let $q \coloneqq |E(G)| - |V(G)|$. First observe that if $k < q$, then by \Cref{lem:pseudo_equivalency}, $(G,k,\ell)$ is a \textsc{No} instance.

\RevisionStart
Now suppose that $k \ge q$. By \Cref{lem:pseudo_equivalency}, it suffices to test whether there exists a set $I \in  \mathcal{I}_B^* \cap \mathcal{I}_\ell$ of size $q$. We construct the matroids $\mathcal{M}^*_B$ and $\mathcal{M}_B^\ell$, and compute
\[
r \coloneqq \max\{\, |I| : I \in \mathcal{I}_B^* \cap \mathcal{I}_\ell \,\}
\]
by matroid intersection.
\RevisionEnd

If $r \ge q$, then by \Cref{lem:exactsize}, there exists a common independent set in $\mathcal{M}^*_B$ and $\mathcal{M}_B^\ell$ of size exactly $q$, and hence \Cref{lem:pseudo_equivalency} gives a \textsc{Yes}-instance. If $r < q$, then no common independent set of size $q$ exists; by \Cref{lem:pseudo_feaMat}, $G$ does not admit an $\ell$-shallow splitting sequence to a pseudoforest, and the instance is a \textsc{No}-instance.

\NewStart
When a component admits a feasible sequence, the common independent set of size $q$ also gives an optimal sequence. Find a matching of its edges to incident slots as in Lemma~\ref{lem:pseudoarboricity_matching}, and use the construction in Lemma~\ref{lem:pseudo_feasPart}. The resulting sequence has exactly $q$ operations, and Lemma~\ref{lem:pseudo_equivalency} gives the matching lower bound. After ignoring isolated vertices, the incidence graph has $2\ell m$ edges, so at most $m$ augmenting-path searches find the matching in $O(\ell m^2)$ time. Listing the corresponding split operations takes $O(n+m)$ time. These steps are within the running-time bound below.
\NewEnd

\RevisionStart
\medskip
\noindent
\textbf{\underline{Running Time:}} Write $n=|V(G)|$ and $m=|E(G)|$. Finding the connected components and testing membership in the target class takes $O(n+m)$ time. For $\ell\ge1$, \Cref{lem:timeformatroid} computes the required common independent set on a component with $n_i$ vertices and $m_i$ edges in $O(n_i+\ell m_i^3)$ time, using its graph representation. Since $\sum_i n_i=n$ and $\sum_i m_i^3\le m^3$, the total running time is $O(n+\ell m^3)$. Including the $\ell=0$ case, the bound is $O(n+m+\ell m^3)$.
This yields a polynomial-time algorithm for \sflsvs.
\RevisionEnd
\end{proof}

\NewStart
\subsection{Minimizing the number of split vertices}
\label{sec:pseudo-splitvertex}

We now consider \lsvpf. The input and budget are as in the forest split-vertex problem, with a pseudoforest as the target. For $\ell=0$, no split is allowed. For $\ell=1$, each split vertex undergoes exactly one operation, so the two objectives coincide and Theorem~\ref{thm:pseudo-total} applies. We next treat $\ell=2$.

To minimize the number of split vertices, we need to keep track of which original vertices receive the edges in the partition above. We therefore extend that argument to a prescribed set $S\subseteq V(G)$ of vertices allowed to split. As in the forest section, a sequence is supported in $S$ if every original vertex that is split belongs to $S$.

Write $n=|V(G)|$ and $m=|E(G)|$, and let $t(G)$ be the number of tree components of $G$, including isolated vertices. Recall that the bicircular matroid $\mathcal M_B(G)$ has rank $n-t(G)$. Hence its dual $\mathcal M_B(G)^*=(E(G),\mathcal I_B^*)$ has rank $q=m-n+t(G)$. A bicircular basis contains every edge of each original tree component; within an original component containing a cycle, its connected components are all unicyclic. The bases of the dual are the complements of bicircular bases.

For $M\subseteq E(G)$, dual independence means that $E(G)\setminus M$ contains a bicircular basis. Equivalently, $G-M$ has the same bicircular rank as $G$. Since $G-M$ retains the entire vertex set, this gives
\begin{equation}
\label{eq:pseudo-dual-independence}
 M\in\mathcal I_B^* \quad\Longleftrightarrow\quad t(G-M)=t(G).
\end{equation}
In particular, $q=0$ exactly when $G$ is already a pseudoforest.

For $M\subseteq E(G)$, an \emph{endpoint assignment} is a map $\phi:M\to V(G)$ satisfying $\phi(uv)\in\{u,v\}$ for every $uv\in M$.

\begin{corollary}
\label{cor:pseudo-prescribed-support}
Let $G$ be a graph, let $\ell\ge0$, and let $S\subseteq V(G)$. There is an $\ell$-shallow splitting sequence to a pseudoforest supported in $S$ if and only if there are a basis $M$ of $\mathcal M_B(G)^*$ and an endpoint assignment $\phi:M\to S$ with $|\phi^{-1}(v)|\le\ell$ for every $v\in S$. Such an assignment gives a sequence of exactly $q=m-n+t(G)$ operations that splits precisely the vertices in $\phi(M)$. Whenever a feasible sequence exists, $q$ is the minimum total number of operations.
\end{corollary}
\begin{proof}
\emph{Forward direction.} Suppose that a feasible sequence supported in $S$ exists. Let $C$ be a connected component of $G$ that is not a pseudoforest. Choose a sequence $\mathcal S$ of minimum length among the feasible sequences on $C$ supported in $S\cap V(C)$.

The proof of Lemma~\ref{lem:pseudo_minimum-pseudoforest-maximal} remains valid with this restriction. Indeed, each identification in that proof decreases the number of descendants of one original vertex and leaves every other descendant count unchanged. By Observation~\ref{obs:merge-descendants}, it therefore gives a shorter $\ell$-shallow sequence still supported in $S\cap V(C)$. Thus the chosen sequence ends in a maximal pseudoforest.

Apply the merge construction in the proof of Lemma~\ref{lem:pseudo_feasPart} to this sequence. It gives a partition $E(C)=Q_C\uplus M_C$, where $(V(C),Q_C)$ is a maximal pseudoforest. The sets $M_v$ constructed there partition $M_C$, and every edge in $M_v$ is incident with $v$. Their sizes satisfy $|M_v|=|\D_{\mathcal S}(v)|-1\le\ell$. Since no vertex outside $S$ is split, $M_v=\emptyset$ for $v\notin S$. Assign each edge in $M_v$ to $v$.

For every original component that is already a pseudoforest, retain all its edges in $Q$ and assign none. The union of these retained edges and the sets $Q_C$ is a bicircular basis of $G$. Its complement $M$ is consequently a dual basis, and the assignments above give the required map $\phi:M\to S$.

\emph{Backward direction.} Let $M$ and $\phi$ satisfy the statement and put $Q=E(G)\setminus M$. Then $Q$ is a bicircular basis, so $(V(G),Q)$ is a pseudoforest. For each $v\in S$, match the edges of $\phi^{-1}(v)$ to distinct slots of $v$ in $A_G^\ell$. Apply the backward construction of Lemma~\ref{lem:pseudo_feasPart}: keep $Q$ on the main copies, and place each assigned edge on a separate leaf copy at its assigned endpoint. Attaching these leaves to $(V(G),Q)$ preserves the pseudoforest property, including on original tree components. The construction uses exactly $|\phi^{-1}(v)|\le\ell$ operations at each $v$ and splits precisely the vertices in $\phi(M)$. Since $M$ is a dual basis, its total length is $|M|=q$.

For the lower bound, let an original connected component $C$ have $n_C$ vertices and $m_C$ edges. If a sequence uses $a_C$ operations on vertices of $C$, its descendants induce a graph with $n_C+a_C$ vertices and $m_C$ edges. A pseudoforest has at most as many edges as vertices, so $a_C\ge\max\{0,m_C-n_C\}$. Summing over the original components gives $m-n+t(G)=q$: tree and unicyclic components contribute zero, and every other component contributes $m_C-n_C$. The construction attains this lower bound.
\end{proof}

By Corollary~\ref{cor:pseudo-prescribed-support}, we must choose a dual-bicircular basis $M$ and an endpoint assignment $\phi$ with at most two assigned edges per vertex, minimizing $|\phi(M)|$. We use two \emph{slots} $v^1,v^2$ at each vertex $v$, where a slot is a formal position for assigning at most one incident edge. These are the vertex-part elements of $A_G^2$. Assigning an edge to a slot at $v$ represents one operation at $v$ in the leaf construction.

The slots, pairs and weights have the same roles as in Section~\ref{sec:forest-splitvertex-two}. Only the edge condition changes: the complement must contain a bicircular basis rather than a spanning forest. The additional algorithmic ingredient is an explicit rational representation of the dual bicircular matroid, given in Lemma~\ref{lem:pseudo-two-representation}.

We use the same ground elements as the forest construction in \Cref{sec:forest-splitvertex-two}, with dual-bicircular independence for the edge choices. For each $e\in E(G)$, each endpoint $v$ of $e$, and each $i\in[2]$, introduce distinct elements $x_{e,v,i}$ and $y_{e,v,i}$. The element $x_{e,v,i}$ records edge $e$, and $y_{e,v,i}$ records slot $v^i$. For each $v\in V(G)$ and $i\in[2]$, introduce $u_{v,i}$ to reserve slot $v^i$ for the unused choice. Put
\[
 X=\{x_{e,v,i}:e\in E(G),\ v\in e,\ i\in[2]\},
 \qquad
 W_{v,i}=\{u_{v,i}\}\cup\{y_{e,v,i}:e\in E(G),\ v\in e\}.
\]
Here $v\in e$ means that $v$ is an endpoint of $e$. Let $Z$ be the disjoint union of $X$ and the blocks $W_{v,i}$. For $I\subseteq Z$, let $E_I=\{e\in E(G):x_{e,v,i}\in I\text{ for some }v,i\}$.

\begin{definition}
\label{def:pseudo-N}
Let $\mathcal I_{\mathrm P}$ consist of the sets $I\subseteq Z$ satisfying the following conditions:
\begin{enumerate}
    \item for every edge $e$, at most one element of the form $x_{e,v,i}$ belongs to $I$;
    \item $t(G-E_I)=t(G)$, or equivalently $E_I\in\mathcal I_B^*$;
    \item $|I\cap W_{v,i}|\le1$ for every $v\in V(G)$ and $i\in[2]$.
\end{enumerate}
Define $\mathcal N_{\mathrm P}=(Z,\mathcal I_{\mathrm P})$.
\end{definition}

\begin{lemma}
\label{lem:pseudo-N-matroid}
The pair $\mathcal N_{\mathrm P}=(Z,\mathcal I_{\mathrm P})$ is a matroid.
\end{lemma}
\begin{proof}
The empty set satisfies the three conditions, and they are preserved under taking subsets. For the second condition, this follows from the hereditary property of $\mathcal M_B(G)^*$ and \eqref{eq:pseudo-dual-independence}.

Let $I,J\in\mathcal I_{\mathrm P}$ with $|I|<|J|$. If $|I\cap X|<|J\cap X|$, then $|E_I|<|E_J|$ by the first condition. Exchange in $\mathcal M_B(G)^*$ gives an edge $e\in E_J\setminus E_I$ with $E_I\cup\{e\}\in\mathcal I_B^*$. Adding to $I$ the element $x_{e,v,i}$ belonging to $J$ preserves the first two conditions and changes no slot block.

Otherwise, $|I\cap X|\ge|J\cap X|$, so $|I\setminus X|<|J\setminus X|$. By the third condition, some slot block contains an element of $J$ and no element of $I$. Adding this element to $I$ preserves the third condition and leaves the edge elements unchanged. In either case, an element of $J\setminus I$ can be added while preserving independence. This proves exchange.
\end{proof}

\begin{definition}
\label{def:pseudo-parity}
For every $e\in E(G)$, endpoint $v$ of $e$, and $i\in[2]$, let $P_{e,v,i}=\{x_{e,v,i},y_{e,v,i}\}$ be an \emph{assignment pair}. For every $v\in V(G)$, let $U_v=\{u_{v,1},u_{v,2}\}$ be an \emph{unused-vertex pair}. Let $\mathcal P_{\mathrm P}$ be the family consisting of these assignment pairs and unused-vertex pairs. They partition $Z$, so $(\mathcal N_{\mathrm P},\mathcal P_{\mathrm P})$ is a matroid parity instance. A collection $\mathcal F\subseteq\mathcal P_{\mathrm P}$ is feasible if $\bigcup_{P\in\mathcal F}P\in\mathcal I_{\mathrm P}$.
\end{definition}

Parity requires an edge element and its chosen slot element to be selected together. Thus selecting $P_{e,v,i}$ assigns $e$ to $v^i$ and represents one split operation. Selecting $U_v$ occupies both slot blocks of $v$ and records that $v$ receives no assigned edge; it represents no split operation. The matroid enforces independence of the elements, while the prescribed pairs enforce these two types of choice.

\begin{lemma}
\label{lem:pseudo-pair-independence}
Let $\mathcal F$ be a collection of assignment pairs, and let $M$ be the set of edges appearing in these pairs. Then $\mathcal F$ is feasible if and only if $M\in\mathcal I_B^*$ and $\mathcal F$ assigns each edge of $M$ exactly once to distinct incident slots.
\end{lemma}
\begin{proof}
Let $I=\bigcup_{P\in\mathcal F}P$. Suppose first that $\mathcal F$ is feasible. The first condition in Definition~\ref{def:pseudo-N} prevents two selected pairs from using the same edge. Thus each edge of $M$ occurs in exactly one pair. The second condition gives $M=E_I\in\mathcal I_B^*$. The third condition prevents two selected $y$-elements from belonging to the same slot block. Hence the assigned slots are distinct, and each is incident with its edge by the definition of $P_{e,v,i}$.

Conversely, assume that $M\in\mathcal I_B^*$ and that each edge is assigned exactly once to distinct incident slots. Then $I$ contains at most one $x$-element for each edge, and $E_I=M\in\mathcal I_B^*$. Each slot block contains at most one selected $y$-element, and no $u$-element is selected. Therefore $I$ satisfies all three conditions in Definition~\ref{def:pseudo-N}, proving feasibility.
\end{proof}

The argument of Lemma~\ref{lem:forest-two-unused} applies to these slot blocks as well. More explicitly, if an assignment uses $v^i$, its $y$-element conflicts with $u_{v,i}$, so $U_v$ cannot be selected. If no assignment uses either slot of $v$, adding $U_v$ puts one element into each empty block and changes no edge element. Thus, for any feasible collection of assignment pairs, the unused-vertex pairs that can be added are exactly those at vertices receiving no assignment. They can all be added together because different vertices have disjoint slot blocks.

Give each assignment pair weight $n+1$ and each unused-vertex pair weight $1$. For a feasible collection $\mathcal F$, let $p(\mathcal F)$ and $h(\mathcal F)$ be its numbers of assignment pairs and unused-vertex pairs, respectively. Its weight is $w(\mathcal F)=(n+1)p(\mathcal F)+h(\mathcal F)$. The comparison in Lemma~\ref{lem:forest-two-weights} depends only on these weights and $0\le h(\mathcal F)\le n$, so it applies here: one additional assignment pair increases the weight by at least $(n+1)-n=1$, regardless of the unused-vertex pairs. Consequently, a maximum-weight feasible collection first maximizes $p(\mathcal F)$ and, among collections with that assignment count, maximizes $h(\mathcal F)$.

\begin{lemma}
\label{lem:pseudo-two-feasibility}
Let $\mathcal F^\star$ be a maximum-weight feasible collection. Then $p(\mathcal F^\star)\le q$. Moreover, $G$ admits a $2$-shallow splitting sequence to a pseudoforest if and only if $p(\mathcal F^\star)=q$.
\end{lemma}
\begin{proof}
The assignment pairs of any feasible collection are themselves feasible by heredity. By Lemma~\ref{lem:pseudo-pair-independence}, they select distinct edges forming an independent set of $\mathcal M_B(G)^*$, whose rank is $q$. Therefore every feasible collection has at most $q$ assignment pairs.

\emph{Forward direction.} Suppose that a $2$-shallow splitting sequence to a pseudoforest exists. Corollary~\ref{cor:pseudo-prescribed-support} gives a dual basis $M$ and an endpoint assignment with at most two edges assigned to each vertex. Assign these edges to distinct slots at their assigned endpoints. Lemma~\ref{lem:pseudo-pair-independence} gives a feasible collection of $|M|=q$ assignment pairs. The weight comparison above shows that $\mathcal F^\star$ has the maximum possible number of assignment pairs. Hence $p(\mathcal F^\star)=q$.

\emph{Backward direction.} Suppose that $p(\mathcal F^\star)=q$. By Lemma~\ref{lem:pseudo-pair-independence}, its assignment pairs select a dual-independent edge set $M$ of size $q$ and assign its edges exactly once to distinct incident slots. Thus $M$ is a dual basis. Define $\phi(e)=v$ whenever $P_{e,v,i}\in\mathcal F^\star$. This is a well-defined endpoint assignment, and the two slots at each vertex give $|\phi^{-1}(v)|\le2$. Corollary~\ref{cor:pseudo-prescribed-support} gives the required splitting sequence.
\end{proof}

\begin{lemma}
\label{lem:pseudo-two-objective}
Let $\mathcal F^\star$ be a maximum-weight feasible collection with $p(\mathcal F^\star)=q$. Then the minimum number of original vertices that must be split in a $2$-shallow splitting sequence to a pseudoforest is $n-h(\mathcal F^\star)$. A sequence attaining this minimum can be constructed from $\mathcal F^\star$ in polynomial time.
\end{lemma}
\begin{proof}
Let $M$ be the selected edge set and let $\phi(e)=v$ whenever $P_{e,v,i}\in\mathcal F^\star$. As in the backward direction of Lemma~\ref{lem:pseudo-two-feasibility}, $M$ is a dual basis and $\phi$ assigns at most two edges to each vertex. Put $S=\phi(M)$.

An unused-vertex pair can be selected only at a vertex outside $S$. Every such pair must belong to $\mathcal F^\star$: otherwise it could be added, preserving feasibility and increasing the weight by $1$. Therefore $h(\mathcal F^\star)=n-|S|$. Corollary~\ref{cor:pseudo-prescribed-support} constructs a sequence splitting exactly the vertices of $S$, with cost $n-h(\mathcal F^\star)$. Reading the assignment and performing its $q\le m$ leaf operations takes polynomial time.

To prove optimality, consider any feasible sequence that splits $r$ original vertices. Apply Corollary~\ref{cor:pseudo-prescribed-support} with its split set as $S$. This gives an assignment of a dual basis to at most $r$ vertices, with at most two assignments per vertex. Choose distinct slots for these assignments and add the unused-vertex pairs at every vertex receiving no assignment. The resulting collection $\mathcal F'$ is feasible, with $p(\mathcal F')=q$ and $h(\mathcal F')\ge n-r$. Since $p(\mathcal F')=p(\mathcal F^\star)$, maximality of the weight gives $h(\mathcal F^\star)\ge h(\mathcal F')$. Thus $n-h(\mathcal F^\star)\le r$, proving optimality.
\end{proof}

The reduction above uses only the independent sets of $\mathcal N_{\mathrm P}$. To compute a maximum-weight parity solution, we now give a rational representation. We first represent the bicircular matroid explicitly, using the standard form of a gain-graph representation~\cite{Zaslavsky2007}, and then use duality.

\begin{lemma}
\label{lem:pseudo-two-representation}
A rational matrix representing $\mathcal N_{\mathrm P}$ can be constructed in polynomial time. It has $q+2n$ rows, $8m+2n$ columns, and polynomial binary encoding length.
\end{lemma}
\begin{proof}
Orient every edge of $G$ arbitrarily and assign distinct primes $\pi_e$ to the edges. Let $D$ be the $n\times m$ matrix whose column for $e=u\!\to\!v$ is $d_e=\mathbf e_u-\pi_e\mathbf e_v$, where $\mathbf e_v$ denotes the unit vector of vertex $v$.

We show that a column set indexed by $F\subseteq E(G)$ is independent exactly when $(V(G),F)$ is a pseudoforest. Different components occupy disjoint row sets, so it suffices to consider one component. For a tree, successively removing a leaf proves independence: its row has exactly one nonzero entry, so the coefficient of its incident edge in any column dependence must be zero.

For a unicyclic component, the same elimination removes all edges outside its cycle. The remaining square cycle matrix has exactly two nonzero terms in its determinant. Their absolute values are products of disjoint subsets of the primes on the cycle, and together these subsets contain every cycle edge. These products are unequal by unique prime factorization, including when one product is empty and equals $1$. The determinant is therefore nonzero, so the cycle columns, and hence all columns of the component, are independent. Finally, a component containing at least two cycles has more edges than vertices. Its columns are dependent because they have nonzero entries only in the rows of that component. This proves that $D$ represents $\mathcal M_B(G)$.

The first $m$ primes have polynomial magnitude and can be generated by a sieve in polynomial time. Thus $D$ has polynomial encoding length. Standard rational linear algebra constructs a representation $A$ of the dual in polynomial time and with polynomial encoding length~\cite{schrijver2003combinatorial}. Explicitly, select a row basis of $D$ and reduce it, after a column permutation, to $[I_r\ K]$, where $r=n-t(G)$. Then $[-K^{\mathsf T}\ I_{m-r}]$ represents the dual; restore the original column order. This gives a $q\times m$ matrix $A$, with column $a_e$ for edge $e$. The standard bounds on rational Gaussian elimination give polynomial encoding length throughout.

For each slot $v^i$, let $b_{v,i}$ be its unit vector in $\mathbb Q^{2n}$. Define a matrix $B$ on ground set $Z$ by
\[
 c_{x_{e,v,i}}=(a_e,0),\qquad
 c_{y_{e,v,i}}=(0,b_{v,i}),\qquad
 c_{u_{v,i}}=(0,b_{v,i}).
\]
The edge columns and slot columns occupy separate coordinate blocks. The edge columns of a set $I$ are independent exactly when no edge is repeated and $E_I\in\mathcal I_B^*$. The slot columns are independent exactly when each block $W_{v,i}$ contributes at most one element. Thus column independence in $B$ is precisely the three conditions in Definition~\ref{def:pseudo-N}, and $B$ represents $\mathcal N_{\mathrm P}$.

There are four endpoint-slot choices per edge, each giving two columns, and two unused-choice columns per vertex. Hence $B$ has $8m+2n$ columns and $q+2n$ rows. Its entries are entries of $A$, zeros, or ones, so its construction time and encoding length are polynomial.
\end{proof}

\begin{theorem}
\label{thm:pseudo-two}
The problem \lsvpf can be solved in polynomial time for $\ell\le2$. For $\ell=2$, a splitting sequence minimizing the number of split vertices can be constructed in polynomial time whenever a feasible sequence exists.
\end{theorem}
\begin{proof}
The cases $\ell=0,1$ were established at the beginning of this subsection. For $\ell=2$, the algorithm has the following steps.
\begin{enumerate}
    \item Compute $q=m-n+t(G)$. If $q=0$, return the empty sequence.
    \item Construct the matrix in Lemma~\ref{lem:pseudo-two-representation}, the pairs in Definition~\ref{def:pseudo-parity}, and the weights above. Apply the polynomial-time algorithm for weighted linear matroid parity~\cite{IwataKobayashi2022} to obtain a maximum-weight feasible collection $\mathcal F^\star$. As in the forest algorithm, the parity algorithm permits any independent union of prescribed pairs.
    \item If $p(\mathcal F^\star)<q$, report infeasibility. If $p(\mathcal F^\star)=q$, return the sequence from Lemma~\ref{lem:pseudo-two-objective}.
\end{enumerate}

Lemma~\ref{lem:pseudo-two-feasibility} proves the rejection criterion, and Lemma~\ref{lem:pseudo-two-objective} proves optimality. When $p(\mathcal F^\star)=q>0$, accept the decision instance exactly when $n-h(\mathcal F^\star)\le k$. The empty-sequence case $q=0$ is accepted for every $k\ge0$.

The representation has polynomial encoding length, and there are $4m+n$ pairs with weights $1$ and $n+1$. Weighted linear matroid parity and reconstruction therefore take polynomial time. All statements apply directly to disconnected graphs, including isolated vertices.
\end{proof}

\subsection{Hardness for \texorpdfstring{$\ell\ge3$}{ell at least 3}}
\label{sec:pseudo-hardness}

We reduce from \textsc{Vertex Cover} on graphs of maximum degree at most three, which is NP-complete~\cite{GareyJohnsonStockmeyer1976}. Let $(G,k)$ be such an instance. For every $v\in V(G)$, add three new vertices $a_v,b_v,c_v$, the triangle $a_vb_vc_va_v$, and the edge $va_v$. Let $H$ be the resulting graph, and retain the budget $k$. The private vertex sets $B_v=\{v,a_v,b_v,c_v\}$ partition $V(H)$, and $\Delta(H)\le4$.

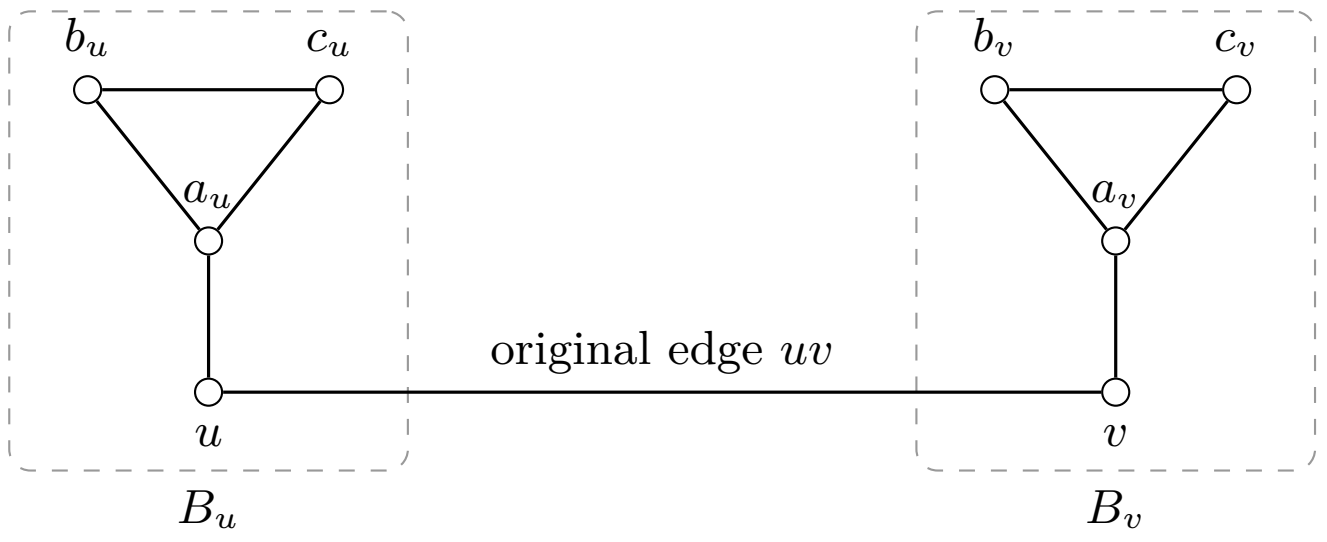
\begin{figure}[htbp]
\centering
\begingroup
\begin{tikzpicture}[font=\small]
\foreach \xx/\vv in {0/u,6/v}{
\begin{scope}[xshift=\xx cm]
 \node[split vertex,label=below:$\vv$] (\vv) at (0,0) {};
 \node[split vertex,label=above:$a_\vv$] (\vv-a) at (0,1) {};
 \node[split vertex,label=above:$b_\vv$] (\vv-b) at (-.8,2) {};
 \node[split vertex,label=above:$c_\vv$] (\vv-c) at (.8,2) {};
 \draw[split edge] (\vv)--(\vv-a)--(\vv-b)--(\vv-c)--(\vv-a);
 \node[draw=black!40,dashed,rounded corners,fit=(\vv)(\vv-a)(\vv-b)(\vv-c),inner sep=12pt,label=below:$B_\vv$] {};
\end{scope}}
\draw[split edge] (u)--node[above]{original edge $uv$}(v);
\end{tikzpicture}
\caption{An uncovered original edge joins two untouched private triangles. This connected subgraph has two cycles and cannot occur in a pseudoforest. Each original vertex has at most three original neighbors and one triangle attachment, so the maximum degree is four.}
\label{fig:pseudo-cover-gadget}
\endgroup
\end{figure}

Figure~\ref{fig:pseudo-cover-gadget} shows the obstruction used in the reverse direction. If no vertex in either block $B_u$ or $B_v$ is split, the original edge $uv$ connects two surviving cycles. Thus the blocks touched by a feasible solution must cover every original edge.

\begin{lemma}
\label{lem:pseudo-hardness-equivalence}
For every $\ell\ge3$, the graph $G$ has a vertex cover of size at most $k$ if and only if $H$ admits an $\ell$-shallow splitting sequence to a pseudoforest that splits at most $k$ original vertices.
\end{lemma}
\begin{proof}
\emph{Forward direction.} Let $C$ be a vertex cover of $G$ with $|C|\le k$. For each $v\in C$, separate its incident edges in $H$ among distinct descendants, each of degree one. This requires $\deg_H(v)-1\le3\le\ell$ operations at $v$ and no operations at other vertices. Thus at most $|C|$ original vertices are split.

Since $G-C$ is independent, each component of $H-C$ consists either of a private triangle or of a vertex of $G-C$ attached to its private triangle. Hence $H-C$ is a pseudoforest. Each new descendant has degree one. Such descendants either attach leaves to components of $H-C$ or form isolated edges with other descendants. The resulting graph is therefore a pseudoforest, proving the forward direction.

\emph{Backward direction.} Let a feasible sequence split the set $X\subseteq V(H)$, with $|X|\le k$. Put $C=\{v\in V(G):B_v\cap X\ne\emptyset\}$. Since the sets $B_v$ are disjoint, $|C|\le|X|\le k$.

Suppose that some edge $uv\in E(G)$ has neither endpoint in $C$. No vertex of $B_u\cup B_v$ is split. Therefore the two private triangles, the attachment edges $ua_u$ and $va_v$, and the edge $uv$ all survive unchanged. They form a connected subgraph containing two cycles in the final graph, contradicting that it is a pseudoforest. Thus $C$ covers every edge of $G$.
\end{proof}

\begin{theorem}
\label{thm:pseudo-hard}
For every fixed $\ell\ge3$, \lsvpf is NP-complete, even on graphs of maximum degree at most four.
\end{theorem}
\begin{proof}
The construction of $H$ is polynomial, and Lemma~\ref{lem:pseudo-hardness-equivalence} proves NP-hardness under the stated degree bound. For membership in NP, a certificate partitions the incident edges at each original vertex among at most $\ell+1$ descendants. Such partitions are realizable by successive exclusive splits, as in Observation~\ref{obs:merge-descendants}. We can verify that at most $k$ originals have more than one descendant and that the resulting graph is a pseudoforest. For fixed $\ell$, the certificate and these checks have polynomial size and running time. Thus the problem belongs to NP.
\end{proof}
\NewEnd

The next section treats cluster graphs. There, Krausz partitions describe feasible outcomes and give a different complexity threshold.

\section{Shallow Splitting to Cluster Graphs}
\label{sec:dynamic-cs}

In this section, we study the problem of transforming a graph into a cluster graph by means of shallow exclusive vertex splitting.

We treat both cost measures in this section. We first develop the Krausz-partition characterization and the total-split results, and then establish the split-vertex results using the same correspondence.

\begin{tcolorbox}[colback=anti-flashwhite,colframe=black!35]
	\noindent\textcolor{red!50!black}{\clsvs}\\
{\bf Fixed parameter:} an integer $\ell\ge 0$.\\
{\bf Input:} A graph~$G$ and a \NewText{nonnegative} integer~$k$.\\
{\bf Question:} Does there exist an $\ell$-shallow splitting sequence \(\mathcal{S}\coloneqq (G_0 = G, G_1, \ldots, G_t)\) with \(t \le k\), such that \(G_t\) is a cluster graph?
\end{tcolorbox}

We show that \clsvs{} is linear-time solvable for \(\ell \le 1\), and NP-hard for every fixed \(\ell \ge 2\).

\subsection{Krausz partitions and hardness for \texorpdfstring{\(\ell \ge 2\)}{ell at least 2}}
In this subsection, we relate \clsvs{} to the classical notion of a Krausz partition. This connection is the basis for both our hardness proof and our algorithmic results. We begin by defining the relevant objects.

\begin{definition}[Krausz \(r\)-partition and Krausz dimension]
A \emph{Krausz \(r\)-partition} of a graph \(G\) is a family of cliques of size at least \(2\) such that every edge of \(G\) belongs to exactly one clique and every vertex of \(G\) belongs to at most \(r\) cliques. When \(r=1\), a \emph{Krausz \(1\)-partition} is a family of pairwise vertex-disjoint cliques of size at least \(2\) such that every edge of \(G\) belongs to exactly one clique. The \emph{Krausz dimension} of \(G\), denoted by \(\mathrm{kz\mbox{-}dim}(G)\), is the minimum integer \(r\) such that \(G\) admits a Krausz \(r\)-partition.
\end{definition}

Krausz partitions are relevant to our setting because each clique in the final cluster graph naturally corresponds to one part of the partition, while splitting a vertex allows it to appear in several parts. The next lemma makes this correspondence precise.

\begin{lemma}
\label{lem:equiv-recognition}
Let \(n\coloneqq |V(G)|\). A graph \(G\) admits an \(\ell\)-shallow splitting sequence of length at most \(n\ell\) to a cluster graph if and only if \(G\) admits a Krausz \((\ell+1)\)-partition.
\end{lemma}

\begin{proof}
\RevisionStart
Suppose first that \(G\) admits an \(\ell\)-shallow splitting sequence \((G_0=G,G_1,\ldots,G_p)\) such that \(G_p\) is a cluster graph. Let \(Q_1,\ldots,Q_q\) be the connected components of \(G_p\) having at least two vertices. Since \(G_p\) is a cluster graph, each \(Q_i\) is a clique. For each \(i\in [q]\), define
\[
P_i \coloneqq \{v\in V(G): \text{some descendant of } v \text{ in } G_p \text{ belongs to } Q_i\}.
\]
We claim that \(\mathcal{P}\coloneqq \{P_1,\ldots,P_q\}\) is a Krausz \((\ell+1)\)-partition of \(G\).
\RevisionEnd

\RevisionStart
No two vertices of \(Q_i\) descend from the same original vertex, so \(|P_i|=|Q_i|\ge 2\). Each \(P_i\) is a clique in \(G\). Indeed, if \(x,y\in P_i\), then \(Q_i\) contains descendants \(x'\) and \(y'\) of \(x\) and \(y\), respectively. Since \(Q_i\) is a clique, \(x'y'\in E(G_p)\). As the splitting is exclusive, \(x'y'\) has the unique original edge \(xy\in E(G)\). Hence \(P_i\) is a clique.
\RevisionEnd

Also, every edge of \(G\) belongs to exactly one part of \(\mathcal{P}\). Let \(uv\in E(G)\). Under exclusive splitting, this edge gives rise to a unique descendant edge \(u'v'\in E(G_p)\). This edge belongs to exactly one connected component \(Q_i\) of \(G_p\). Thus \(u,v\in P_i\). If another part $P_j$ contained both $u$ and $v$, the clique $Q_j$ would contain a second descendant edge of $uv$, contradicting exclusivity. Hence every edge belongs to exactly one part.

\RevisionStart
Now let \(v\in V(G)\), and let \(t\) be the number of descendants of \(v\) in \(G_p\). Since descendants of the same original vertex are pairwise non-adjacent, no two of them can lie in the same clique \(Q_i\). Hence \(v\) belongs to at most \(t\) parts of \(\mathcal{P}\), that is, \(\mathrm{freq}_{\mathcal{P}}(v)\le t\); descendants that are isolated in \(G_p\) contribute no part. Since the splitting sequence is \(\ell\)-shallow, we have \(t\le \ell+1\). Therefore \(\mathcal{P}\) is a Krausz \((\ell+1)\)-partition of \(G\).
\RevisionEnd

\RevisionStart
For the converse, suppose that \(G\) admits a Krausz \((\ell+1)\)-partition \(\mathcal{P}=\{P_1,\ldots,P_m\}\). We show that \(G\) admits an \(\ell\)-shallow splitting sequence of length at most \(n\ell\) that transforms \(G\) into a cluster graph.
\RevisionEnd

\RevisionStart
For each vertex \(v\in V(G)\), let \(t=\mathrm{freq}_{\mathcal{P}}(v)\), and let \(P_{i_1},\ldots,P_{i_t}\) be the parts containing \(v\). If \(t=0\), then \(v\) is isolated, and we leave it unchanged. If \(t\ge 1\), using \(t-1\) exclusive splittings, we replace \(v\) by \(t\) descendants \(v^{i_1},\ldots,v^{i_t}\), one corresponding to each part containing \(v\). After doing this for every vertex of \(G\), we define the edges of the resulting graph \(G'\) as follows. For each edge \(uv\in E(G)\), let \(P_i\) be the unique part of \(\mathcal{P}\) containing \(uv\). We then add the edge \(u^iv^i\), where \(u^i\) and \(v^i\) are the descendants of \(u\) and \(v\) corresponding to the part \(P_i\). This construction is well-defined since every edge of \(G\) belongs to exactly one part of \(\mathcal{P}\). For each \(i\in [m]\), let
\[
Q_i\coloneqq \{v^i:v\in P_i\}.
\]
Since \(P_i\) is a clique in \(G\), every two vertices of \(Q_i\) are adjacent in \(G'\). Thus \(Q_i\) induces a clique in \(G'\). Moreover, by construction, every edge of \(G'\) has both endpoints in some \(Q_i\), and hence there are no edges between \(Q_i\) and \(Q_j\) for distinct \(i,j\). Therefore \(G'\) is a disjoint union of cliques, that is, a cluster graph.
\RevisionEnd

\RevisionStart
It remains to verify the shallowness and the size bound. For each vertex \(v\in V(G)\), the number of descendants of \(v\) is exactly \(\max\{1,\mathrm{freq}_{\mathcal{P}}(v)\}\le \ell+1\). Hence the splitting sequence is \(\ell\)-shallow.
Also, the number of splittings performed on \(v\) is \(\max\{0,\mathrm{freq}_{\mathcal{P}}(v)-1\}\). Therefore the total number of splittings is
\[
\sum_{v\in V(G)}  \max\{0,\mathrm{freq}_{\mathcal{P}}(v)-1\}\le \sum_{v\in V(G)}\ell=n\ell.
\]
\RevisionEnd

\RevisionStart
By applying this process, we obtain the desired splitting sequence.
Thus \(G\) admits an \(\ell\)-shallow splitting sequence of length at most \(n\ell\) to a cluster graph.
\RevisionEnd
\end{proof}

We now use the above equivalence to derive hardness for \(\ell \ge 2\). For this, we reduce from the following decision problem, which is known to be NP-complete for every fixed \(r \ge 3\)~\cite{DBLP:conf/wg/HlinenyK97}.

\begin{tcolorbox}[colback=anti-flashwhite,colframe=black!35,title=Krausz $r$-Partition]
{\bf Fixed parameter:} an integer $r\ge 3$.\par
{\bf Input:} a graph $G$.\par
{\bf Question:} does $G$ admit a Krausz $r$-partition?
\end{tcolorbox}

\begin{theorem}\label{thm:cluster_NP_hard}
\clsvs{} is NP-hard for every fixed \(\ell \ge 2\).
\end{theorem}

\begin{proof}
\RevisionStart
Fix \(\ell \ge 2\). Given an instance \(G\) of Krausz \((\ell+1)\)-Partition, output the instance \((G,k)\) of \clsvs{}, where $k=|V(G)|\ell$. By Lemma~\ref{lem:equiv-recognition}, this is a yes-instance if and only if \(G\) admits a Krausz \((\ell+1)\)-partition. The same equivalence holds for the empty graph, which is a yes-instance of both problems. Since deciding whether a graph admits a Krausz \(r\)-partition is NP-hard for every fixed \(r \ge 3\), this polynomial-time reduction proves the theorem.
\RevisionEnd
\end{proof}

We now turn to the complementary case \(\ell \le 1\). As we shall see, Krausz \(2\)-partitions are much more rigid, and this structural rigidity leads to a linear-time algorithm.

\subsection{Linear-time algorithms for \texorpdfstring{$\ell\le1$}{ell at most 1}}

In this subsection, we give a linear-time algorithm for \clsvs{} when \(\ell \le 1\).
Observe that when \(\ell=0\), the problem simply asks whether the input graph is a cluster graph.
Equivalently, this is the same as checking whether the graph contains no induced \(P_3\), which can be tested in $O(|V(G)|+|E(G)|)$ time by checking that every connected component is a clique.
Hence, for the rest of this subsection, we assume that \(\ell=1\).

We also assume throughout that the input graph \(G\) is connected.
If \(G\) is disconnected, then the problem can be solved independently on each connected component, and the optimum value is the sum of the optimum values over the components.

\RevisionStart
By \Cref{lem:equiv-recognition}, a graph \(G\) admits a \(1\)-shallow splitting sequence to a cluster graph if and only if \(G\) admits a Krausz \(2\)-partition.
Moreover, the sequence constructed from a Krausz \(2\)-partition in the proof of \Cref{lem:equiv-recognition} uses exactly one split for each vertex belonging to two parts of the partition.
This motivates the following notion of cost.
\RevisionEnd

\begin{definition}[Cost of a Krausz \(r\)-partition]
Let \(\mathcal{P}\) be a Krausz \(r\)-partition of a graph \(G\).
The \emph{cost} of \(\mathcal{P}\), denoted by \({\tt cost}(\mathcal{P})\), is
\[
{\tt cost}(\mathcal{P})
=
\sum_{v \in V(G)} \max\{0,\mathrm{freq}_{\mathcal{P}}(v)-1\},
\]
where
$
\mathrm{freq}_{\mathcal{P}}(v)
\coloneqq
\bigl|\{P \in \mathcal{P} : v \in V(G[P])\}\bigr|
$
denotes the number of parts of \(\mathcal{P}\) that contain \(v\).
\end{definition}

When \(\mathcal{P}\) is a Krausz \(2\)-partition, every vertex belongs to at most two parts, and therefore
\[
{\tt cost}(\mathcal{P})
=
|\{v \in V(G) : \mathrm{freq}_{\mathcal{P}}(v)=2\}|.
\]

For \(\ell=1\), it suffices to find a Krausz \(2\)-partition that minimizes ${\tt cost}(\mathcal{P})$. We now relate this optimization problem to line graphs. For a partition $\mathcal{P}$, a vertex that belongs to more than one clique is called a \textit{heavy vertex} of $\mathcal{P}$.

\begin{definition}[Line graph]
Let \(H\) be a graph.
The \emph{line graph} of \(H\), denoted by \(L(H)\), is the graph whose vertex set is \(E(H)\), where two vertices are adjacent if and only if the corresponding edges of \(H\) share an endpoint.
\end{definition}

The following characterization is classical and goes back to Krausz; see, for example,
Beineke~\cite{Beineke1970} or Hlin\v{e}n\'y and Kratochv\'il~\cite{DBLP:conf/wg/HlinenyK97}.

\begin{proposition}[Krausz characterization]
\label{prop:krauszcharacterization}
A graph \(G\) admits a Krausz \(2\)-partition if and only if \(G\) is a line graph.
Equivalently, \(\mathrm{kz\mbox{-}dim}(G)\le 2\).
\end{proposition}

The preceding proposition characterizes feasibility. To optimize the number of splits, we next study the cost of a Krausz \(2\)-partition. For this, we use the notion of a \emph{root graph}: a graph \(H\) is a root graph of a line graph \(G\) if \(L(H)\cong G\). Except for one exceptional case, a connected line graph has a unique connected root graph up to isomorphism. The following classical theorem is due to Whitney~\cite{Whitney1932}; see also Hemminger~\cite{Hemminger1972} for a short proof.

\begin{proposition}[Whitney's Isomorphism Theorem]
\label{prop:whitneytheorem}
Let \(H_1\) and \(H_2\) be connected graphs such that \(L(H_1)\cong L(H_2)\).
Then \(H_1\cong H_2\), unless \(\{H_1,H_2\}=\{K_3,K_{1,3}\}\).
\end{proposition}

For a connected graph \(H\), let
$
I(H)\coloneqq \{x\in V(H)\mid \deg_H(x)\ge 2\}.
$
We call an edge \(xy\in E(H)\) \emph{internal} if \(x,y\in I(H)\).

\begin{definition}[Krausz \(2\)-partition of \(L(H)\) defined by the stars of \(H\)]
Let \(H\) be a simple graph. For each vertex \(x\in V(H)\) with \(\deg_H(x)\ge 2\), let
\[
K_x \coloneqq \{e\in E(H): e \text{ is incident with } x\}.
\]
Viewing \(V(L(H))\) as \(E(H)\), each \(K_x\) induces a clique in \(L(H)\). The family
\[
\mathcal{K}(H)\coloneqq \{K_x : x\in V(H),\ \deg_H(x)\ge 2\}
\]
is called the \emph{Krausz \(2\)-partition of \(L(H)\) defined by the stars of \(H\)}.
\end{definition}

\begin{lemma}
\label{lem:krausz-root}
Let \(G\) be a connected graph and  \(\mathcal P=\{P_1,\dots,P_m\}\) be a Krausz \(2\)-partition of \(G\). Then there exists a connected graph \(H'\) such that \(L(H')\cong G\), and, under this identification, \(\mathcal P\) coincides with the Krausz \(2\)-partition of \(L(H')\) defined by the stars of \(H'\).
\end{lemma}

\begin{proof}
\RevisionStart
If \(G\cong K_1\), then \(\mathcal P\) is empty; taking \(H'=K_2\) proves the claim. Hence assume that \(|V(G)|\ge 2\). From \(G\) and \(\mathcal{P}\), we construct a graph \(H'\) as follows. For each part \(P_i\in \mathcal P\), create a vertex \(x_i\) in \(H'\). For each vertex \(v\in V(G)\), define
$I_v \coloneqq \{i\in [m] : v\in P_i\}.$
Since \(\mathcal{P}\) is a Krausz \(2\)-partition, every vertex of \(G\) belongs to at most two parts of \(\mathcal P\), and hence \(|I_v|\le 2\). Moreover, since \(G\) is connected, every vertex of \(G\) is incident with some edge, and every edge belongs to exactly one part of \(\mathcal P\). Therefore \(v\) belongs to at least one part, and so \(|I_v|\in\{1,2\}\) for every \(v\in V(G)\).
\RevisionEnd

Now fix \(v\in V(G)\). If \(I_v=\{i,j\}\) with \(i\neq j\), then we add the edge \(x_i x_j\) to \(H'\). This edge is well defined, and it is uniquely determined by \(v\), since for distinct \(i,j\in [m]\), the intersection \(P_i\cap P_j\) contains at most one vertex. Indeed, if two distinct vertices \(u,v\) belonged to both \(P_i\) and \(P_j\), then the edge \(uv\) would belong to both parts, contradicting the definition of a Krausz partition.

If \(I_v=\{i\}\), then we create a new vertex \(y_v\) and add the edge \(x_i y_v\) to \(H'\). Again, this edge is uniquely associated with \(v\).

Thus every vertex \(v\in V(G)\) gives rise to a unique edge \(e_v\in E(H')\). This defines a bijection
\[
\varphi:V(G)\to E(H').
\]

We claim that \(\varphi\) is an isomorphism from \(G\) to \(L(H')\). Let \(u,v\in V(G)\). Since \(\mathcal P\) is a Krausz partition, every edge of \(G\) belongs to exactly one part, and hence
\[
uv\in E(G)
\quad\Longleftrightarrow\quad
\text{there exists } i\in [m] \text{ such that } u,v\in P_i.
\]
By construction, this happens if and only if the edges \(e_u\) and \(e_v\) of \(H'\) are both incident with the vertex \(x_i\). Equivalently,
\[
uv\in E(G)
\quad\Longleftrightarrow\quad
\varphi(u)\text{ and }\varphi(v)\text{ are adjacent in }L(H').
\]
Hence \(L(H')\cong G\).

Since \(G\) is connected and \(L(H')\cong G\), the graph \(L(H')\) is connected. Therefore all edges of \(H'\) lie in a single connected component of \(H'\). Removing any isolated vertices if necessary, we may assume that \(H'\) is connected.

Finally, we show that \(\mathcal P=\{P_1,\dots,P_m\}\) coincides, under the identification \(L(H')\cong G\), with the Krausz \(2\)-partition of \(L(H')\) defined by the stars of \(H'\).  For each \(i\in[m]\), the degree of \(x_i\) in \(H'\) is exactly the number of vertices \(v\in V(G)\) with \(v\in P_i\), that is, $\deg_{H'}(x_i)=|P_i|$.
Since \(P_i\) is a clique of size at least \(2\), it follows that \(\deg_{H'}(x_i)\ge 2\). Fix \(i\in [m]\), and let
\[
S_i \coloneqq \{e\in E(H') : e \text{ is incident with } x_i\}.
\]
Since every two edges in \(S_i\) share the endpoint \(x_i\), the set \(S_i\) induces a clique in \(L(H')\). We claim that, under the bijection \(\varphi:V(G)\to E(H')\), we have $\varphi(P_i)=S_i$.

Let \(v\in V(G)\), and let \(e_v=\varphi(v)\in E(H')\). If \(v\in P_i\), then by construction either \(e_v=x_ix_j\) for some \(j\neq i\), or \(e_v=x_i y_v\). In either case, \(e_v\) is incident with \(x_i\), and hence \(e_v\in S_i\). Therefore, $v\in P_i \implies e_v\in S_i$.

Conversely, suppose that \(e_v\in S_i\). Then \(e_v\) is incident with \(x_i\). By construction, every edge of \(H'\) is of the form \(x_ax_b\) or \(x_a y_w\), and the indices appearing in this edge are exactly the parts of \(\mathcal P\) containing the corresponding vertex of \(G\). Hence \(e_v\) being incident with \(x_i\) implies that \(v\in P_i\). Thus,
$
e_v\in S_i \implies v\in P_i.
$

Therefore \(\varphi(P_i)=S_i\). Since \(S_i\) is exactly the clique of \(L(H')\) corresponding to the star centered at \(x_i\), it follows that \(P_i\) coincides with this clique under the isomorphism \(L(H')\cong G\).

Since this holds for every \(i\in [m]\), the family \(\mathcal P=\{P_1,\dots,P_m\}\) is exactly the Krausz \(2\)-partition of \(L(H')\) defined by the stars of \(H'\), viewed in \(G\) via the isomorphism \(L(H')\cong G\).
\end{proof}

\begin{lemma}
\label{lem:cost-root}
Let \(G\) be a connected graph with \(G\not\cong K_3\), and suppose that \(G\) admits a Krausz \(2\)-partition. Let \(H\) be the unique connected graph up to isomorphism such that \(L(H)\cong G\). Then, for every Krausz \(2\)-partition \(\mathcal P\) of \(G\),
\[
{\tt cost}(\mathcal P)
=
|\{xy\in E(H): x,y\in I(H)\}|.
\]
\end{lemma}

\begin{proof}
Let \(\mathcal P\) be a Krausz \(2\)-partition of \(G\). By \Cref{lem:krausz-root}, there exists a connected graph \(H'\) such that \(L(H')\cong G\) and, under this identification, \(\mathcal P\) coincides with the Krausz \(2\)-partition of \(L(H')\) defined by the stars of \(H'\). Since \(G\not\cong K_3\), \Cref{prop:whitneytheorem} implies that \(H'\cong H\).

\RevisionStart
Now let \(v\in V(G)\), and let \(e=xy\in E(H')\) be the edge corresponding to \(v\). Recall that, by definition, the Krausz \(2\)-partition of \(L(H')\) defined by the stars of \(H'\) consists of the sets
\[
K_z=\{f\in E(H') : f \text{ is incident with } z\}
\qquad\text{for } z\in V(H') \text{ with } \deg_{H'}(z)\ge 2.
\]
Since \(e=xy\), the edge \(e\) is incident only with the two vertices \(x\) and \(y\). Therefore, among all parts of the form \(K_z\), the only ones that can contain \(e\) are \(K_x\) and \(K_y\). Moreover,
\[
e\in K_x \iff \deg_{H'}(x)\ge 2,
\qquad
e\in K_y \iff \deg_{H'}(y)\ge 2.
\]
Equivalently, under the identification \(V(L(H'))=E(H')\), the vertex \(v\) belongs to the part \(K_x\) if and only if \(x\in I(H')\), and similarly \(v\) belongs to the part \(K_y\) if and only if \(y\in I(H')\). It follows that the number of parts of \(\mathcal P\) containing \(v\) is exactly the number of endpoints of \(e\) that belong to \(I(H')\). In particular:
\begin{itemize}
    \item if neither \(x\) nor \(y\) belongs to \(I(H')\), then \(v\) belongs to no part of \(\mathcal P\);
    \item if exactly one of \(x\) and \(y\) belongs to \(I(H')\), then \(v\) belongs to exactly one part of \(\mathcal P\);
    \item if both \(x\) and \(y\) belong to \(I(H')\), then \(v\) belongs to exactly two parts of \(\mathcal P\).
\end{itemize}
Hence the heavy vertices of \(\mathcal P\) are precisely the vertices of \(G\) corresponding to internal edges of \(H'\). Since \(H'\cong H\), the claimed equality follows.
\RevisionEnd
\end{proof}

\begin{figure}[htbp]
\centering
\begingroup
\begin{tikzpicture}[font=\small]
\foreach \i in {0,1,2,3}{\node[split vertex] (h\i) at (\i,0) {};}
\draw[split edge] (h0)--(h1) (h2)--(h3);
\draw[split edge,line width=1.5pt] (h1)--node[above]{$e$}(h2);
\node at (1.5,-.65) {$H=P_4$};
\draw[split arrow] (3.5,0)--node[above]{$L$}(4.4,0);
\foreach \i in {0,1,2}{\node[split vertex] (g\i) at (5+\i,0) {};}
\draw[split edge] (g0)--(g1)--(g2);
\node[above] at (g1) {$v_e$};
\node at (6,-.65) {$G=P_3$};
\draw[split arrow] (7.4,0)--node[above]{split $v_e$}(8.9,0);
\foreach \i/\xx in {0/9.4,1/10.2,2/10.8,3/11.6}{\node[split vertex] (s\i) at (\xx,0) {};}
\draw[split edge] (s0)--(s1) (s2)--(s3);
\node at (10.5,-.65) {$K_2\uplus K_2$};
\begin{scope}[yshift=-2.6cm]
 \node at (1,1.2) {exception: $G=K_3$};
 \foreach \i/\xx/\yy in {0/0/0,1/1/0,2/.5/.7}{\node[split vertex] (t\i) at (\xx,\yy) {};}
 \draw[split edge] (t0)--(t1)--(t2)--(t0);
 \node at (4.6,1.2) {root $K_3$: cost $3$};
 \foreach \i/\xx/\yy in {0/4/0,1/5/0,2/4.5/.7}{\node[split vertex] (r\i) at (\xx,\yy) {};}
 \draw[split edge,line width=1.5pt] (r0)--(r1)--(r2)--(r0);
 \node at (9,1.2) {root $K_{1,3}$: cost $0$};
 \node[split vertex] (q) at (9,.2) {};
 \foreach \i/\xx/\yy in {0/8.2/0,1/9/.7,2/9.8/0}{\node[split vertex] (q\i) at (\xx,\yy) {};\draw[split edge] (q)--(q\i);}
\end{scope}
\end{tikzpicture}
\caption{The cost is the number of internal root edges, shown thick. Each such edge corresponds to a line-graph vertex belonging to two star cliques. The triangle has two nonisomorphic connected roots, so it is handled separately.}
\label{fig:cluster-root}
\endgroup
\end{figure}
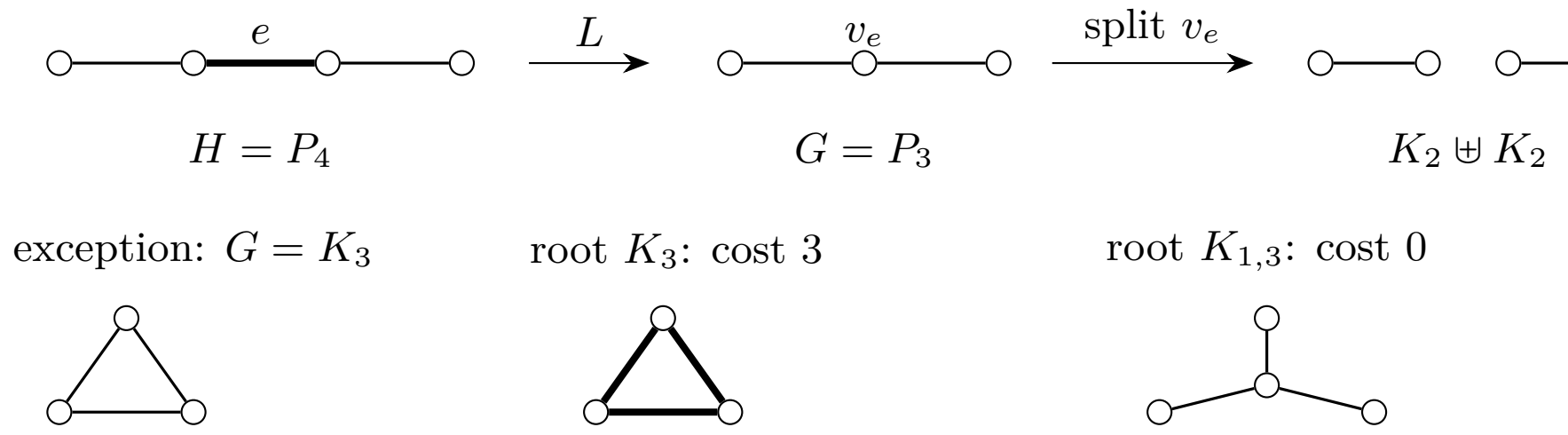

For example, the root $H=P_4$ has one internal edge. Its line graph is $P_3$, whose middle vertex must be split once to obtain two disjoint edges (Figure~\ref{fig:cluster-root}). The exception $G=K_3$ matters: the roots $K_3$ and $K_{1,3}$ have three and zero internal edges, respectively. The claw root gives the optimum zero, since $G$ is already a clique.

\RevisionStart
The following lemma relates vertex splitting to the cost of a Krausz partition. It shows that, when the graph is not already a cluster graph, one can perform a split that decreases the cost by exactly one. \NewText{Its proof uses the same splitting construction as Gaikwad et al.~\cite{GAIKWAD2026116197}, adapted to Krausz $r$-partitions.} This describes one step of the partition-based construction: separate one clique membership from the remaining memberships at a vertex. The following corollary counts the total cost directly.
\RevisionEnd

\begin{samepage}
\begin{lemma}
\label{lem:cost-drop}
Let \(G\) be a graph that is not a cluster graph and \(\mathcal{P}\) be a Krausz \(r\)-partition of \(G\) with \({\tt cost}(\mathcal{P})=k\). Then there exists a vertex \(u \in V(G)\) with \(\mathrm{freq}_{\mathcal{P}}(u)=s\ge 2\) such that, by splitting \(u\) into two vertices \(u_1\) and \(u_2\), we obtain a graph \(G'\) satisfying the following:
\begin{enumerate}
    \item \(G'\) admits a Krausz \(r\)-partition \(\mathcal{P}'\),
    \item \({\tt cost}(\mathcal{P}') = k-1\),
    \item \(\mathrm{freq}_{\mathcal{P}'}(u_1)=1\) and \(\mathrm{freq}_{\mathcal{P}'}(u_2)=s-1\).
\end{enumerate}
\end{lemma}
\end{samepage}

\begin{proof}
Let \(\mathcal{P}=\{P_1,\ldots,P_t\}\). Since \(G\) is not a cluster graph, \(\mathcal{P}\) cannot be a Krausz \(1\)-partition. Hence there exists a vertex \(u \in V(G)\) that belongs to at least two clusters of \(\mathcal{P}\). Let $u$ be the vertex such that $\mathrm{freq}_{\mathcal{P}}(u)=s\ge 2$
and, after renumbering if necessary, assume that $u \in P_1,\ldots,P_s
\quad\text{and}\quad
u \notin P_{s+1},\ldots,P_t$.

We obtain \(G'\) from \(G\) by splitting \(u\) into two vertices \(u_1\) and \(u_2\), where
$N_{G'}(u_1)\coloneqq P_1\setminus\{u\}$, and $N_{G'}(u_2)\coloneqq N_G(u)\setminus \bigl(P_1\setminus\{u\}\bigr)$,  and all adjacencies among the vertices of \(V(G)\setminus\{u\}\) remain unchanged. Now define a family \(\mathcal{P}'=\{P'_1,\ldots,P'_t\}\) by $P'_1 \coloneqq (P_1\setminus\{u\})\cup\{u_1\}$, $P'_i \coloneqq (P_i\setminus\{u\})\cup\{u_2\}
\quad\text{for } 2\le i\le s$, and $P'_i \coloneqq P_i
\quad\text{for } i>s$.

Each \(P'_i\) is a clique in \(G'\), and every edge of \(G'\) belongs to exactly one member of \(\mathcal{P}'\). Thus \(\mathcal{P}'\) is a Krausz \(r\)-partition of \(G'\). Moreover, $\mathrm{freq}_{\mathcal{P}'}(u_1)=1
\quad\text{and}\quad
\mathrm{freq}_{\mathcal{P}'}(u_2)=s-1.$

\RevisionStart
For every vertex \(v\in V(G)\setminus\{u\}\), its membership in the clusters is unchanged, so $\mathrm{freq}_{\mathcal{P}'}(v)=\mathrm{freq}_{\mathcal{P}}(v).$
The contribution $s-1$ of $u$ is replaced by contributions $0$ and $s-2$. All other contributions are unchanged, so ${\tt cost}(\mathcal P')={\tt cost}(\mathcal P)-(s-1)+(s-2)=k-1$.
This completes the proof.
\RevisionEnd
\end{proof}

\RevisionStart
\begin{corollary}
\label{cor:krauszandsplitting}
Let \(G\) be a graph and let \(k\) be a non-negative integer. Then \(G\) admits an \(\ell\)-shallow splitting sequence of length at most \(k\) to a cluster graph if and only if \(G\) admits a Krausz \((\ell+1)\)-partition \(\mathcal{P}\) with \({\tt cost}(\mathcal{P})\le k\).
\end{corollary}
\RevisionEnd

\begin{proof}
\RevisionStart
Suppose first that \(G\) admits an \(\ell\)-shallow splitting sequence of length \(t\le k\) to a cluster graph. By the proof of \Cref{lem:equiv-recognition}, the nonsingleton components of the final graph yield a Krausz \((\ell+1)\)-partition \(\mathcal{P}\) of \(G\). For each \(v\in V(G)\), let \(d(v)\) be its number of final descendants. Then \(\mathrm{freq}_{\mathcal{P}}(v)\le d(v)\), and each split increases the number of descendants of exactly one original vertex by one. Therefore
\[
{\tt cost}(\mathcal{P})
=\sum_{v\in V(G)}\max\{0,\mathrm{freq}_{\mathcal{P}}(v)-1\}
\le\sum_{v\in V(G)}(d(v)-1)=t\le k.
\]
\RevisionEnd

\RevisionStart
Conversely, suppose that \(G\) admits a Krausz \((\ell+1)\)-partition \(\mathcal{P}\) with \({\tt cost}(\mathcal{P})\le k\). The construction in the proof of \Cref{lem:equiv-recognition} leaves isolated vertices unchanged and gives each other vertex one descendant for each part containing it. It produces a cluster graph using exactly \({\tt cost}(\mathcal{P})\le k\) splits. Every original vertex has at most \(\ell+1\) descendants, so the sequence is \(\ell\)-shallow.
\RevisionEnd
\end{proof}

\begin{theorem}
\label{thm:cluster-l1-poly}
There is an $O(|V(G)|+|E(G)|)$-time algorithm for \clsvs{} when $\ell=1$. It either reports infeasibility or computes the optimum number of splits.
\end{theorem}

\begin{proof}
Let \(G\) be a connected graph.
We first test whether \(G\) is a line graph and, if so, reconstruct a connected graph \(H\) such that \(L(H)\cong G\).
This can be done in linear time~\cite{Lehot1974,Roussopoulos1973}.
If \(G\) is not a line graph, then by \Cref{prop:krauszcharacterization}, \(G\) does not admit a Krausz \(2\)-partition.
Hence, by \Cref{lem:equiv-recognition}, \(G\) does not admit a \(1\)-shallow splitting sequence to a cluster graph.
Therefore, the answer is \textsc{No}.

\RevisionStart
Assume now that \(G\) is a line graph.
If \(G\cong K_3\), then \(G\) is already a clique, and hence already a cluster graph.
Thus we set \(s=0\), which is the optimum value.
\RevisionEnd

So assume that \(G\not\cong K_3\).
Let \(H\) be the unique connected graph up to isomorphism such that \(L(H)\cong G\), and define
$s \coloneqq |\{xy\in E(H)\mid x,y\in I(H)\}|$.
We claim that \(s\) is the optimum number of splits.

\RevisionStart
Let \(\mathcal{P}\) be any Krausz \(2\)-partition of \(G\).
By \Cref{lem:cost-root}, we have ${\tt cost}(\mathcal{P})=s$.
Hence \(G\) admits a Krausz \(2\)-partition of cost \(s\). By \Cref{cor:krauszandsplitting}, it follows that \(G\) admits a \(1\)-shallow splitting sequence of length at most \(s\). Therefore the optimum number of splits is at most \(s\).
\RevisionEnd

\RevisionStart
Conversely, suppose that \(G\) admits a \(1\)-shallow splitting sequence of length \(t\) to a cluster graph.
By \Cref{cor:krauszandsplitting}, \(G\) admits a Krausz \(2\)-partition \(\mathcal{P}'\) with
${\tt cost}(\mathcal{P}')\le t$.
Applying \Cref{lem:cost-root} to \(\mathcal{P}'\), we obtain
$s={\tt cost}(\mathcal{P}')\le t$.
Thus every feasible solution has size at least \(s\). Since we have already shown that a solution of size at most \(s\) exists, it follows that \(s\) is the optimum number of splits.
\RevisionEnd

Hence, given an instance \((G,k)\) of \clsvs{} with \(\ell=1\), we answer \textsc{Yes} if and only if \(G\) is a line graph and the above value \(s\) satisfies \(s\le k\).
For a disconnected graph, process each connected component and add its optimum; if any component is infeasible, so is $G$. Isolated vertices contribute zero. The componentwise running times sum to $O(|V(G)|+|E(G)|)$, including for the empty graph.
\end{proof}

\subsection{Minimizing the number of split vertices}
For $\ell=0$, recognition suffices. For $\ell=1$, the two objectives coincide, so the preceding linear-time algorithm also minimizes the number of split vertices. For $\ell\ge2$, the Krausz correspondence yields hardness for this objective as well.

\begin{lemma}
\label{lem:equiv-recognitionset}
Let \(n\coloneqq |V(G)|\). A graph \(G\) admits an \(\ell\)-shallow splitting sequence to a cluster graph in which at most \(n\) original vertices are split if and only if \(G\) admits a Krausz \((\ell+1)\)-partition.
\end{lemma}

\begin{proof}
The bound of $n$ on split original vertices is automatic. Every $\ell$-shallow sequence on $n$ original vertices has length at most $n\ell$, because each original vertex and its descendants account for at most $\ell$ operations. The statement therefore follows directly from Lemma~\ref{lem:equiv-recognition}.
\end{proof}

\begin{theorem}
\lsvc{} is NP-hard for every fixed \(\ell \ge 2\).
\end{theorem}

\begin{proof}
\RevisionStart
Fix \(\ell \ge 2\). By Lemma~\ref{lem:equiv-recognitionset}, a graph \(G\) admits an \(\ell\)-shallow splitting sequence to a cluster graph using at most \(|V(G)|\) split original vertices if and only if \(G\) admits a Krausz \((\ell+1)\)-partition. Since deciding whether a graph admits a Krausz \(r\)-partition is NP-hard~\cite{DBLP:conf/wg/HlinenyK97} for every fixed \(r \ge 3\), it follows by taking \(r=\ell+1\) and setting the split-vertex budget to $|V(G)|$ that \lsvc{} is NP-hard for every fixed \(\ell \ge 2\).
\RevisionEnd
\end{proof}

This completes the results for the three target classes. We conclude by discussing the classification and further directions.

\par\medskip
\noindent\textbf{AI disclosure.}
We used GPT-6 to improve the writing and presentation of this paper. The proofs are due to the authors. The authors reviewed and edited the resulting text and take full responsibility for the paper's content, accuracy, and originality.

\bibliography{\ManuscriptPath bibliography}

\end{document}